\documentclass[a4paper,11pt]{article}
\usepackage[margin=0.8in]{geometry}

\usepackage{graphicx} 
\usepackage{float}
\usepackage{epstopdf}

\usepackage{amssymb,amsmath}

\usepackage[colon,authoryear]{natbib}
\pdfoutput=1

\begin{document}

\title{Origin and continuation of 3/2, 5/2, 3/1, 4/1 and 5/1 resonant periodic orbits in the circular and elliptic restricted three-body problem}
 \author{Kyriaki I. Antoniadou and Anne-Sophie Libert\vspace{0.25cm}	\\
				\small{NaXys, Department of Mathematics, University of Namur, 8 Rempart de la Vierge, 5000 Namur, Belgium} \\  \small{kyriaki.antoniadou@unamur.be}} 

\date{}
\maketitle

\begin{center}
\large{
Accepted for publication in Celestial Mechanics and Dynamical Astronomy. The final authenticated version is available online at: http://dx.doi.org/10.1007/s10569-018-9834-8}
\end{center}

\begin{abstract}
We consider a planetary system consisting of two primaries, namely a star and a giant planet, and a massless secondary, say a terrestrial planet or an asteroid, which moves under their gravitational attraction. We study the dynamics of this system in the framework of the circular and elliptic restricted three-body problem, when the motion of the giant planet describes circular and elliptic orbits, respectively. Originating from the circular family, families of symmetric periodic orbits in the 3/2, 5/2, 3/1, 4/1 and 5/1 mean-motion resonances are continued in the circular and the elliptic problems. New bifurcation points from the circular to the elliptic problem are found for each of the above resonances and thus, new families, continued from these points are herein presented. Stable segments of periodic orbits were found at high eccentricity values of the already known families considered as whole unstable previously. Moreover, new isolated (not continued from bifurcation points) families are computed in the elliptic restricted problem. The majority of the new families mainly consist of stable periodic orbits at high eccentricities. The families of the 5/1 resonance are investigated for the first time in the restricted three-body problems. We highlight the effect of stable periodic orbits on the formation of stable regions in their vicinity and unveil the boundaries of such domains in phase space by computing maps of dynamical stability. The long-term stable evolution of the terrestrial planets or asteroids is dependent on the existence of regular domains in their dynamical neighbourhood in phase space, which could host them for long time spans. This study, besides other celestial architectures that can be efficiently modelled by the circular and elliptic restricted problems, is particularly appropriate for the discovery of terrestrial companions among the single-giant planet systems discovered so far.
\end{abstract}

{\bf keywords} Periodic orbits -- Circular family -- Mean-motion resonance  -- Orbital stability -- Three-body problem -- High eccentricity

\section{Motivation}
The restricted three-body problem (RTBP) has long been employed as a model to provide hints for the dynamics of a secondary body of negligible mass moving under the gravitational attraction of the two primaries. In particular, some problems whose dynamics can be tackled with such a modelling include systems like star-planet-asteroid, planet-satellite-spacecraft and binary stars-planet. 

Over the past decades, a tremendous increase in the number of discovered exoplanets\footnote{The reader may refer to {\em exoplanet.eu} \citep{enc} and {\em exoplanets.org} \citep{exorg}} has been accomplished by ground-based telescopes or missions. This, in turn, triggered a targeted quest for ``exo-Earths'', whose habitability has, among others \citep[see e.g.][]{kast,kop14}, the long-term orbital stability as a prerequisite. 

Several works have studied the stability of planetary systems with hypothetical terrestrial planets. In particular, \citet{sandor07} realized, with the use of chaotic indicators, a catalogue of maps of dynamical stability (DS-maps) of hypothetical terrestrial planets in the habitable zone of 15 known exoplanetary systems with one giant. They explored the phase space projected on planes where the semi-major axis of the terrestrial and the eccentricity of the giant varied. They took into account various values of the mass parameter and the giant evolved on elliptic orbits taking eccentricity values up to 0.5 starting either from the pericentre or the apocentre. The hypothetical terrestrial planet evolved only on circular orbits and started always from the pericentre. \citet{erdi12} explored the phase space of a similar planetary configuration, but restricted the motion of the giant (Jupiter or Neptune) to circular orbits. Higher order internal and external resonances were studied by DS-maps in the vicinity of the exact value of the resonance. The massless body evolved on orbits with all possible eccentricity values and the initial positions of the giant were restricted to specific values.

Incited by this ongoing exploration, \citet{kiaasl} have recently studied the dynamics of a system consisting of a star, an inner terrestrial planet and an outer giant planet in coplanar configurations and provided hints for possible existence and survival of terrestrial planets in 93 single-giant planet systems, based on the extensive computation of DS-maps. These maps have been realized for diverse eccentricity values of the terrestrial planet (0.02, 0.1, 0.3 and 0.5) and all possible values for the eccentricity of the giant planet. The configurations of the giant and the terrestrial planets, starting both from pericentre and apocentre, have been exhaustively explored. The maps have revealed additional regions of stability for the terrestrial planet. These domains are associated with families of stable periodic orbits. In particular, regarding the 2/1 mean-motion resonance (MMR), new families of symmetric and asymmetric periodic orbits in the elliptic problem were presented therein.

Resonant periodic orbits can be of particular importance in the dynamics of our Solar System, too, as the invariant manifolds could be applied directly to the trajectory design process for missions, such as those to Europa \citep[see e.g.][]{davis,broad}. The generation and continuation of families of periodic orbits has already provided hints for the underlying dynamics of many problems. For instance, the unperturbed planar case has been studied by \citet{Bruno1994}. \citet{broucke1968} studied the RTBP with Earth-Moon masses, while \citet{kot05}, \citet{kotvoy05} and \citet{vk05} studied the exterior resonances with Neptune for the planar and spatial RTBP. Recently, \citet{vta} studied inclined asymmetric periodic solutions in exterior MMRs in the Sun--Neptune--trans-Neptunian object system.

In the present work, we aim to justify the existence of islands of stability revealed in the neighbourhood of the 3/2, 5/2, 3/1, 4/1 and 5/1 MMRs by \citet{kiaasl}, via the respective stable periodic orbits of the elliptic problem, when the eccentricity values of both the inner and the outer body vary within the range $[0,1]$. These interior MMRs (apart from 5/1) were studied in the past either with the fixed points of the averaged Hamiltonian or with the computation of families of periodic orbits of the original system. In particular, for the restricted problem, we may cite:
\begin{itemize}
	\item for 3/2 MMR \citep{fetsukla,moomo93,mife95,hadjvoy00,avk11};
	\item for 5/2 MMR \citep{fetsukla92,moomo93};
	\item for 3/1 MMR \citep{fetsukla92,hadj92,hadj93,moomo93};
	\item for 4/1 MMR \citep{fetsukla92,hadj93,moomo93}.
\end{itemize}

We herein present a complete view of the families in the circular RTBP (CRTBP) and the elliptic RTBP (ERTBP). We validate the results already reported above and provide new bifurcation points from the CRTBP to the ERTBP and therefore, new families for every MMR. We also continue the already known unstable families to higher eccentricity values and find stable segments. We additionally present isolated (not linked with bifurcation points) families for each MMR studied. The majority of the new families is stable. Moreover, regarding 5/1 MMR, all of the results are new, since, to our knowledge, 5/1 MMR has only been studied for the general TBP (GTBP) by \citet{mbf06}. Let us note that all of the rest above-mentioned MMRs have also been studied for the planar GTBP by \citet{mbf06} and by \citet{av13} with respect to their vertical stability \citep{hen}, thus, both for the planar and the spatial GTBP.
 
The technical details regarding the computation of the periodic orbits are provided in Sect. \ref{meth}. The schemes followed regarding the generation of the periodic orbits from the circular family and their continuation to the CRTBP and ERTBP are given in Sect. \ref{ori}. The computation of the linear stability and the DS-maps  are discussed in Sect. \ref{stab}. The families of periodic orbits being classified by MMR are provided in Sect. \ref{res}, where for each MMR we additionally visualise the regular domains in phase space with suitable DS-maps in the neighbourhood of stable periodic orbits. We discuss and summarize our results in Sect. \ref{con}.

\section{Methodology}\label{meth}

The position and stability of periodic orbits in phase space can unveil the regular and chaotic domains in phase space. The periodic orbits correspond to the fixed or periodic points of a Poincar\'e map, the idea of which was introduced by \citet{pnc} and its properties were discussed by \citet{birk}. Therefore, the computation of families of periodic orbits seems necessary for the study of planetary systems like the above-mentioned ones (star-planet-asteroid, planet-satellite-spacecraft and binary stars-planet), whose evolution may additionally, be associated with MMRs. 

In this paper, we consider the following celestial bodies and interior MMRs (but the results can be applied to any case modelled by the RTBP): a star, $P_0$, an inner massless body, $P_1$, and an outer planet, $P_2$, with point masses $m_0$, $m_1=0$ and $m_2=0.001$, respectively, normalized to unity ($m_0+m_1+m_2=1$), so that $m_0=0.999$. The mass parameter of our system is $\mu=\frac{m_2}{m_0+m_2}=0.001$. Throughout the present study, subscript 1 (2) refers to the inner (outer) body.

We consider a rotating frame of reference, $Oxy$, where the bodies $P_0$ and $P_2$ are located on the $Ox$-axis. $O$ is the centre of mass of $P_0$ and $P_2$, the axis $Ox$ is defined by the direction $P_0-P_2$ and the axis $Oy$ is vertical to $Ox$. Consequently, in the rotating frame, the position of the system is given by the coordinates $x'$ (for $P_2$) and $x$, $y$ (for $P_1$), while the rotation of the frame is defined by the angular velocity $\dot\theta$ of the $Ox$-axis with respect to the inertial frame.

The CRTBP describes the motion of $P_1$, when the other two bodies evolve on circular orbits, while in the ERTBP the primaries move on elliptic orbits. In the CRTBP, we have 2 degrees of freedom, since the star and the giant planet are fixed on the $Ox$-axis at $-\mu$ and $1-\mu$. Whereas in the ERTBP, we have a non-autonomous problem of $2+1$ degrees of freedom, since the angular velocity of rotation of the $Oxy$ frame is not constant and the positions of the star and giant planet on the rotating $Ox$-axis are not fixed.

For the ERTBP, the Lagrangian of the massless body in the rotating frame is
\begin{equation} \label{lagr}
\begin{array}{c}
L= 0.5[(\dot x -\dot \theta y)^2+(\dot y +\dot\theta x)^2]+\frac{1-\mu}{r_1}+\frac{\mu}{r_2}
\end{array}
\end{equation}
where $r_1=\sqrt{(x+\mu r)^2+y^2}$, $r_2=\sqrt{[x-(1-\mu) r]^2+y^2}$ and $r=x'/(1-\mu)$ is the distance between the primaries. For the CRTBP, we have $r=1$ and $\dot\theta=1$. For the theory of the RTBPs see e.g. \citet{sze}. 

In all cases, the system obeys the fundamental symmetry (\citealt{hen97})
\begin{equation} \label{EqSymmetry}
\Sigma: (t,x,y) \rightarrow (-t,x,-y).
\end{equation}

Let us consider an orbit $\textbf{X}(t)=(x'(t), x(t), y(t),\dot x'(t), \dot x(t), \dot y(t))$, which is a set of positions and velocities in the phase space of $P_1$ (primed quantities) and $P_2$. This orbit is periodic if it fulfils the periodic conditions $\textbf{X}(0) = \textbf{X}(T)$, where $T$ is the orbit's period satisfying $t = kT$, with $k \ge 1$ being an integer and particularly the following conditions:
\begin{equation}\begin{array}{lll}
x'(T)=x'(0),& &\dot x'(T)=\dot x'(0)=0,\\
x(T)=x(0),& &\dot x(T)=\dot x(0),\\
y(T)=y(0),& &\dot y(T)=\dot y(0).\\
\end{array}
\end{equation}

A periodic orbit is symmetric if it invariant under $\Sigma$ and asymmetric otherwise. The asymmetric orbit is mapped by $\Sigma$ to its mirror image. Given the Lagrangian of the system and the respective equations of motion, the system remains invariant under certain periodicity conditions, which determine whether the periodic orbit is symmetric or asymmetric \citep[see e.g.][]{avk11}. In this work, we deal exclusively with symmetric periodic orbits, where the longitude of pericentre is $\varpi_i=0$ or $\pi$, $i=1,2$. 

We always set the giant planet, $P_2$, at a semi-major axis $a_2=1$ and the gravitational constant $G=1$. The mean-motion ratio would then be $\frac{n_2}{n_1}=\left(\frac{a_1}{a_2}\right)^{-3/2}\approx\frac{p+q}{p}$, where $p, q \in \mathbb{Z}^*$ and $q$ is the order of the MMR. 

The periodic orbits correspond to the fixed points (or stationary solutions) of an averaged Hamiltonian and depend on the resonant angles 
\begin{equation}\begin{array}{l}
\theta_1=p\lambda_1-(p+q)\lambda_2+q\varpi_1, \\
\theta_2=p\lambda_1-(p+q)\lambda_2+q\varpi_2, \\
\end{array}
\end{equation}
where $\lambda_i=M_i+\varpi_i$ is the mean longitude and $M_i$ the mean anomaly ($i=1,2$). The stationary solutions where $\dot\theta_i=0$ ($i=1,2$) are also called apsidal corotation resonances (ACRs) in the works of \citet{femibe06} and \citet{mbf06}. Therefore, an ACR corresponds to a periodic orbit in the rotating frame, which in turn showcases the exact location of an MMR in phase space. 

When $q$ is odd (i.e. in 3/2, 5/2 and 4/1 MMRs in our study) we use the pair $(\theta_1,\theta_2)$, when $q=2$ (in 3/1 MMR) we use the pair $(\theta_3,\theta_1)$, where
\begin{equation}
\theta_3=\lambda_1-3\lambda_2+\varpi_1+\varpi_2,
\end{equation}
and when $q=4$ (in 5/1 MMR) we use the pair $(\theta_4,\theta_1)$, where 
\begin{equation}
\theta_4=\lambda_1-5\lambda_2+3\varpi_2+\varpi_1.
\end{equation}

By letting the resonant bodies be aligned ($\Delta\varpi$=0) or anti-aligned ($\Delta\varpi=\pi$) and by considering also the mean anomalies, we get four symmetric configurations distinguished by the pairs of resonant angles. Those 4 configurations are the following: $(0,0), (0,\pi), (\pi,0)$ and $(\pi,\pi)$. When the families of periodic orbits are presented in the ERTBP, we give a negative value to $e_1$ ($e_2$) when the first (second) argument in brackets librates about $\pi$ and keep the positive value otherwise.

\begin{figure}[h]\centering
\includegraphics[width=7.25cm]{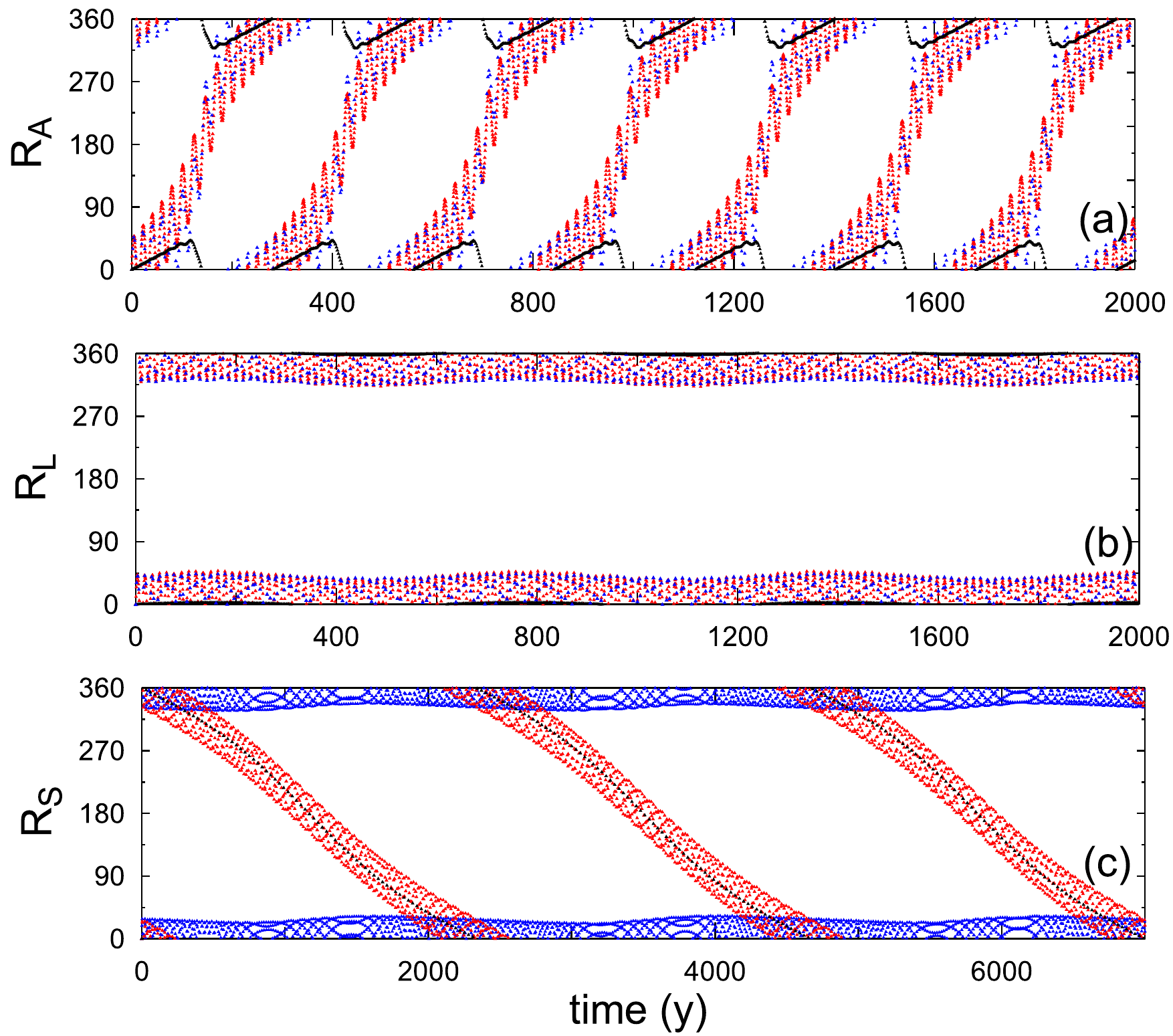}
\caption{Evolution of the resonant angles, $\theta_1$ (blue), $\theta_2$ (red) and the apsidal difference, $\Delta\varpi$ (black) for the 3 cases observed in the DS-maps. Initial conditions are derived from Fig. \ref{32_00pp0}a. {\bf a} Apsidal difference oscillation, $R_A$, when $e_1=0.48$ and $e_2=0.22$. {\bf b} MMR, $R_L$, when $e_1=0.71$ and $e_2=0.39$.  {\bf c} Secondary resonance, $R_S$, when $e_1=0.52$ and $e_2=0.02$}
\label{R}
\end{figure}

In the vicinity of a periodic orbit that is stable, both the resonant angles and the apsidal difference, $\Delta\varpi=(\theta_2-\theta_1)/q$, librate about 0 or $\pi$ if the orbit is symmetric or around other values if the orbit is asymmetric. As the amplitude of the oscillation tends to zero, the orbit changes from quasi-periodic to periodic. $\Delta\varpi$ is a measure of geometric asymmetry and $\Delta M=M_2-M_1$ is a measure of dynamic asymmetry. 
In true resonance (like the MMR, denoted by $R_L$ in Fig. \ref{R}b), two proper frequencies of a system become commensurable. 

Inside an MMR, when we consider the average of the frequencies between the rotation of $\theta_2$ ($\omega_{2R}$) and the libration of $\theta_1$ ($\omega_{1L}$), we have $\omega_{2R}/\omega_{1L}=s_2/s_1$, where $s_i \in \mathbb{Z}^*$, known as a secondary resonance \citep{momo93}, denoted by $R_S$ (see Fig. \ref{R}c). 

For the non-resonant orbits, when only the apsidal difference oscillates about 0 or $\pi$ and the rest resonant angles rotate, we have the apsidal resonance \citep{murray,morby}, where the close encounters are avoided as well, since the phases are protected \citep{malho02}. In this case, the amplitude of one proper frequency becomes zero (or gets close to zero), so both eccentricities are dominated by the other proper frequency. Far from the centre of resonance, this purely secular regime is separated from near-resonant domains by narrow chaotic motion and the existence of topological separatrix between librations and circulations. Apsidal resonance will be further discussed in Sect. \ref{stab} (Fig. \ref{02}).

However, inside an MMR, during the transition between two configurations, no separatrix crossing is being involved and the passage is merely kinematical. The apsidal difference can oscillate about 0 (or $\pi$) or circulate, while the rest resonant angles rotate. For instance, in Fig. \ref{R}a, $\Delta\varpi$ oscillates  during the passage from the configuration $(0,0)$ to $(0,\pi)$. This behaviour will be called apsidal difference oscillation, $R_A$, in the following.

A periodic orbit corresponds to the fixed points on a Poincar\'e surface of section and its multiplicity equals to the number of points with which it is depicted by on the Poincar\'e map. If we consider the section $y=0$, a periodic orbit is symmetric with respect to the $x$-axis, when it has two perpendicular crossings with that axis, i.e. $y(T)=y(0)=0$ and $\dot x(T)=\dot x(0)=0$. 

The periodic orbits in the ERTBP are represented by a point in the 3D phase space 
\begin{equation}\begin{array}{lll}
	x'(T)=x'(0),& x(T)=x(0), & \dot{y}(T)=\dot{y}(0), 
\end{array}\end{equation}
where $x'$ denotes the position of $P_2$ on the $Ox$-axis, and is related to the eccentricity of $P_2$, namely $x'(0)$=$a_2(1-e_2)$ or $x'(0)$=$a_2(1+e_2)$ when $P_2$ is at pericentre ($\varpi_2=0$) or apocentre ($\varpi_2=\pi$), respectively, while $x$ and $\dot y$ correspond to the position and velocity of $P_1$. 

In the CRTBP, $x'$ is constant and equates to $1-\mu$ . Therefore, a symmetric periodic orbit with respect to the $x$-axis satisfies the initial conditions in the 2D phase space
\begin{equation}\begin{array}{ll}
x(T)=x(0), & \dot{y}(T)=\dot{y}(0). 
\end{array}\end{equation}

\section{Origin and bifurcation of periodic orbits}\label{ori}

Let us assume that the massive bodies $P_0$ and $P_2$ (perturbed case with $\mu=0.001$) are moving on circular orbits around their common centre of mass.  When $P_1$ evolves on a circular orbit, then, in the rotating frame, $Oxy$, its orbit seems circular as well and is called \textit{circular periodic orbit}. This motion of the massless body $P_1$ is a Keplerian orbit in the inertial frame. Therefore, the radius of its orbit corresponds to $a_1$ and the coordinate $x$. All the circular orbits of $P_1$ in the rotating frame are symmetric and periodic, whilst the elliptic ones are periodic in the rotating frame, as long as they are resonant, and they may be symmetric or asymmetric.

\begin{figure}[H]\centering
\includegraphics[width=12cm]{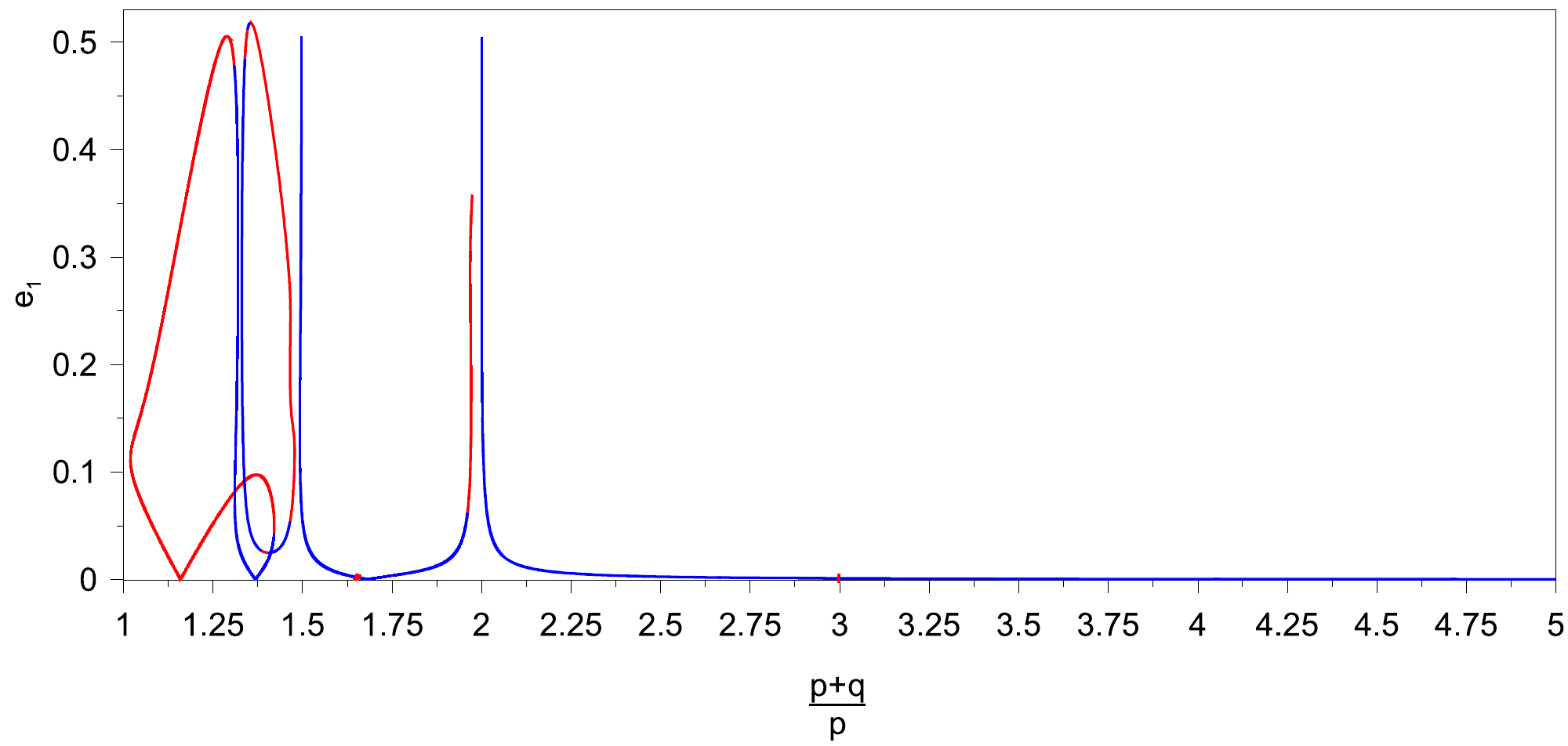}
\caption{The circular family (of circular stable (blue) periodic orbits with $e_{1,2}\approx 0$) for $\mu=0.001$ as the mean-motion ratio ($\frac{p+q}{p}$) varies.  As the eccentricity, $e_1$, of the small body increases, the continuation of the circular family to the elliptic families of the CRTBP is showcased. At first-order MMRs, the gaps at the circular family are evident. At second-order MMRs, the circular periodic orbits become unstable (red)}
\label{circ1}
\end{figure}

Along the \textit{circular} family (of circular periodic orbits with $e_{1,2}\approx 0$) (see Fig. \ref{circ1} when $e_1\approx 0$) the MMR varies. At rational values of the mean-motion ratio, namely $\frac{p+q}{p}$, where the periodic orbits are of period $T=2\pi(a_1^{-3/2}-1)^{-1}$ (when $a_2=1.0$), there bifurcate two branches of resonant families of symmetric elliptic periodic orbits in the CRTBP; one being stable (depicted by blue colour) and one unstable (depicted by red), due to the Poincar\'e-Birkhoff fixed point theorem. One corresponds to the location of $P_1$ at pericentre (denoted hereinafter by $I$) and one to its location at the apocentre (denoted by $II$). 

The circular family consists of stable (blue coloured) periodic orbits with the exception of the neighbourhood of mean-motion ratios which correspond to second-order ($q=2$) MMRs (e.g. 3/1, 5/3, etc.) where instability is observed (see the red segments when $e_1\approx 0$ in Fig. \ref{circ1}). At first-order ($q=1$) MMRs (e.g. 2/1, 3/2 etc.) the circular family breaks (perturbed case where $\mu\neq 0$) and a gap is formed. The circular family is then continued smoothly to the CRTBP (see the stable and unstable branches ($I$ and $II$) that are emanating from the circular family when $e_1>0$ in Fig. \ref{circ1}).

Along a family of the CRTBP the period $T$ varies. In the non-autonomous ERTBP, the period $T$ is an integer multiple of the period of the primaries $T_0=\frac{2\pi}{|\frac{p+q}{p}-1|}$, i.e. $T=k T_0$. 

There are two \textit{Schemes} of continuation to the ERTBP:
\begin{description}

	\item[\textit{Scheme I}-- From the CRTBP:] \hfill \\
	A periodic orbit of the CRTBP can be continued to the ERTBP mono-parametrically forming families, along which the eccentricity, $e_2$, of the primary $P_2$ varies (increases from 0), when its period gets equal to $T=k T_0$ or $T=k T_0/m$, where $m$ is the multiplicity of the generated periodic orbit. For the bifurcation points, $B$, we use the following notation: $B^{\frac{p+q}{p}}_{F,\#}$, where $\#$ stands for the number of bifurcation points along a certain family ($F$) in the MMR ($\frac{p+q}{p}$). 	\\
	
\item[\textit{Scheme II} -- From the circular family:]\hfill \\
	At the ends of the small unstable segments of the circular family, where $q=2$, (second-order MMRs, e.g. 3/1, 5/3, etc.), we have bifurcation points from which doubly symmetric periodic orbits emanate directly to the ERTBP. These periodic orbits have a period equal to twice the period of the bifurcation point from which they are generated, i.e. $T=2\;T_0=\frac{4\pi}{|\frac{p+q}{p}-1|}$. The number of the periodic orbits that are generated within this Scheme depends on the magnitude of the perturbation (i.e. the masses of the celestial bodies) and they can be two or four \citep[see e.g.][for examples of applications in the GTBP]{hadj06}. When $q>2$, we have bifurcation points that generate periodic orbits described $q$ times, i.e. they have a period $T=q\;T_0$.	
\end{description}

%
%
%

It has been shown that there also exist families that do not bifurcate from periodic orbits and are isolated. \citet{av12} computed spatial isolated symmetric families in 2/1 MMR in 3D-GTBP resulting by foldings of the spatial families as the mass of the inner or outer body increased from zero. \citet{voyhadj05,vkh09, av16} presented planar asymmetric isolated families in 2/1 MMR in GTBP resulting by collision bifurcations as the planetary mass ratio varied. In this paper, we present some new isolated families of symmetric periodic orbits in the ERTBP.

\section{Linear stability and DS-maps}\label{stab}

With regards to the linear stability of the periodic orbits, we compute the conjugate eigenvalues in reciprocal pairs of the monodromy matrix of the variational equations of the system. The periodic orbit is called linearly stable iff all of the eigenvalues $(\lambda_1=1/\lambda_2)$ and $(\lambda_3=1/\lambda_4)$ lie on the unit circle. 

For the CRTBP one pair is always the unit pair, $\lambda_1=\lambda_2=1$, due to the existence of the Jacobi integral, sometimes called the integral of relative energy. When the other pair lies on the real axis, the periodic orbit is unstable. When the other pair lies on the complex plane but is located on the unit circle, the periodic orbit is stable.

For the ERTBP, we may have 4 different cases: stability, simple, double and complex instability. Additionally, we can define the stability indices $b_1=\lambda_1+\lambda_2$ and $b_2=\lambda_3+\lambda_4$. If $|b_i|<2$ ($i=1,2$), the two pairs of eigenvalues are complex conjugate lying on the unit circle. If $|b_i|>2$, six different types of instability arise (the eigenvalues may either be real or complex) \citep{Broucke1969,marchal90} and in case $|b_i|=2$, the pairs of eigenvalues equal to $1$ or $-1$. In this study, we always depict the stable (unstable) periodic orbits by blue (red) colour.

In Hamiltonian systems, the stable periodic orbits are surrounded by invariant tori, where the motion is regular and quasi-periodic, whereas in the neighbourhood of unstable periodic orbits the motion is irregular, since homoclinic webs are formed. In order to unveil the phase space of the periodic orbits, we compute DS-maps by estimating the evolution stability while using as index the de-trended Fast Lyapunov Indicator (DFLI) (\citealt{voyatzis08}), defined as 
\begin{equation}
DFLI(t)=\frac{1}{t}\frac{\eta(t)}{\eta(0)},
\end{equation}
where $\mathbf{\eta}$ is the deviation vector computed after numerical integration of the variational equations along the orbit. This chaotic indicator remains almost constant over time when the orbit is regular and takes a value $DFLI(t)<10$ (dark coloured regions in the DS-maps). DFLI increases exponentially, when the orbit is chaotic (pale coloured regions). In this study, we have chosen $t_{max}=250 Ky$ and we stop the numerical integration when $DFLI(t)>10^{30}$ and classify the orbits as chaotic. We use the white colour, in order to showcase the orbits for which the numerical integration has failed at $t<t_{max}$, since a very small time step was reached, due to very close encounters.

The stable periodic orbits constitute the backbone of the stability domains in phase space and via dynamical analyses we can provide the regions, where the newly discovered exoplanets locked in MMRs should be ideally hosted in favour of their long-term stability \citep[e.g.][]{a16}. Therefore, the knowledge of the families in the ERTBP is important for the dynamical vicinities of terrestrial planets trapped in MMR with giant planets. What is more, exoplanetary systems on highly eccentric orbits may survive collisions and close encounters even when the planetary orbits intersect, if the bodies evolve close to an exact MMR (periodic orbit) \citep[e.g.][]{av16}. 

In general, two coplanar Keplerian orbits can intersect if the criterion
\begin{equation}
a_1^2(1-e_1^2)+a_2^2(1-e_2^2)-2a_1 a_2 (1-e_1 e_2 \cos\Delta \varpi) \leq 0 
\label{ColCri}
\end{equation}
is satisfied \citep{kho99}. Collision between the planetary bodies can occur when the equality holds. When studying the disturbing function of the averaged Hamiltonian, the collisions correspond to its singularities. Additionally to the points going to infinity, peaks may be apparent and they are related to close encounters between the planets. Herein, the collision and the close encounters of the planets are represented by dashed grey lines and curves, respectively. 

\begin{figure}[H]\centering
\includegraphics[width=15.cm]{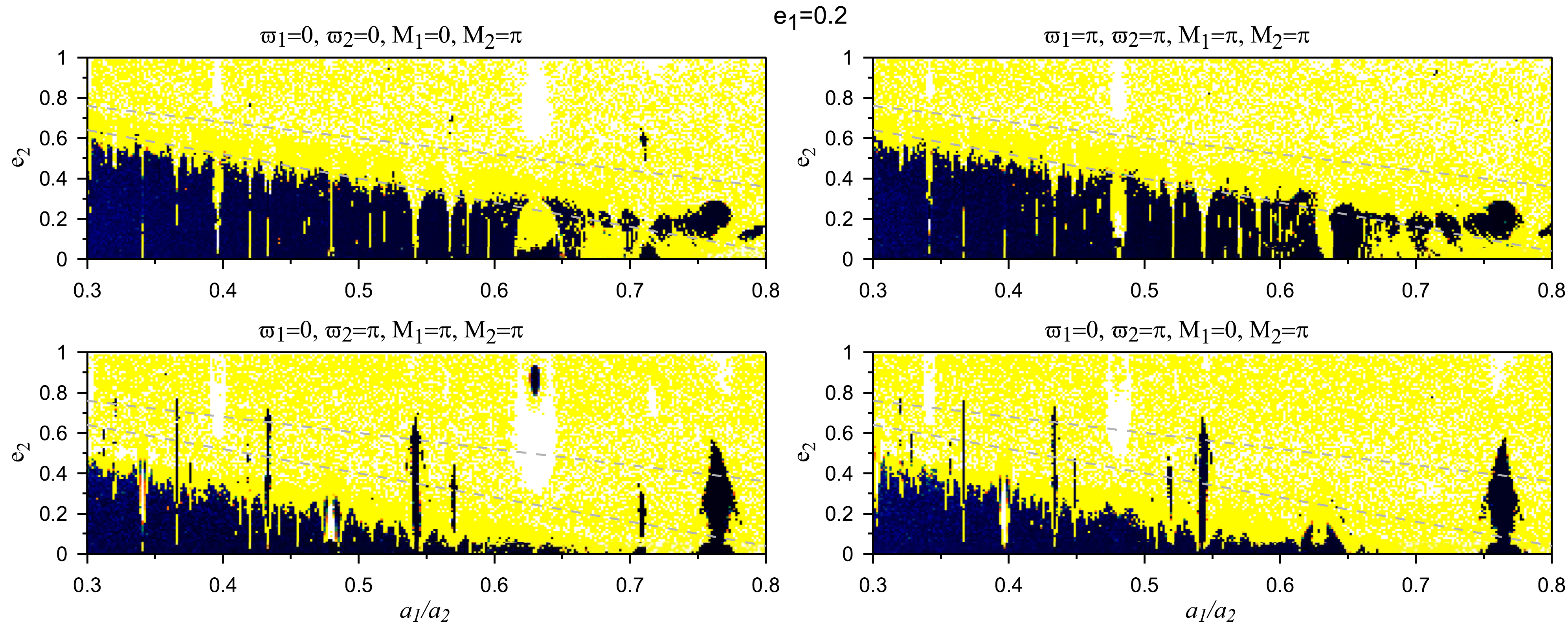}\\ \vspace{0.5em}
\includegraphics[width=2.8cm,height=0.6cm]{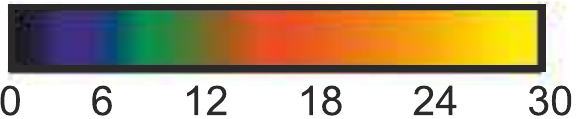} 
\caption{DS-maps on the plane ($a_1/a_2,e_2$) when $e_1=0.2$ and $a_2=1.0$. The values of the angles are noted above each panel. The dashed bold grey lines indicate the intersection of the elliptic orbits. The coloured bar corresponds to the $log (\rm{DFLI})$; dark (pale) colours showcase regular (chaotic) orbital evolution. The MMRs investigated herein are located at: $a_1=0.7631$ (3/2 MMR), $a_1=0.5428$ (5/2 MMR), $a_1=0.4807$ (3/1 MMR), $a_1=0.3968$ (4/1 MMR) and $a_1=0.3419$ (5/1 MMR). The grid of the initial conditions is $300\times100$}
\label{02}
\end{figure}

Likewise in \citet{kiaasl}, in Fig. \ref{02} we showcase the regions of stability of the ERTBP that become apparent when $P_1$ has an eccentricity equal to $e_1=0.2$ for the four different symmetric configurations on the plane ($a_1/a_2,e_2$) together with the collision lines. It is evident that islands of stability and tongues are not broken by the collision lines, since they are built about stable periodic orbits (MMRs) and the phases are protected \citep[e.g.][]{morby}. At those domains both the resonant angles and the apsidal difference librate. At the non-resonant ($\frac{p+q}{p}$ is not a rational number) regular orbits below the collision lines, the close encounters are not in effect, since an apsidal resonance is observed. Secondary resonances at low values of $e_2$ are observed. 

\section{Results}\label{res}

The aim of this section is to provide the families of periodic orbits that belong to the CRTBP and the ERTBP for the 3/2, 5/2, 3/1, 4/1 and 5/1 MMRs. Detailed DS-maps are computed on the planes $(e_1,e_2)$ and ($\varpi_2,e_2$), in order for the extent of the regular domains existing in each MMR to be identified. These domains are linked with the libration (or rotation) of the resonant angles, $\theta_i$, ($i=1,...,4$) --the choice of $i$ depending on the order of the MMR--, and the oscillation/circulation of the apsidal difference, $\Delta\varpi$. These DS-maps are guided by stable periodic orbits existing in the symmetric configurations of the ERTBP. Results on the families of periodic orbits and the respective DS-maps are classified by MMR in the above-mentioned order. 
\clearpage
\subsection{3/2 MMR}

\begin{figure}[H]\centering
$\begin{array}{cp{-1.5cm}c}
\includegraphics[width=6.0cm]{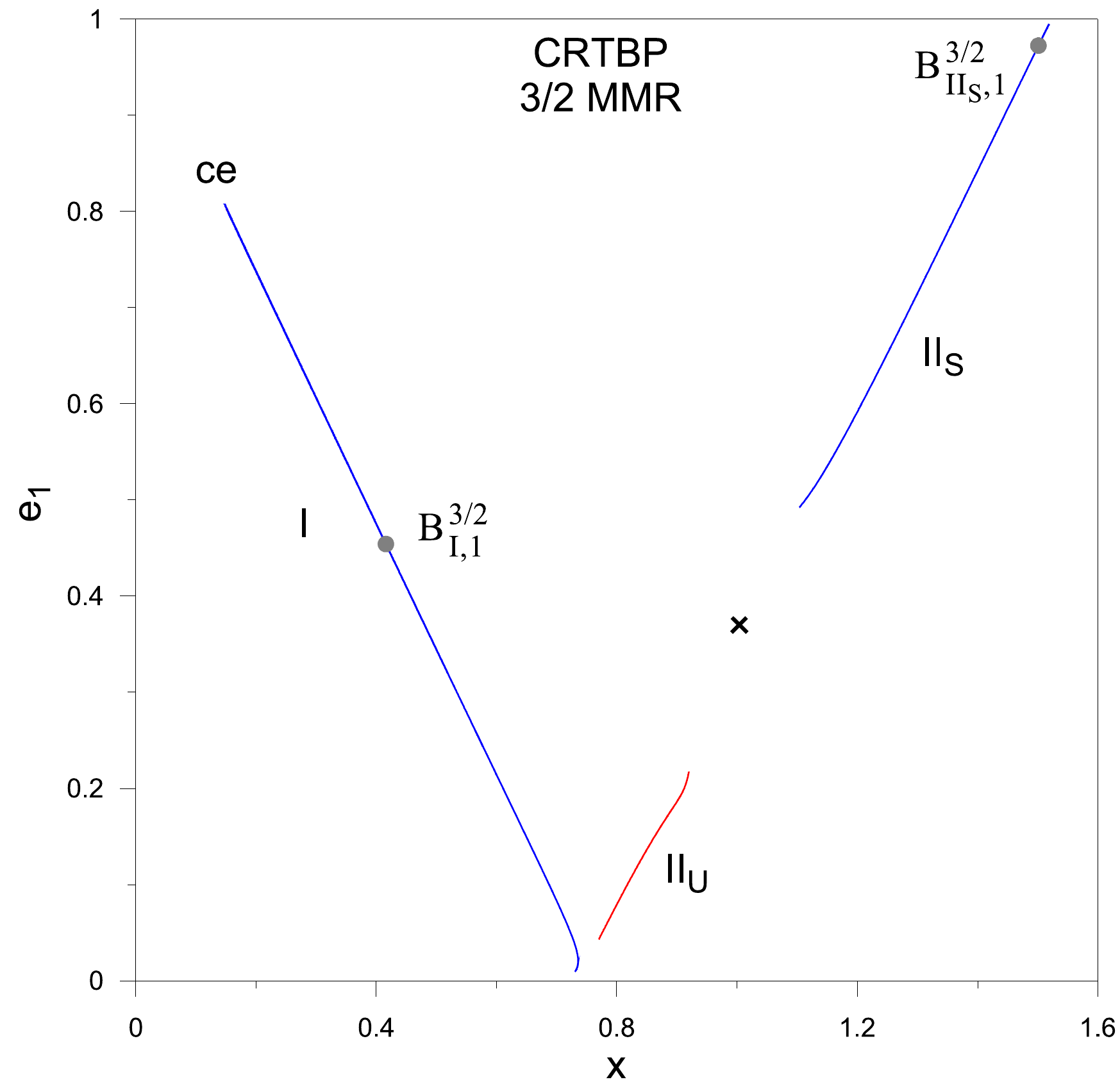}  & \qquad&
\includegraphics[width=5.8cm]{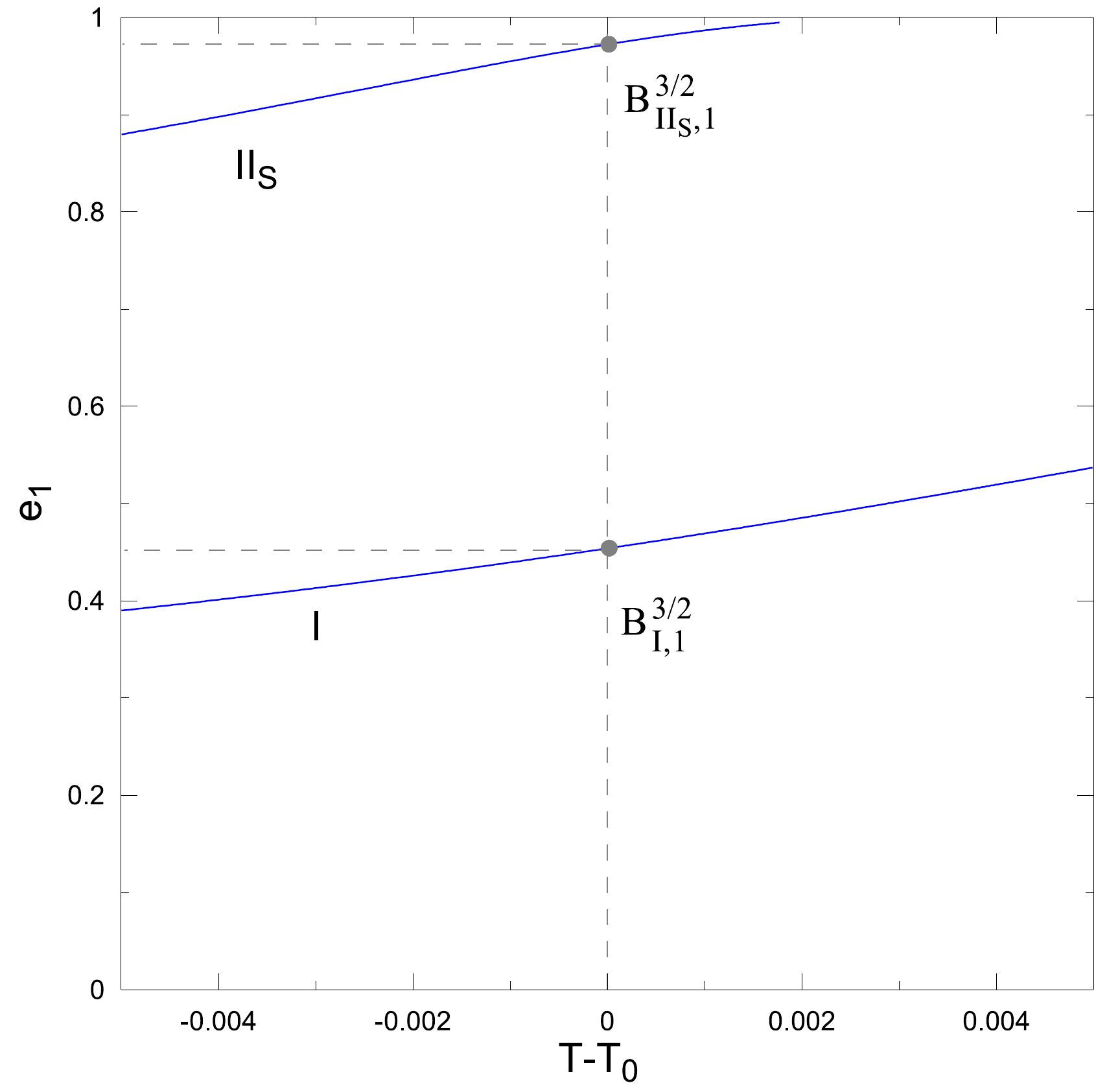} \\
\textnormal{(a)} & \qquad & \textnormal{(b)} 
\end{array} $
$\begin{array}{c}
\includegraphics[width=10cm]{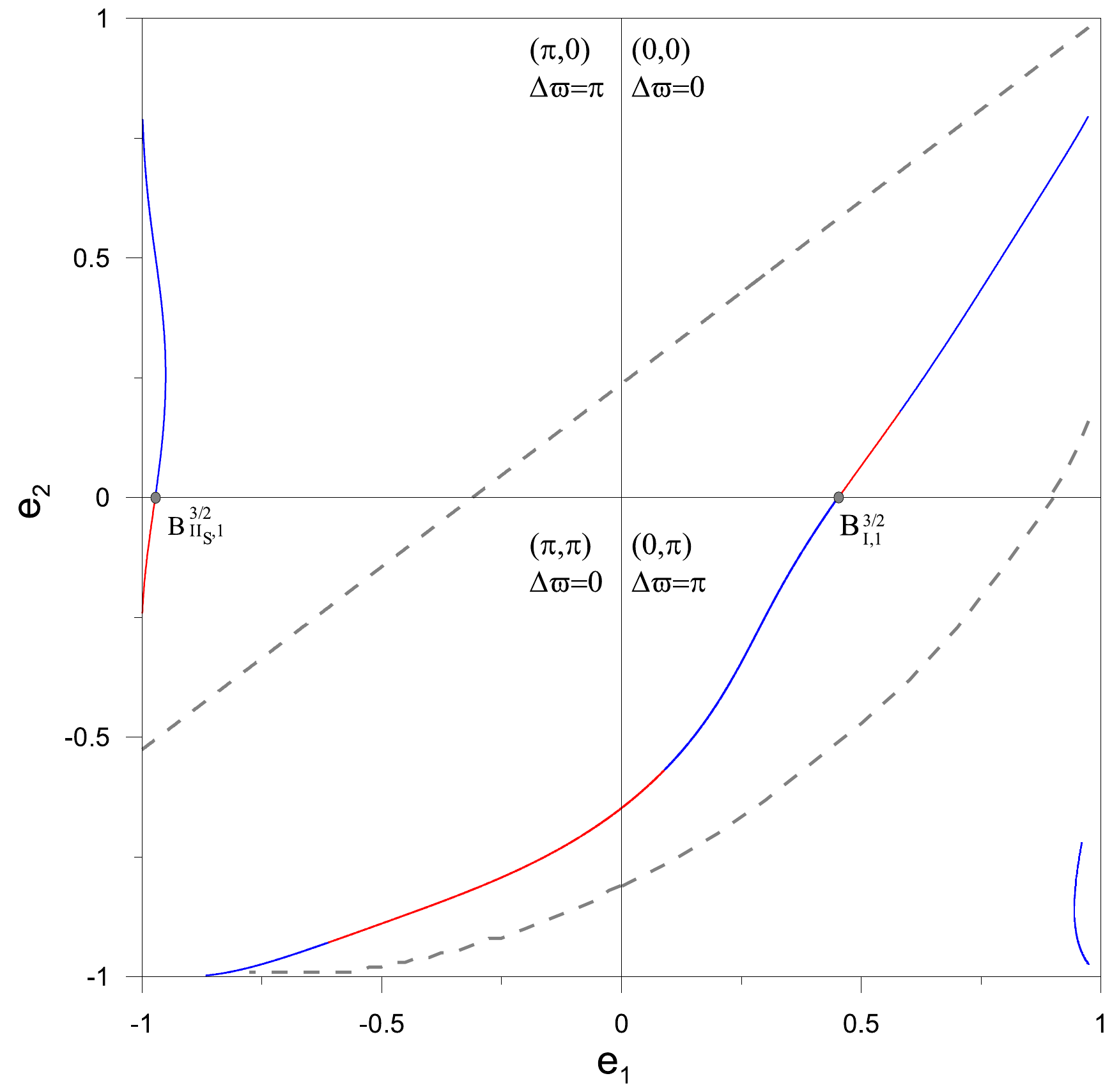} \\
\textnormal{(c)} 
\end{array} $
\caption{{\bf a} Families of periodic orbits in 3/2 MMR of the CRTBP projected on the plane $(x,e_1)$. Along the family $I$, $P_1$ is initially located at pericentre, whereas along $II_U$ and $II_S$ it begins at apocentre. The collision between $P_1$ and $P_2$ is denoted by the symbol ``$\times$'' and the close encounters by ``ce''. Blue (red) stands for stable (unstable) periodic orbits. The bifurcation points from the CRTBP to the ERTBP are also shown. {\bf b} Justification of existence of bifurcation points in the families of CRTBP in 3/2 MMR, where $T=T_0=4\pi$, that generate periodic orbits in the ERTBP. {\bf c} Families of symmetric periodic orbits in 3/2 MMR of the ERTBP presented in four quadrants in correspondence with the four different symmetric configurations on the plane $(e_1,e_2)$. Apart from the apsidal difference, $\Delta\varpi$, being noted, the angles in brackets represent the pair of resonant angles ($\theta_1,\theta_2$). The dashed grey line and curve depict the collision between $P_1$ and $P_2$ and the close encounters, respectively}
\label{32_all}
\end{figure}

In Fig. \ref{32_all}a, we present the families of periodic orbits in 3/2 MMR in the CRTBP, which are continued smoothly from the circular family, due to the gap at $a_1=\frac{3}{2}^{-2/3}$. At $x\approx 0.763143$ the two branches are formed, $I$ ($P_1$ at pericentre) and $II$ ($P_1$ at apocentre). The family $I$ is stable (blue) but could not be continued for higher values of $e_1$, since there exists a region of close encounters (denoted by ``ce''). The branch $II$ consists of two families, one unstable (red) ($II_U$) and one stable ($II_S$), which are divided by the region where collisions (denoted by the symbol ``$\times$'') between $P_1$ and $P_2$ take place. 

Along those families there exist bifurcation points whenever the period of the periodic orbits, $T$, equates to $T_0=4\pi$. In Fig. \ref{32_all}b, we justify the existence of two such points, $B^{3/2}_{I,1}$ and $B^{3/2}_{II_S,1}$. The former was studied by \citet{avk11} and the latter had not been reported.

In Fig. \ref{32_all}c, we provide the families of symmetric periodic orbits in the ERTBP. The two families emanating from $B^{3/2}_{I,1}$ are continued at the configurations $(\theta_1,\theta_2)=(0,0)$ and $(0,\pi)$. Along the family belonging to $(0,\pi)$ there is a change of the configuration when $P_1$ reaches $e_1=0$ and the family then evolves in the configuration $(\pi,\pi)$. There are segments of stable periodic orbits in each of these configurations. While these families were previously studied by \citet{avk11}, from the bifurcation point $B^{3/2}_{II_S,1}$, two new families are computed herein: one being stable in the configuration $(\pi,0)$ and one being unstable in the configuration $(\pi,\pi)$. All of the above-mentioned families were continued by following \textit{Scheme I}. Finally, we computed an isolated family which consists of highly eccentric stable periodic orbits for both $P_1$ and $P_2$ and belongs to the configuration $(0,\pi)$.

\begin{figure}[H]
\begin{center}
$\begin{array}{cp{-2cm}c}
\includegraphics[width=6.0cm]{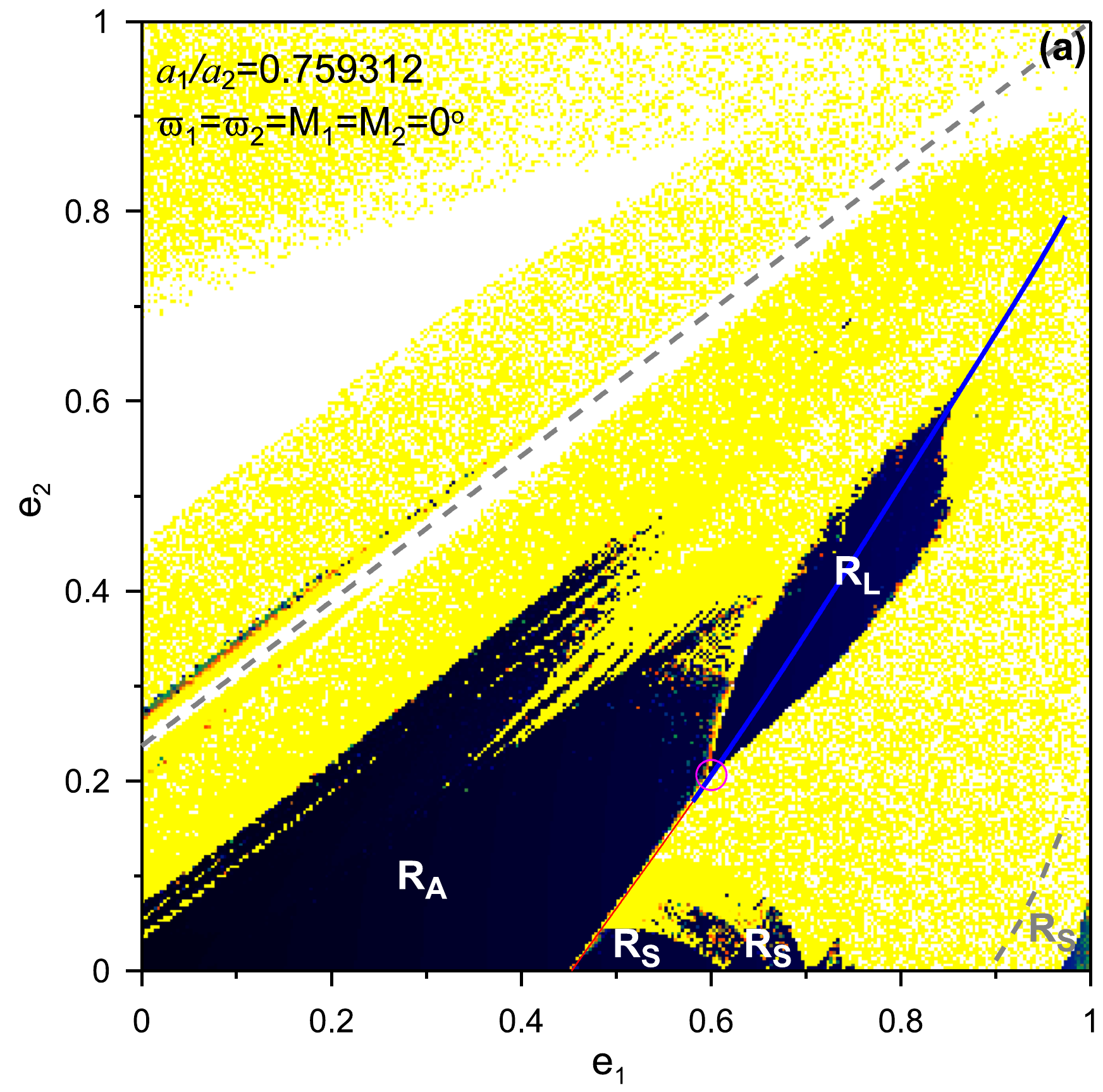} & \qquad&  \includegraphics[width=6.0cm]{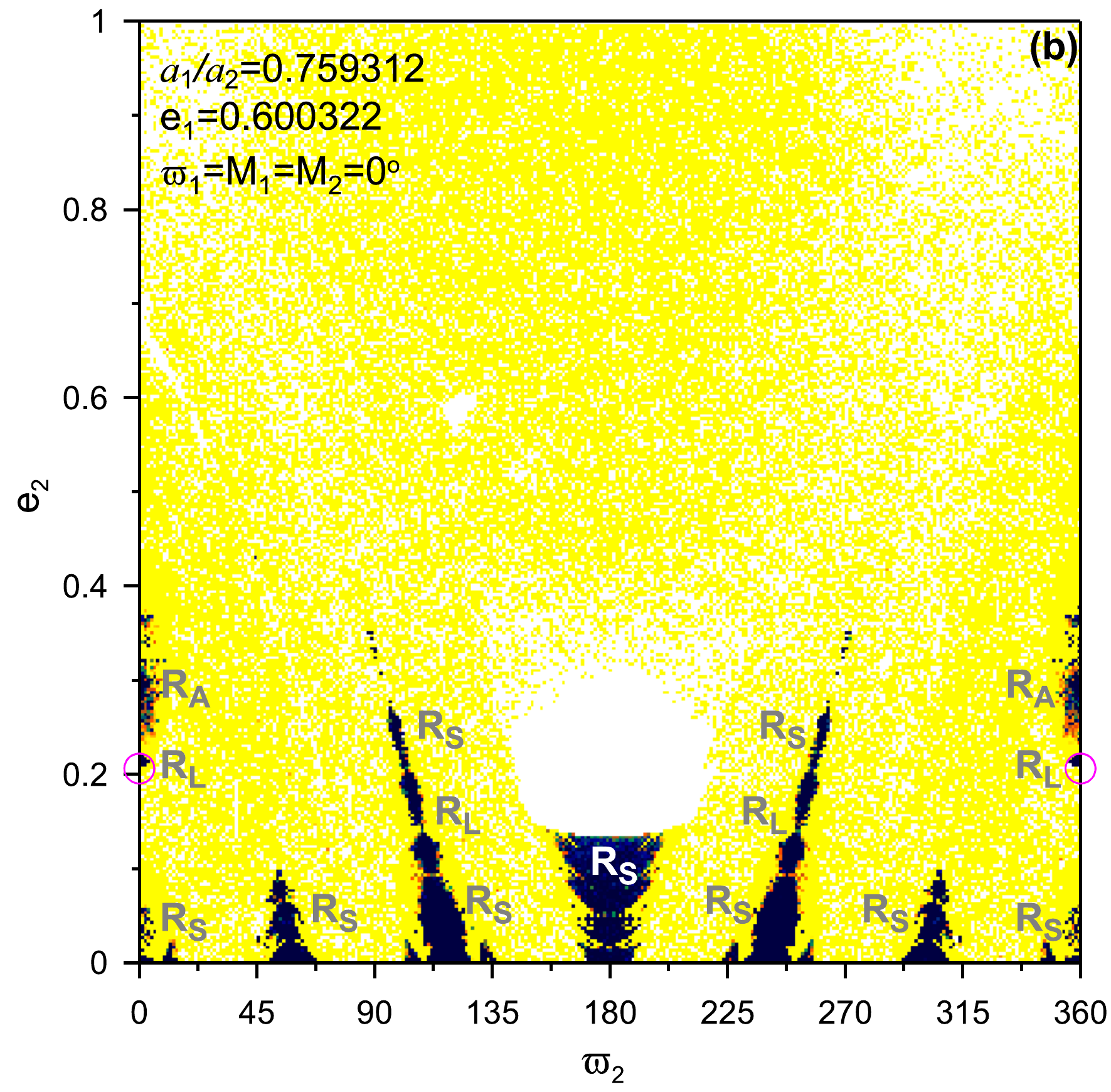}\\
\includegraphics[width=6cm]{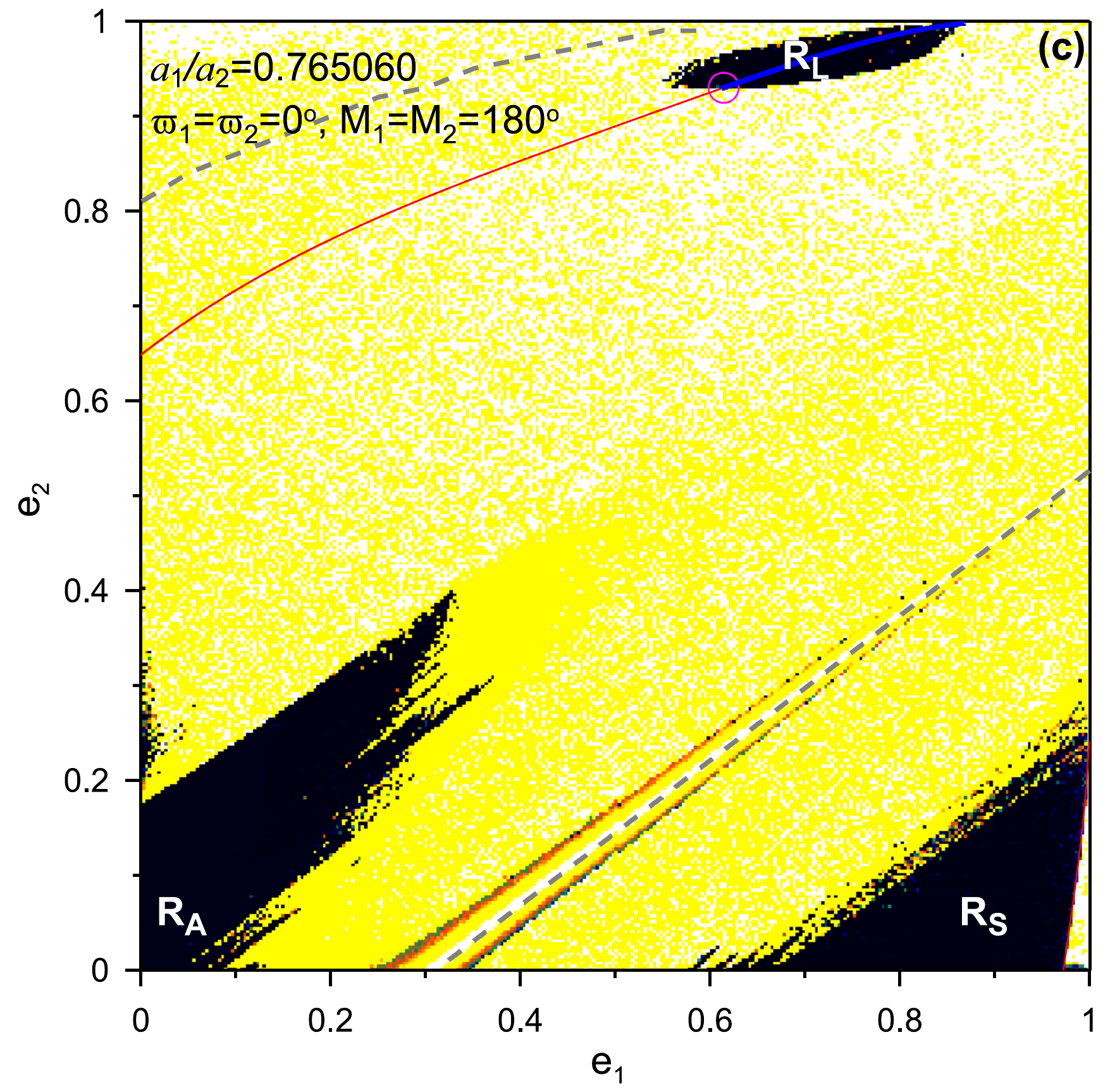} & \qquad& \includegraphics[width=6cm]{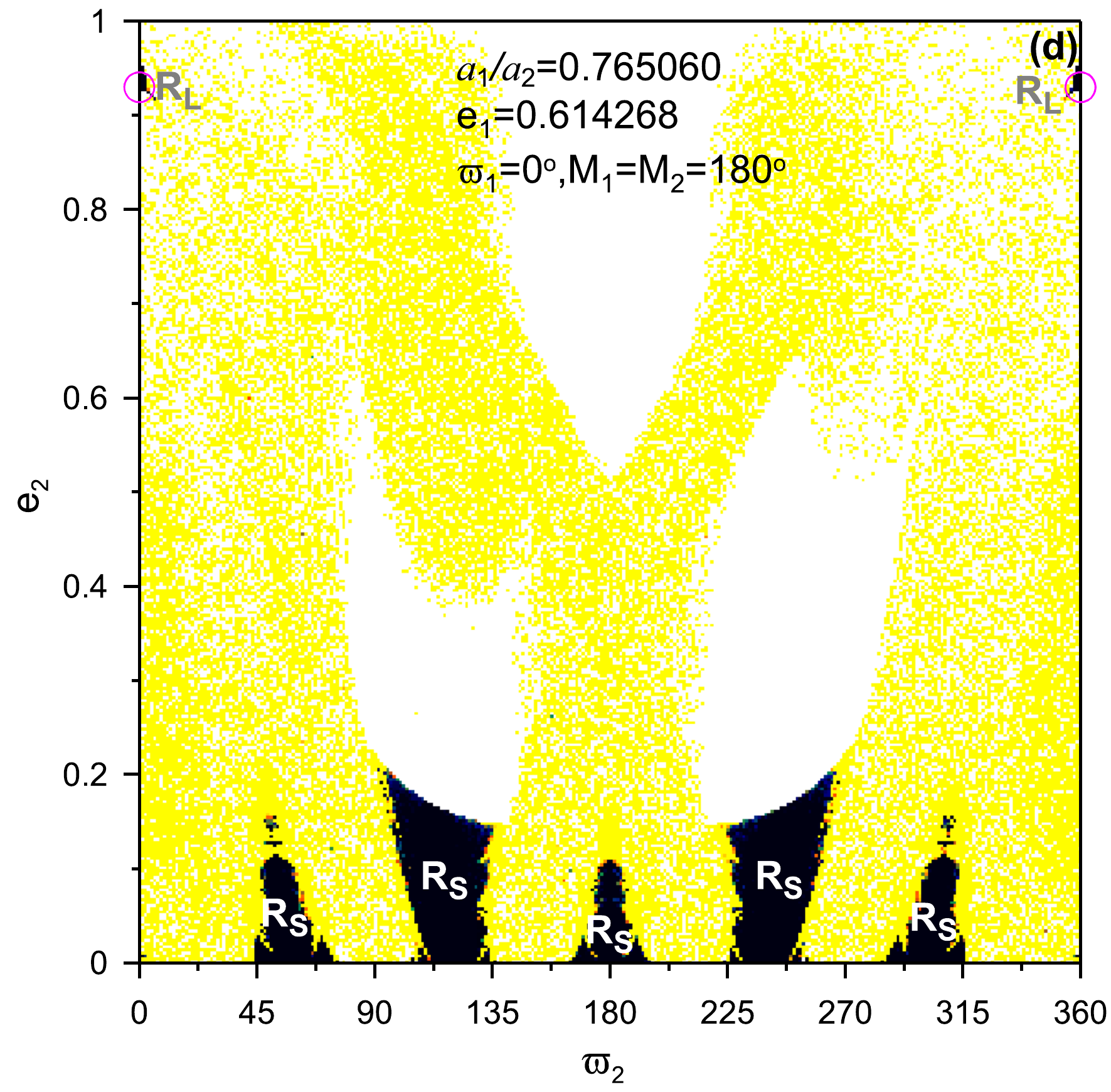} \\
\end{array} $
\end{center}
\caption{DS-maps on the planes \textbf{a} ($e_1,e_2$) and \textbf{b} ($\varpi_2,e_2$) yielded by a stable periodic orbit (magenta coloured circle) that belongs to the stable (bold blue) segment of the family in the configuration ($\theta_1,\theta_2$)=($0,0$) in 3/2 MMR (see Fig. \ref{32_all}c). The unstable periodic orbits are coloured with red. Similarly to the presentation of Fig. \ref{32_all}c, the dashed grey line and curve depict the region where the collisions and close encounters take place, respectively. The orbital elements of the periodic orbit that remain fixed during the computation of each map are accordingly noted down on each plot. The islands of stability are located at certain regions, which exhibit the following attributes: $R_L$: Libration of all resonant angles and apsidal difference (MMR), $R_S$: Secondary resonance inside the MMR, $R_A$: Apsidal difference oscillation. In the panels \textbf{c} and \textbf{d}, the choice of the periodic orbit is made from the configuration ($\theta_1,\theta_2$)=($\pi,\pi$) (see Fig. \ref{32_all}c). The grid of the initial conditions on each DS-map is $300\times300$}
\label{32_00pp0}
\end{figure}

In Fig. \ref{32_00pp0}, we present two DS-maps on the planes $(e_1,e_2)$ and $(\varpi_2,e_2)$ (two columns) for two different configurations (two rows). The selected stable periodic orbit that guides the computation each time is depicted by a magenta coloured circle. 

In Figs. \ref{32_00pp0}a,b, we are guided by a stable periodic orbit within the stable (blue) segment of the family evolving in the configuration ($\theta_1,\theta_2$)=($0,0$), while in Figs. \ref{32_00pp0}c,d, we select a stable periodic orbit from the configuration ($\pi,\pi$). In the regular domains that are built about the selected stable periodic orbit we specify $R_L$, where the resonant angles, $\theta_1$, $\theta_2$ and the apsidal difference, $\Delta\varpi$, librate about the angles of the respective configuration. In panels a and b about 0, 0 and 0 and in panels c and d, about $\pi$, $\pi$ and 0. When $R_A$ is observed, $\Delta\varpi$ oscillates about 0. The secondary resonance, $R_S$, that is observed, is 1/1 where $\theta_1$ librates about 0 or $\pi$ (unless otherwise stated-see below). 

However, when $\varpi_2$ varies (panels b and d), we can observe libration of the above angles about a different combination of 0 or $\pi$, as we approach a stable periodic orbit of another configuration. Particularly, in Fig. \ref{32_00pp0}b, when $\varpi_2$ is around $\pi/3$ (or $5\pi/3$ symmetrically) and $\pi$ in 1/1 secondary resonance, $\theta_1$ librates about $\pi$. Whereas when $\varpi_2$ is around $2\pi/3$ (or $4\pi/3$ symmetrically) there is a region $R_L$ with libration of $(\theta_1,\theta_2)$ and $\Delta\varpi$ about $(0,\pi)$ and $\pi$, respectively. Moreover, when the $R_L$ of the configuration $(0,\pi)$ is observed, 1/1 secondary resonances are also apparent; for $e_2<0.1$, $\theta_1$ librates about 0 and for $e_2>0.2$, $\theta_2$ librates about $\pi$. Additionally, in Fig. \ref{32_00pp0}d, when $\varpi_2$ is around $\pi/3$ (or $5\pi/3$ symmetrically) and $\pi$ in 1/1 secondary resonance, $\theta_1$ librates about $0$. Whereas when $\varpi_2$ is around $2\pi/3$ (or $4\pi/3$ symmetrically), $\theta_1$ librates about $\pi$.

\begin{figure}[H]
\begin{center}
$\begin{array}{cp{-2cm}c}
\includegraphics[width=6cm]{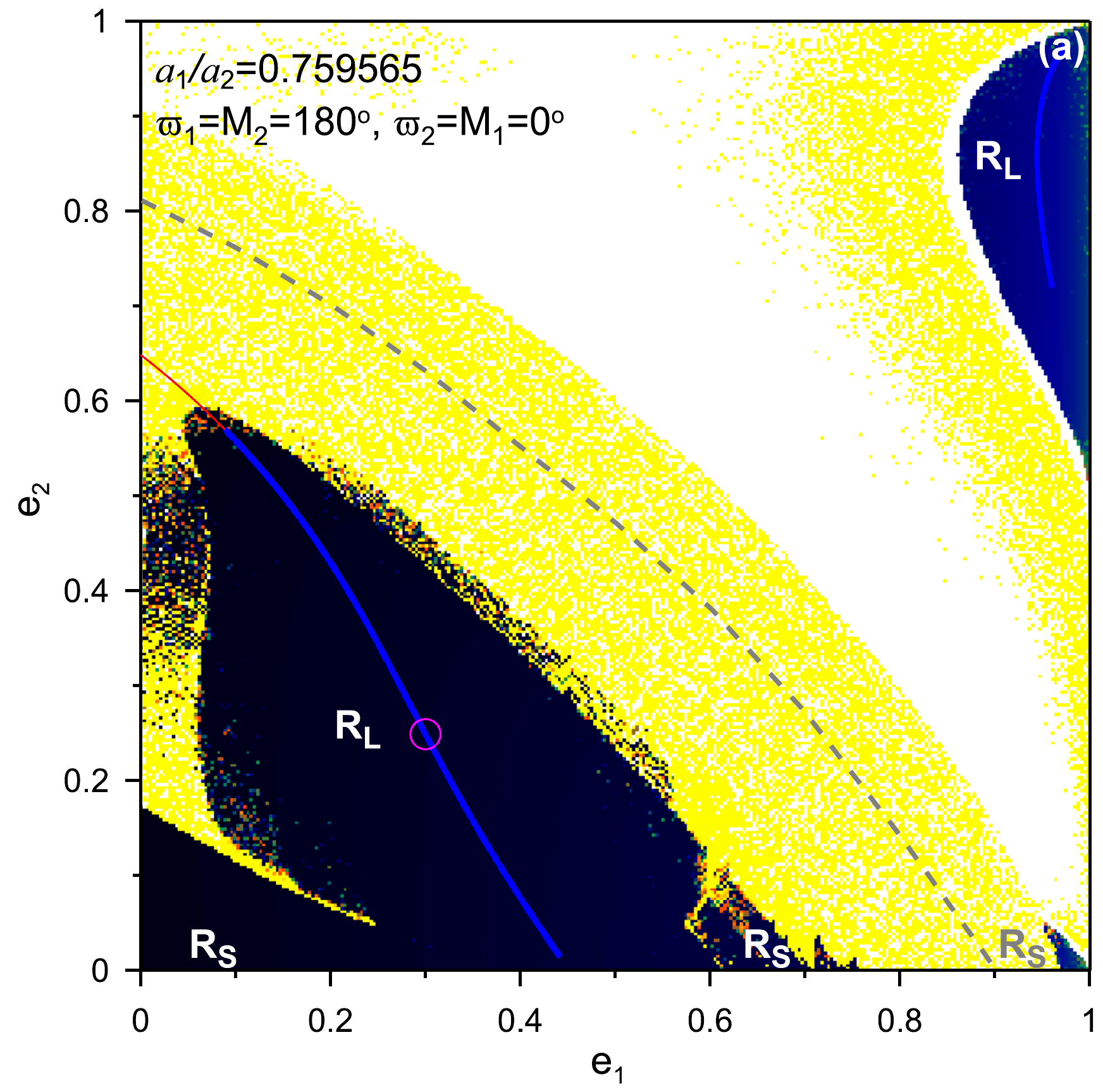} & \qquad&  \includegraphics[width=6cm]{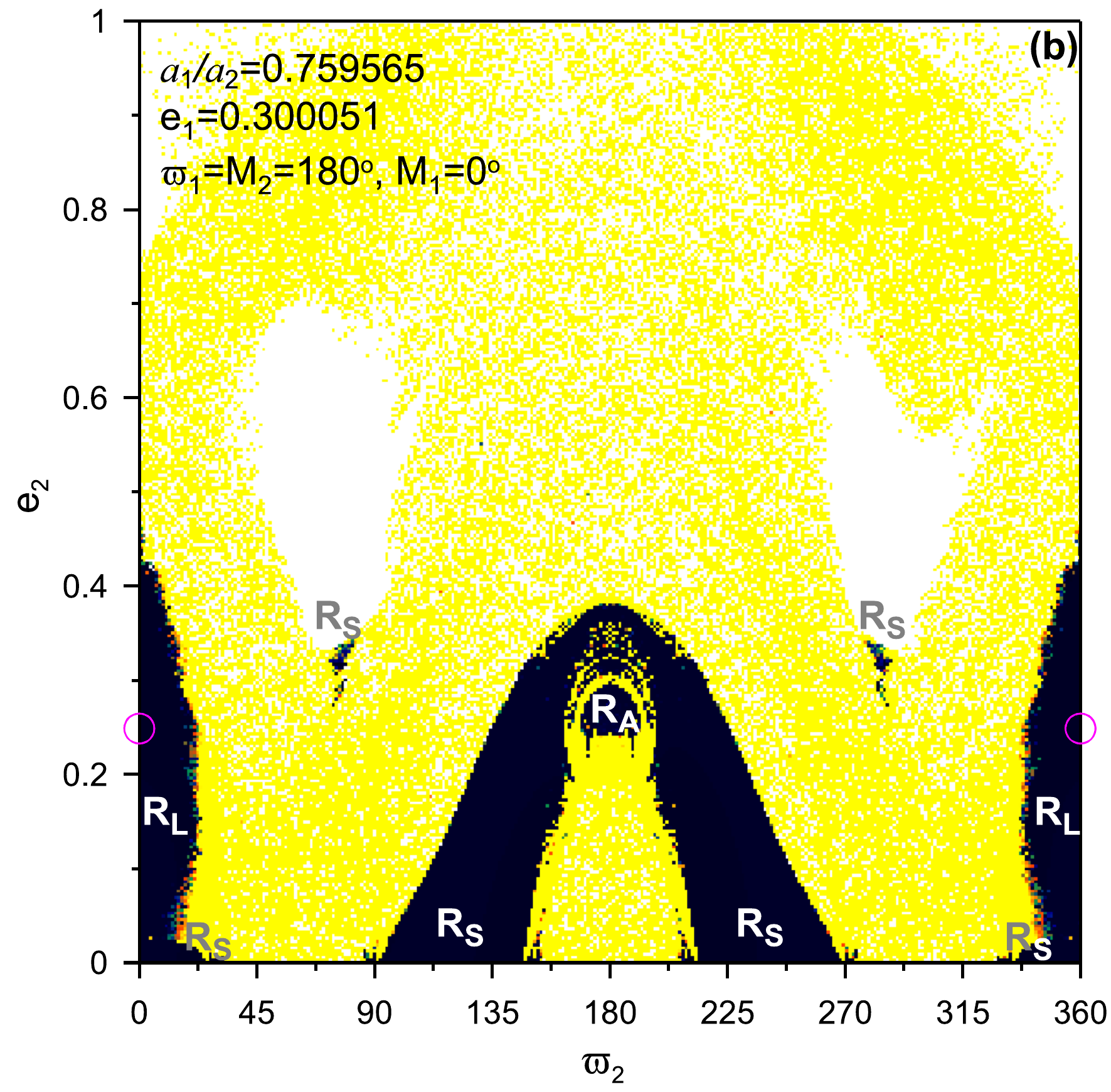}\\
\includegraphics[width=6cm]{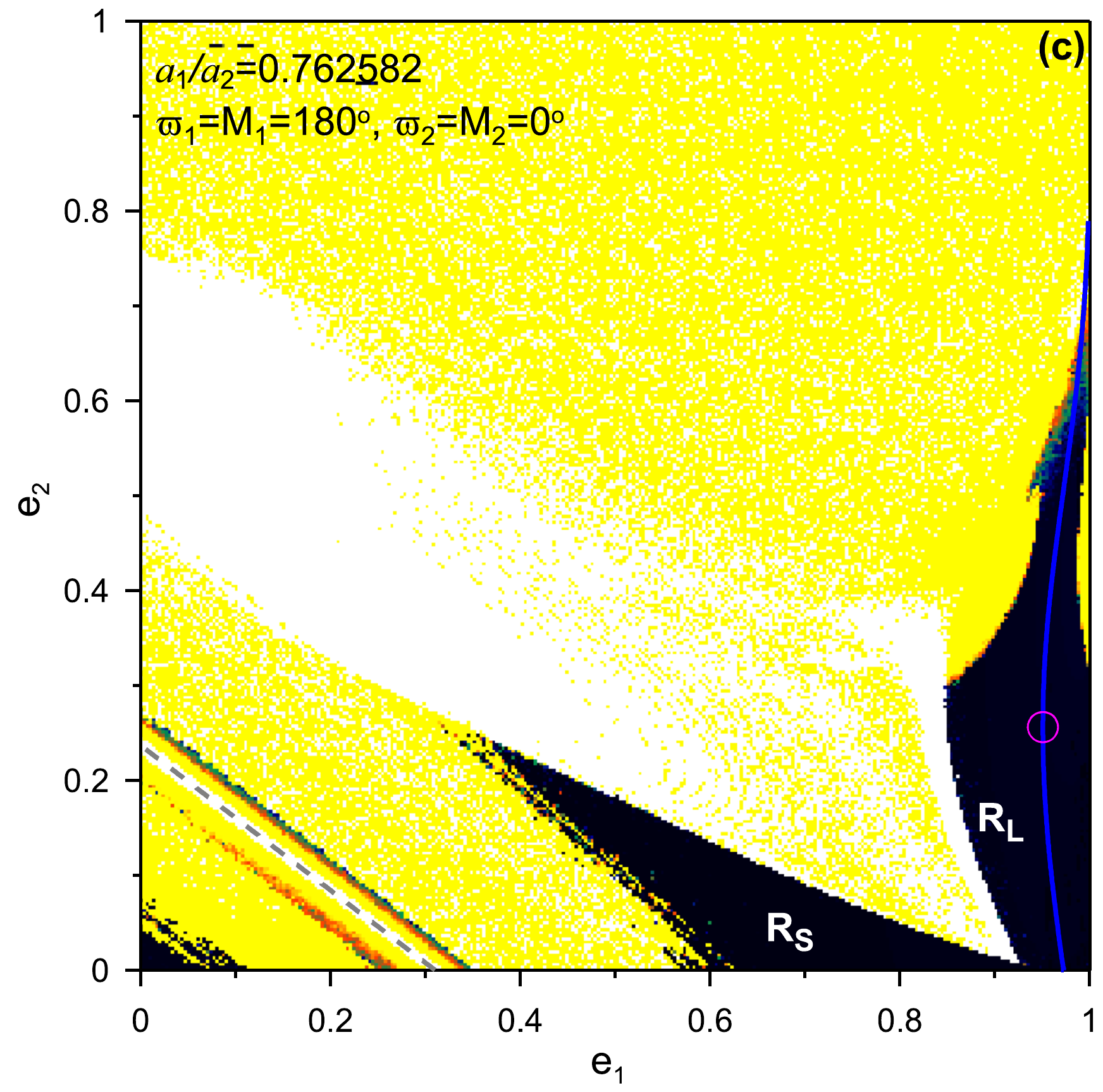} & \qquad& \includegraphics[width=6cm]{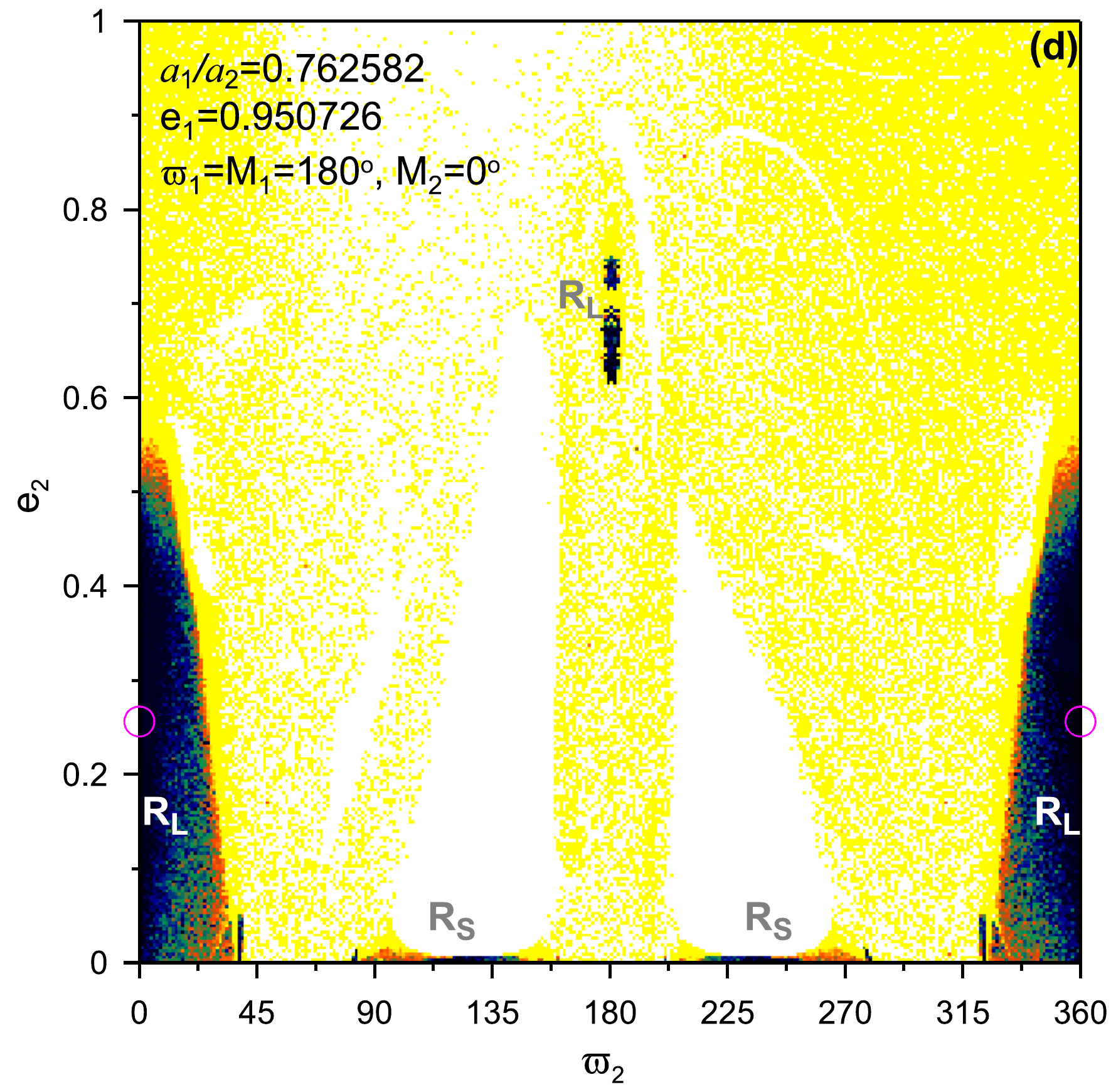}  
\end{array} $
\end{center}
\caption{DS-maps guided by stable periodic orbits of the 3/2 MMR (see Fig. \ref{32_all}c) in the configuration ($\theta_1,\theta_2$)=($0,\pi$) on the planes \textbf{a} ($e_1,e_2$) and \textbf{b} ($\varpi_2,e_2$). Accordingly, in the panels {\bf c} and {\bf d}, the choice of the periodic orbit is made from the configuration ($\theta_1,\theta_2$)=($\pi,0$). Presented as in Fig. \ref{32_00pp0}}
\label{32_0pp0}
\end{figure}

In Fig. \ref{32_0pp0}, we present DS-maps, likewise Fig. \ref{32_00pp0}. In Figs. \ref{32_0pp0}a,b, we are guided by a stable periodic orbit within the stable segment of the family evolving in the configuration ($0,\pi$), while in Figs. \ref{32_0pp0}c,d, we select a stable periodic orbit from the configuration ($\pi,0$). In Figs. \ref{32_0pp0}a,b, whenever an $R_L$ is observed, $(\theta_1,\theta_2)$ and $\Delta\varpi$ librate about $(0,\pi)$ and $\pi$, respectively. When an 1/1 secondary resonance is apparent, $\theta_1$ librates about 0. In Figs. \ref{32_0pp0}c,d, when the $R_L$ is related to the selected periodic orbit, $(\theta_1,\theta_2)$ and $\Delta\varpi$ librate about $(\pi,0)$ and $\pi$, respectively. When an 1/1 secondary resonance is apparent, $\theta_1$ librates about $\pi$. In  Fig. \ref{32_0pp0}d, when $\varpi_2$ is near $\pi$, we get an $R_L$ associated with another stable periodic orbit of the configuration ($0,0$). 

\clearpage
\subsection{5/2 MMR}

\begin{figure}[H]\centering
$\begin{array}{cp{-1.5cm}c}
\includegraphics[width=6.0cm]{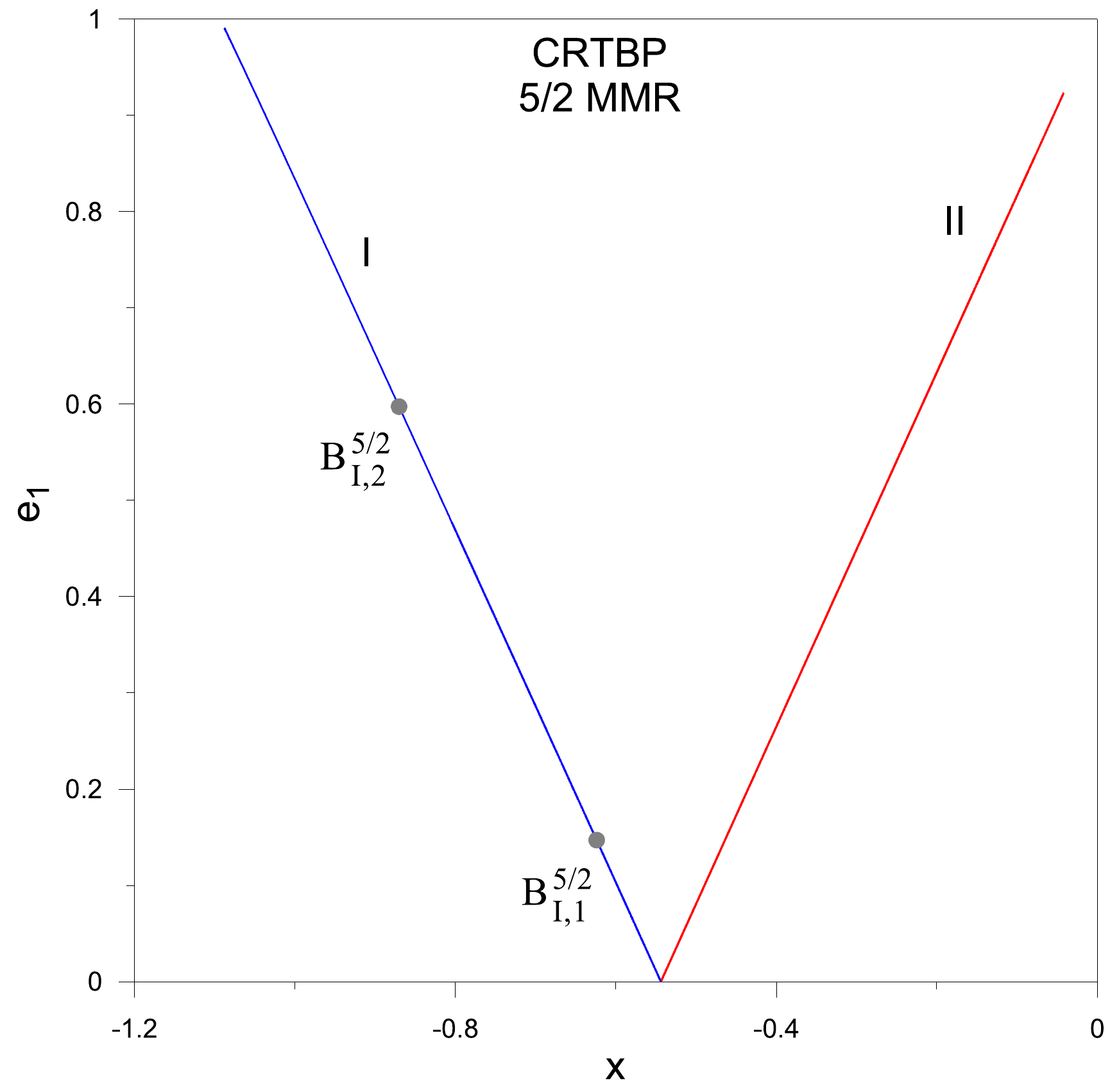}  & \qquad&
\includegraphics[width=5.9cm]{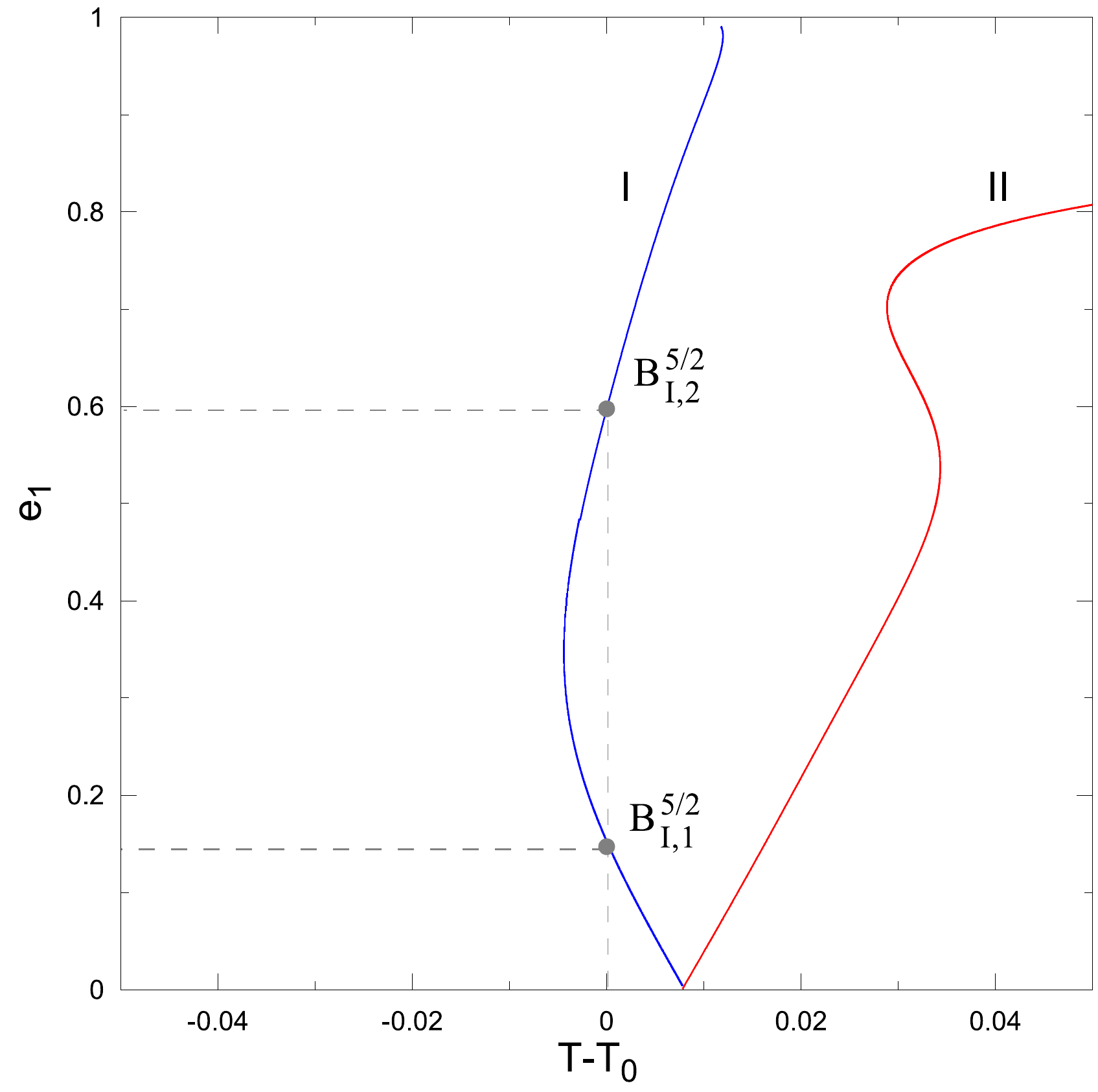} \\
\textnormal{(a)} & \qquad & \textnormal{(b)} 
\end{array} $
$\begin{array}{c}
\includegraphics[width=10cm]{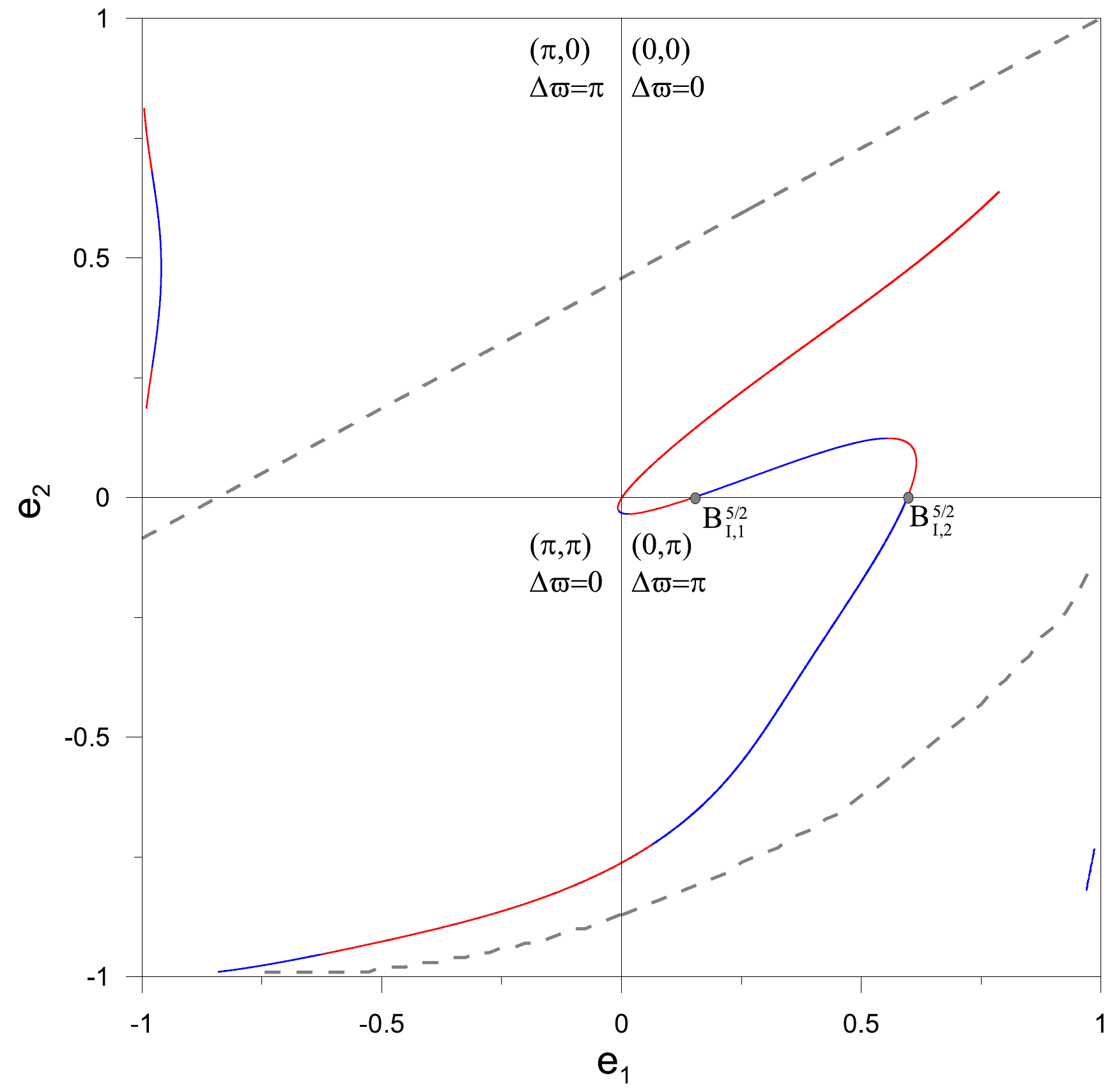} \\
\textnormal{(c)} 
\end{array} $
\caption{{\bf a} Families of periodic orbits in 5/2 MMR of the CRTBP presented as in Fig. \ref{32_all}a.  {\bf b} Justification of existence of bifurcation points in the families of CRTBP in 5/2 MMR, where $T=T_0=4\pi/3$, that generate periodic orbits in the ERTBP. {\bf c} Families of periodic orbits in 5/2 MMR of the ERTBP, presented as in Fig. \ref{32_all}c}
\label{52_all}
\end{figure}

In Fig. \ref{52_all}a, we present the families of periodic orbits in 5/2 MMR in the CRTBP which are generated by the bifurcation point of the circular family with $a_1=\frac{5}{2}^{-2/3}$. At $x\approx 0.542884$  two branches are formed: $I$, which consists of stable periodic orbits and $II$, where the orbits are unstable. The latter family could not be continued for higher values of the eccentricity of the small body, $e_1$. 

Along the family $I$ there exist two bifurcation points, $B^{5/2}_{I,1}$ and $B^{5/2}_{I_,2}$, when the period of the periodic orbits, $T$, equates to $T_0=4\pi/3$. In Fig. \ref{52_all}b, we justify their existence. 

In Fig. \ref{52_all}c, we provide the families of symmetric periodic orbits in the ERTBP. From the bifurcation point $B^{5/2}_{I,1}$, two new families are computed: one stable (configuration $(\theta_1,\theta_2)=(0,0)$) and one unstable (configuration $(0,\pi)$). The two families emanating from $B^{5/2}_{I,2}$ are continued at the configurations $(0,0)$ (unstable branch) and $(0,\pi)$ (stable branch). In the configuration $(0,0)$, the unstable branch generated by $B^{5/2}_{I,2}$ joins smoothly with the stable branch generated by $B^{5/2}_{I,1}$.
Similarly to 3/2 MMR, along the stable family belonging to $(0,\pi)$ and generated by $B^{5/2}_{I,2}$, there is a change of the configuration, when $P_1$ reaches $e_1=0$ and the family then evolves in the configuration $(\pi,\pi)$. There are segments of stable periodic orbits in each of these configurations. The families emanating from $B^{5/2}_{I,2}$ were previously studied by \citet{fetsukla92}, but for low values of the eccentricity of Jupiter, namely $e_2\leq0.25$. All of the above-mentioned families were continued by following \textit{Scheme I}.
Along the unstable family belonging to $(0,\pi)$ and generated by $B^{5/2}_{I,1}$, there is a change of the configuration when $P_1$ reaches $e_1=0$ and the family then evolves in the configuration $(\pi,\pi)$. During this change of the configuration a small stable segment exists. Then, the family becomes unstable again, meets the point $(e_1,e_2)=(0,0)$ and then evolves to the configuration $(0,0)$.
Finally, in the configuration $(\pi,0)$, we computed a family which possesses a broad segment of stable periodic orbits where the small body is highly eccentric, $e_1>0.96$. This family could not be continued for lower values of the eccentricity of the planet, $e_2$. We additionally computed an isolated family which consists of highly eccentric stable periodic orbits for both the planet and the small body and belongs to the configuration $(0,\pi)$.

\begin{figure}[H]
\begin{center}
$\begin{array}{cp{-2cm}c}
\includegraphics[width=6.0cm]{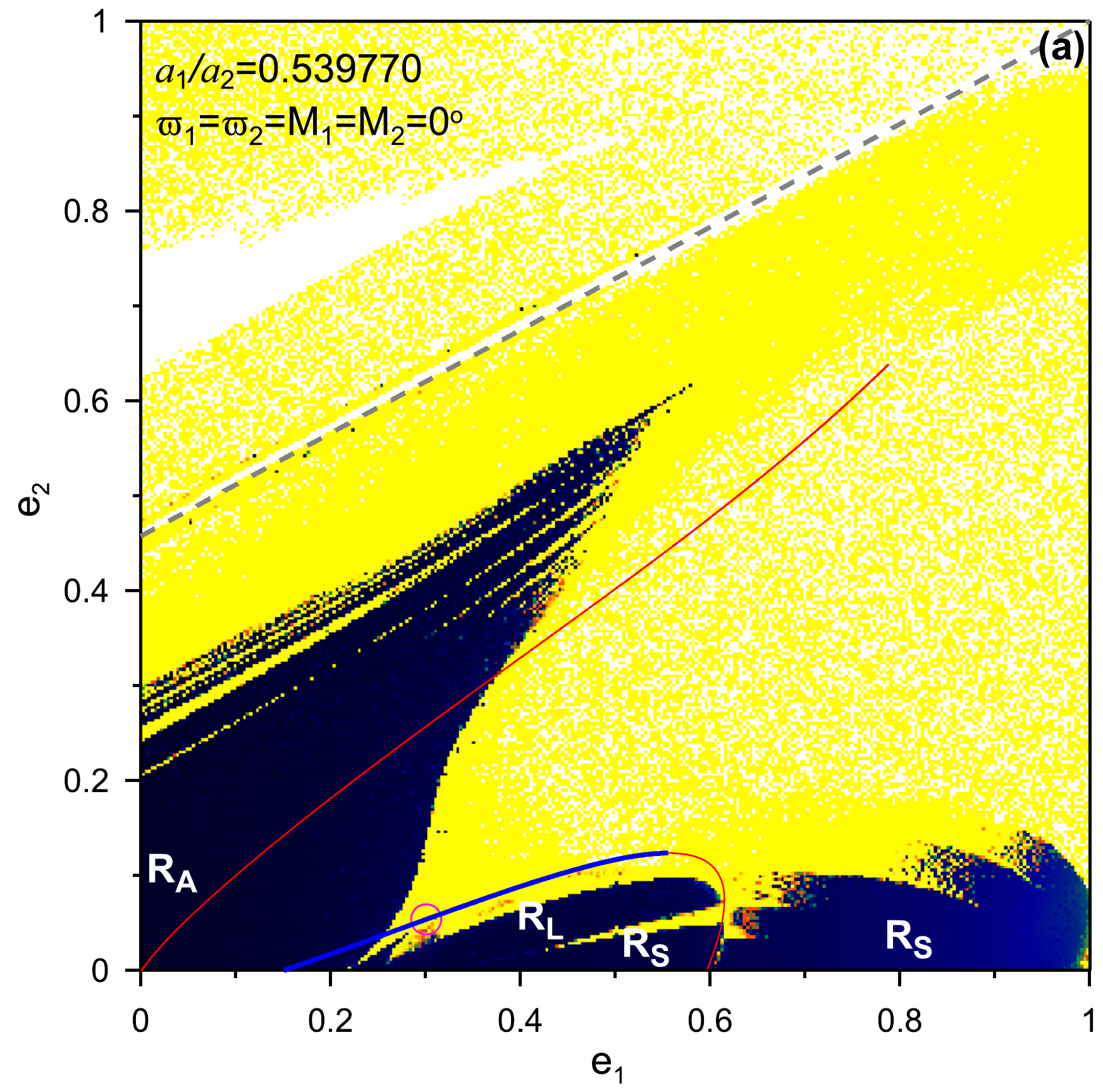} & \qquad&  \includegraphics[width=6.0cm]{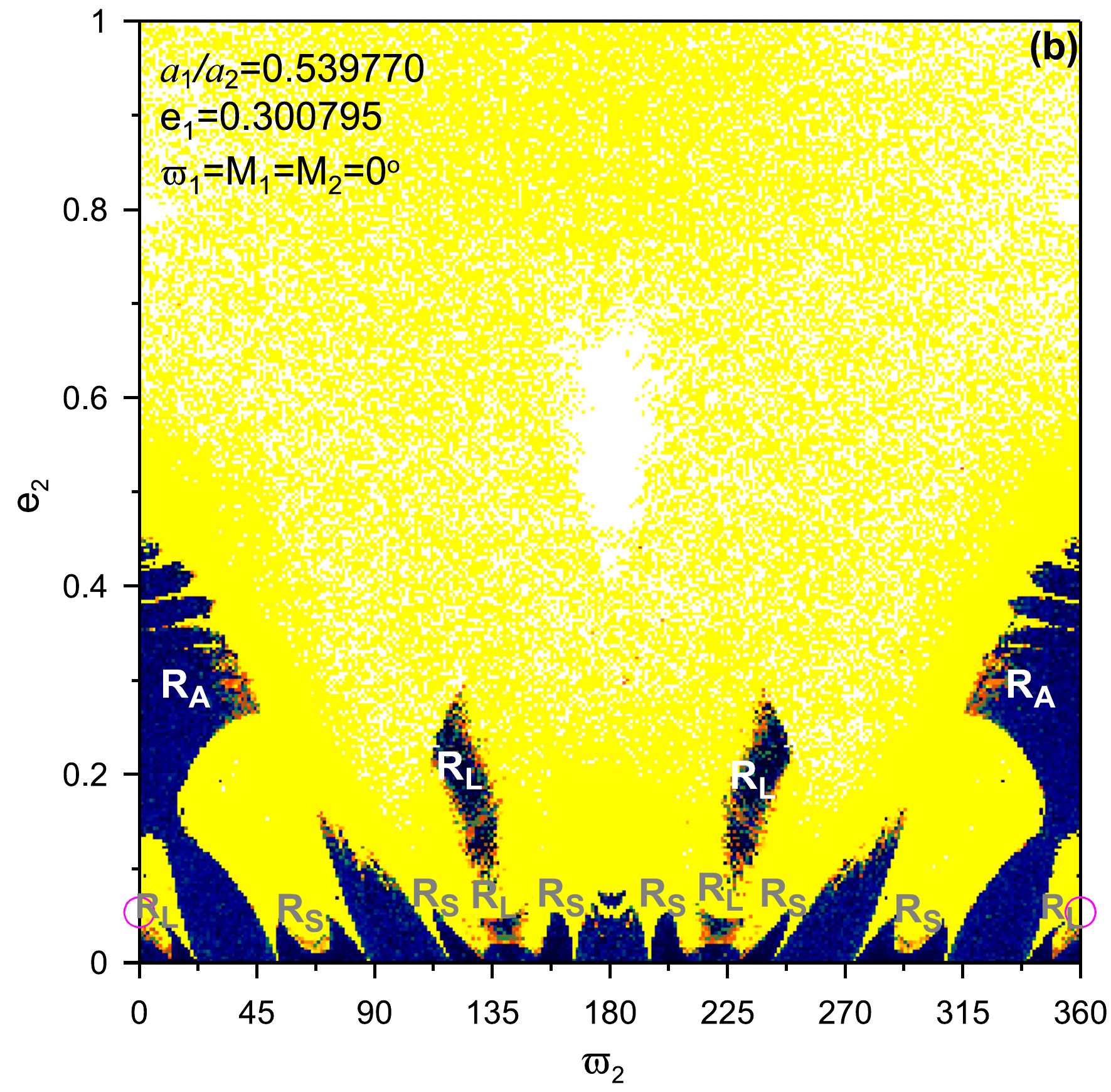}\\
\includegraphics[width=6cm]{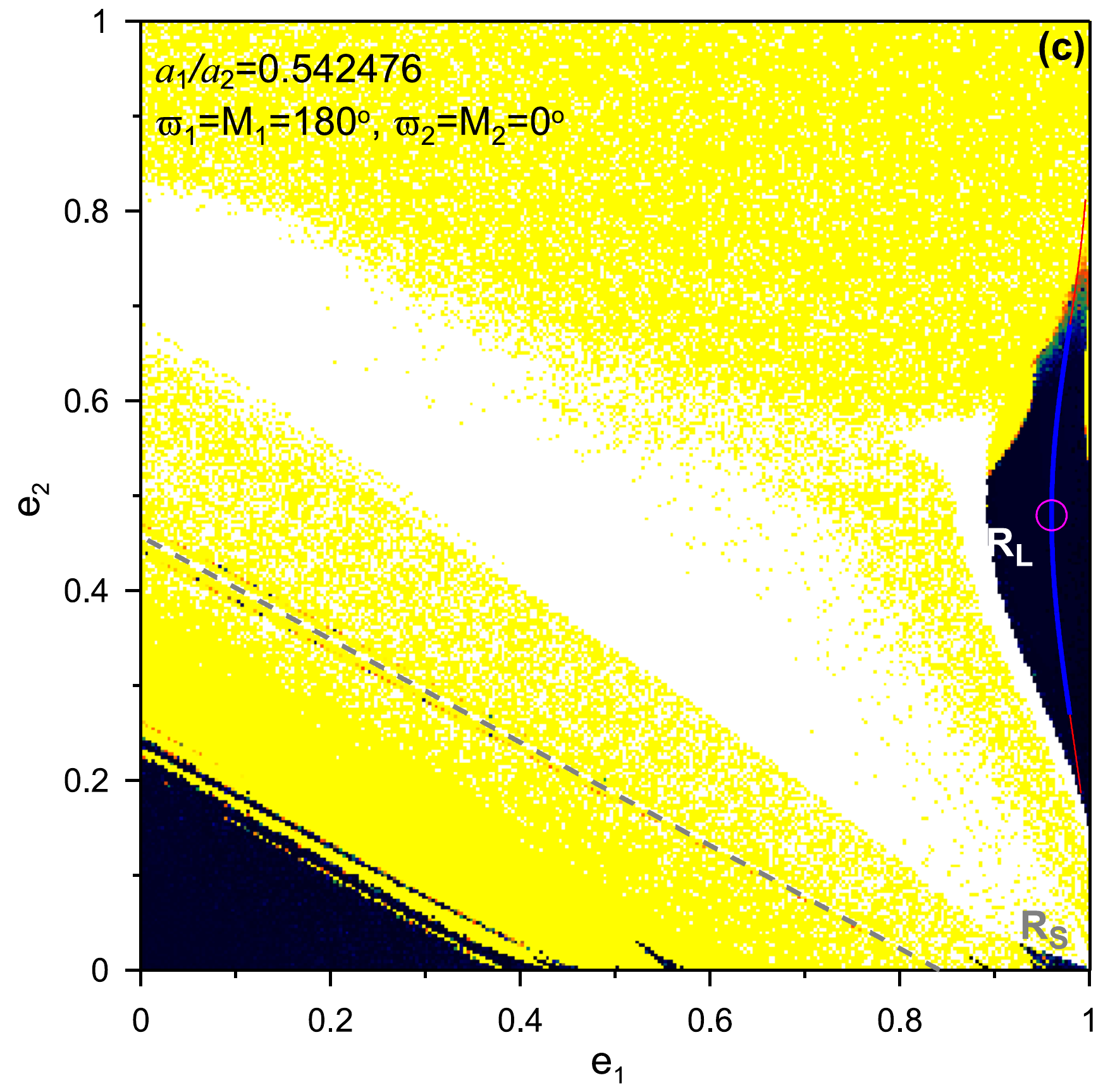} & \qquad& \includegraphics[width=6cm]{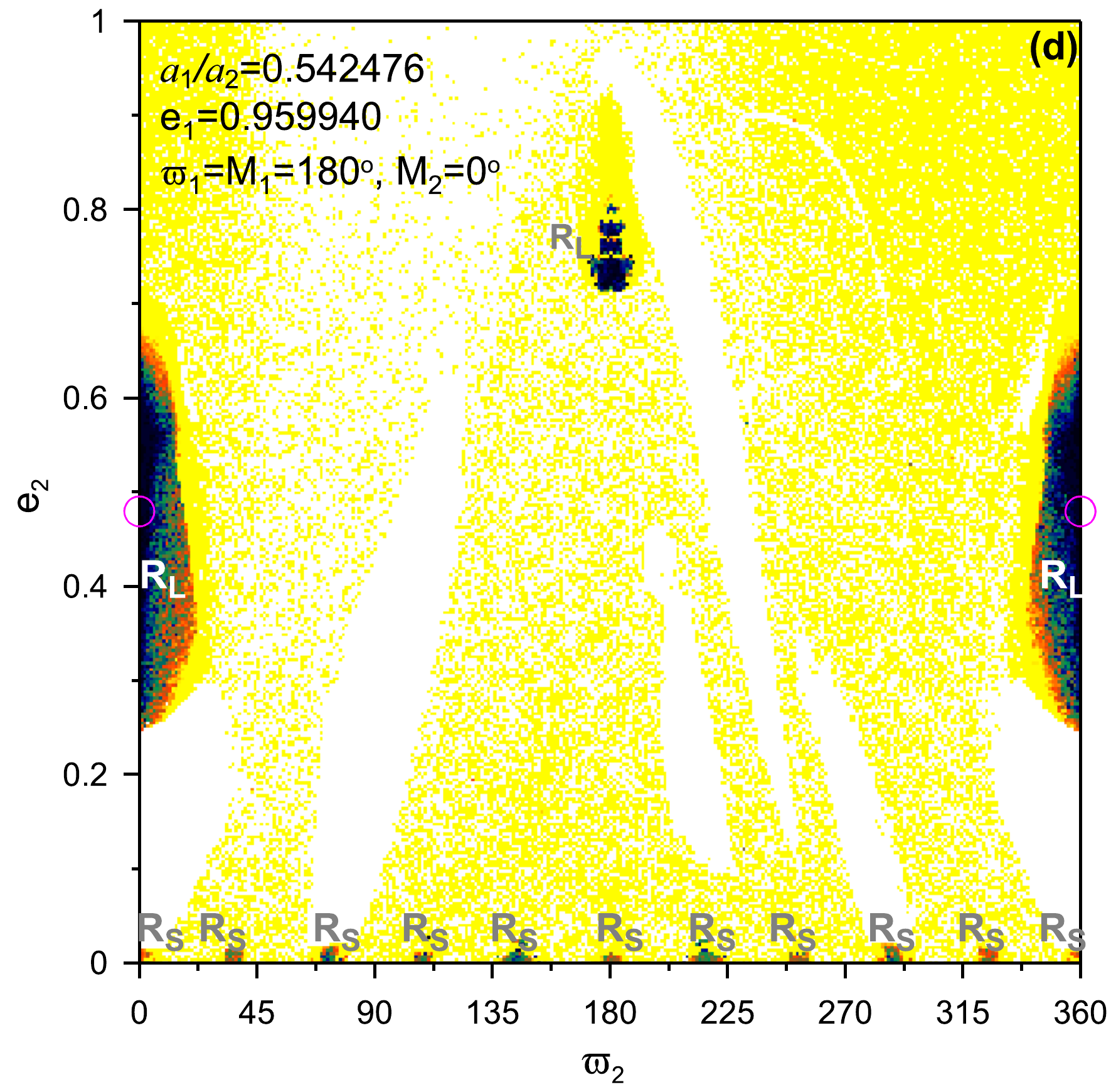} \\
\end{array} $
\end{center}
\caption{DS-maps guided by stable periodic orbits of the 5/2 MMR (see Fig. \ref{52_all}c) in the configuration ($\theta_1,\theta_2$)=($0,0$) on the planes \textbf{a} ($e_1,e_2$) and \textbf{b} ($\varpi_2,e_2$). Accordingly, in the panels {\bf c} and {\bf d}, the choice of the periodic orbit is made from the configuration ($\theta_1,\theta_2$)=($\pi,0$). Presented as in Fig. \ref{32_00pp0}}
\label{52_00p0}
\end{figure}

Likewise 3/2 MMR, in Fig. \ref{52_00p0}, we present two DS-maps on the planes $(e_1,e_2)$ and $(\varpi_2,e_2)$ (two columns) for two different configurations (two rows). In Figs. \ref{52_00p0}a,b, we are guided by a stable periodic orbit (magenta circle) of the configuration ($\theta_1,\theta_2$)=($0,0$), while in Figs. \ref{52_00p0}c,d, the configuration is ($\pi,0$). In the regular domains that are built about the selected stable periodic orbit, we observe an $R_L$, where the resonant angles, $\theta_1$, $\theta_2$ and the apsidal difference, $\Delta\varpi$, librate about the angles of the respective configuration. When $R_A$ is observed, $\Delta\varpi$ oscillates about 0. The secondary resonance, $R_S$, that is observed, is 3/1, where $\theta_1$ librates about 0 (panels a, b and d (see below)) or $\pi$ (panels c and d). Additionally, in Fig. \ref{52_00p0}b, we observe that an $R_L$ is taking place when $\varpi_2$ is around $7\pi/10$ (or $13\pi/10$ symmetrically) and corresponds to a stable periodic orbit of the configuration ($\theta_1,\theta_2$)=($0,\pi$). In Fig. \ref{52_00p0}d, in $R_S$ $\theta_1$ librates about $\pi$, when $\varpi_2$ is near 0, $2\pi/5$ and $4\pi/5$ (or $2\pi$, $8\pi/5$ and $6\pi/5$ symmetrically) and about 0, when $R_S$ is noted at the rest angles' values ($\pi/5$, $3\pi/5$ (or $9\pi/5$, $7\pi/5$ symmetrically) and $\pi$).

\begin{figure}[H]
\begin{center}
$\begin{array}{cp{-2cm}c}
\includegraphics[width=6.0cm]{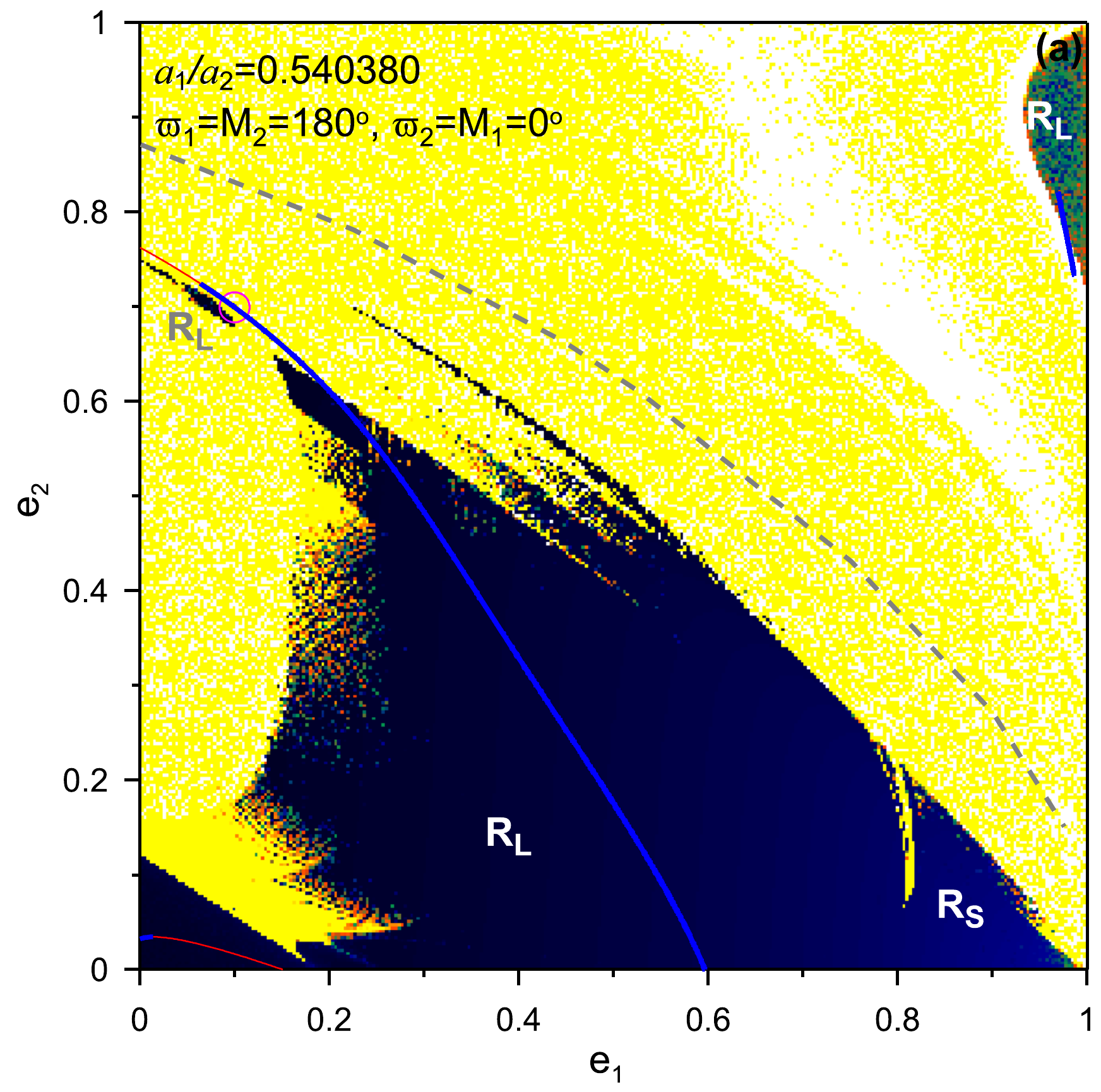} & \qquad&  \includegraphics[width=6.0cm]{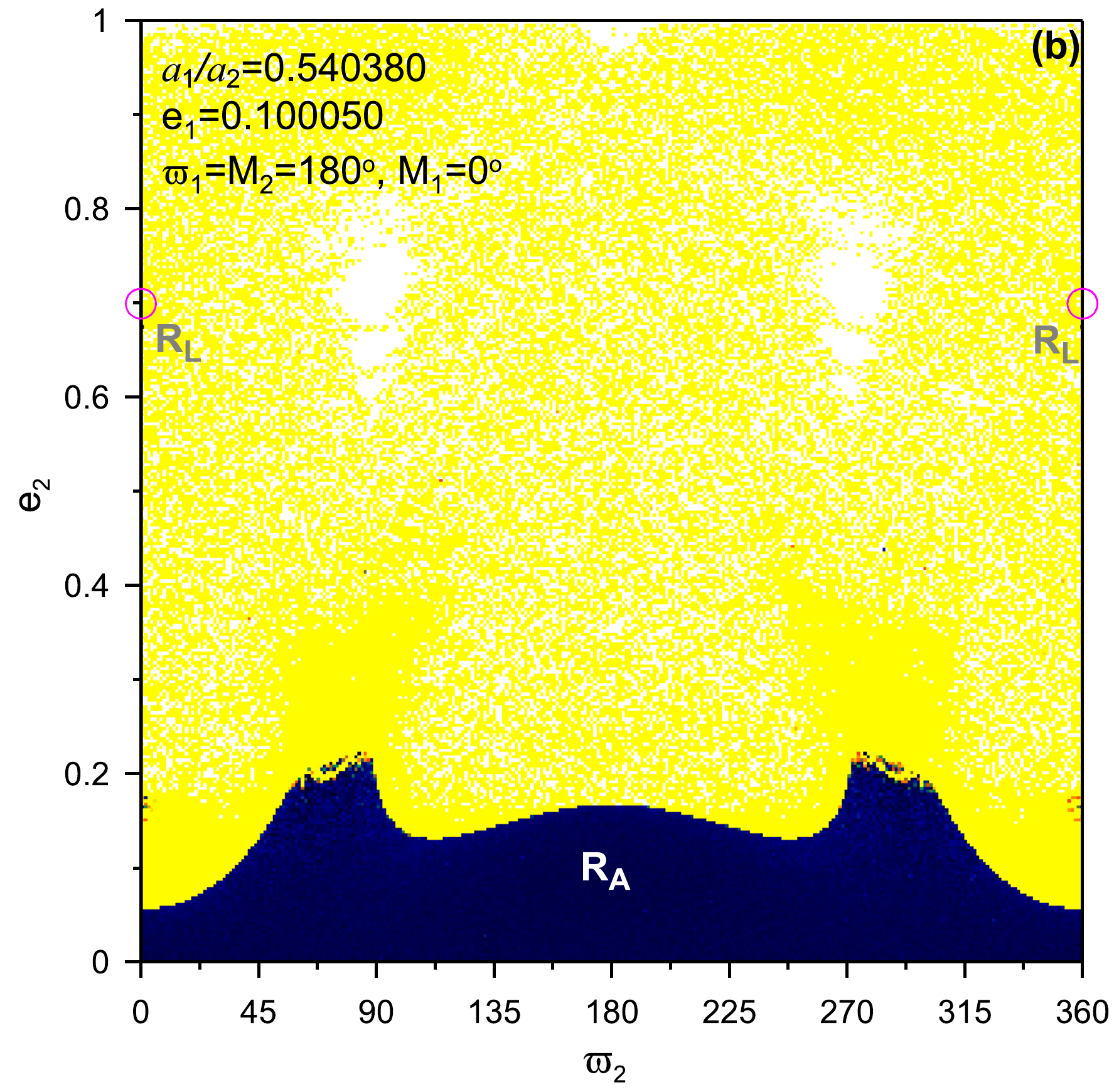}\\
\includegraphics[width=6cm]{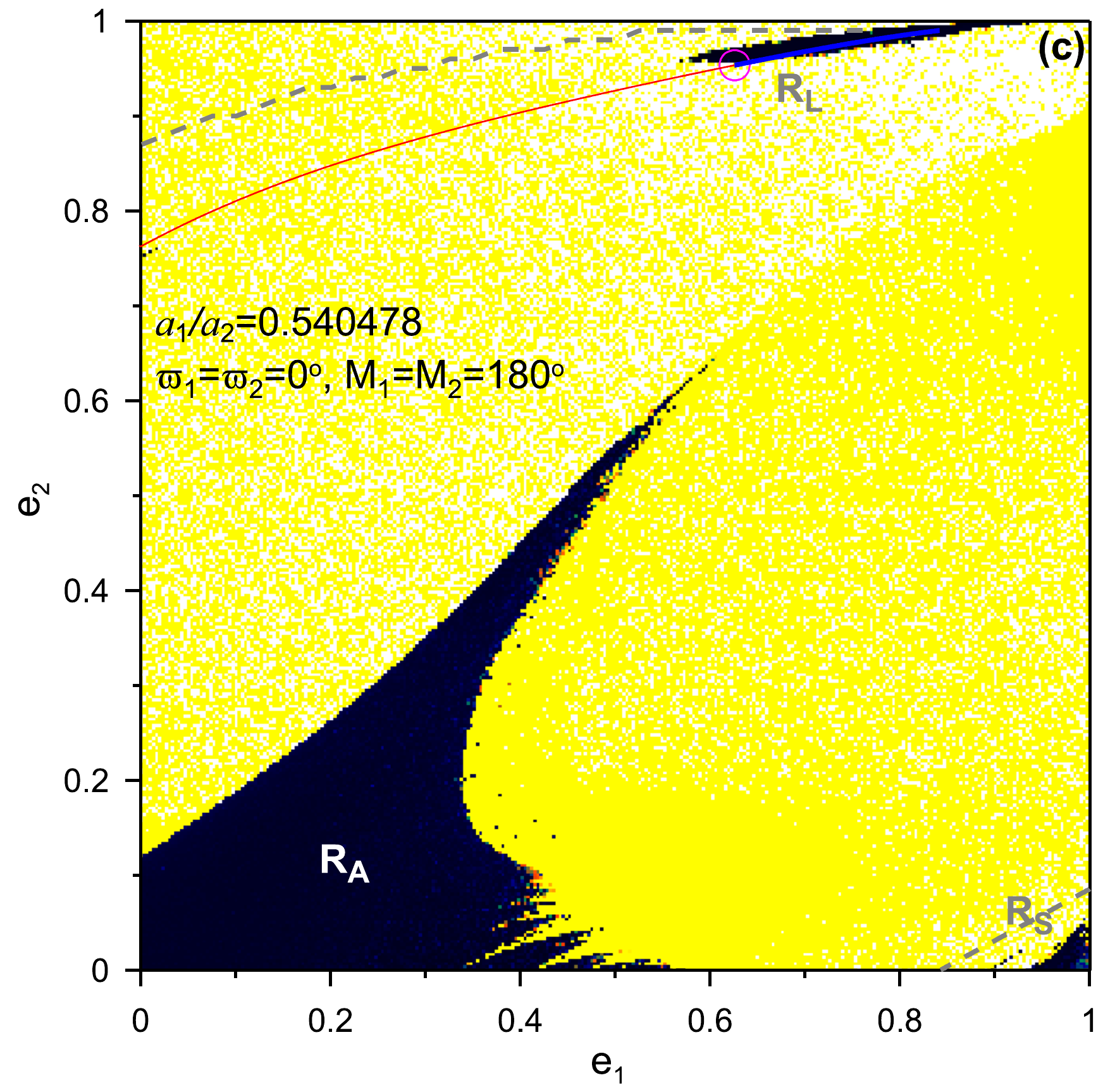} & \qquad& \includegraphics[width=6cm]{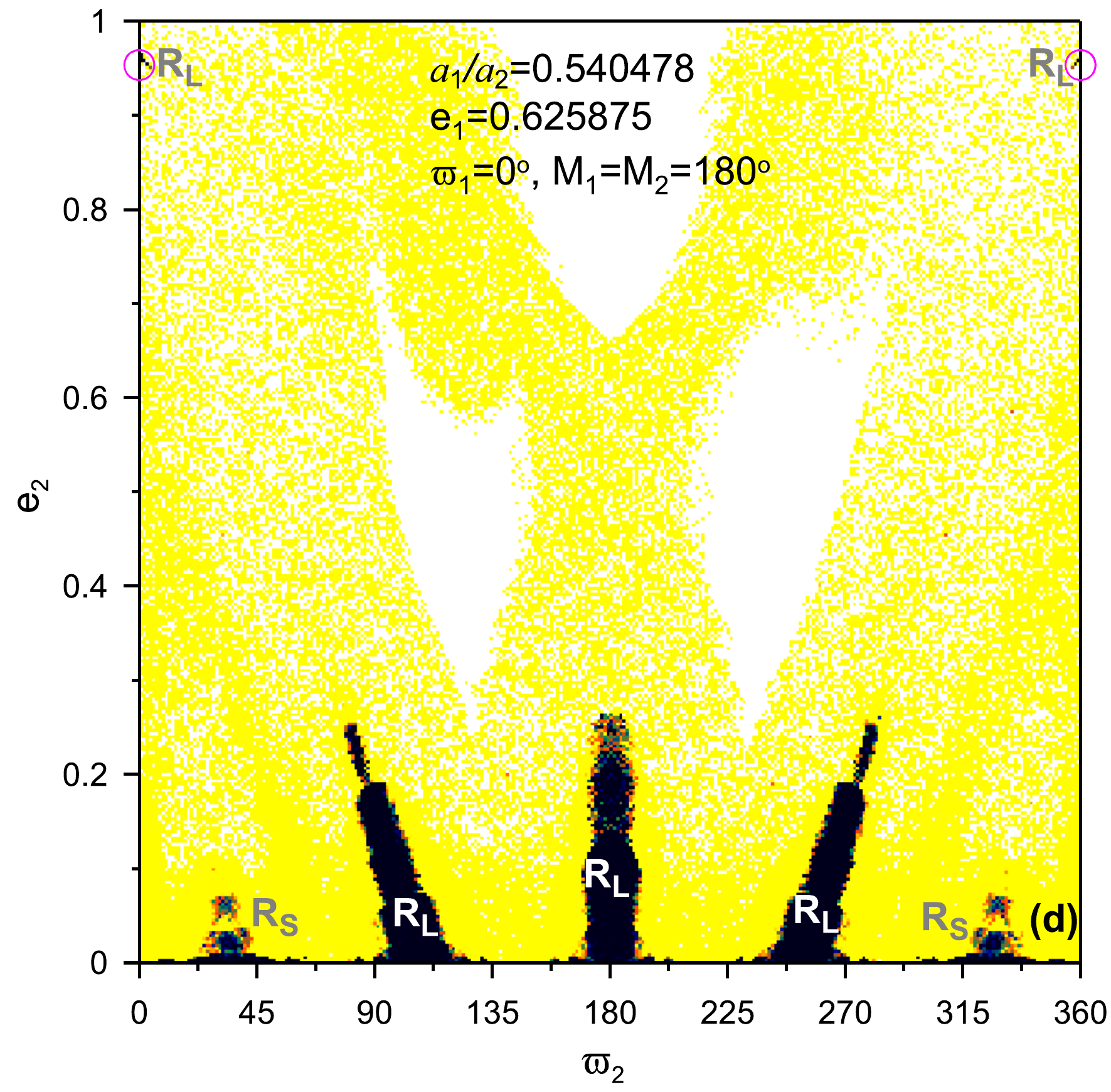}  
\end{array} $
\end{center}
\caption{DS-maps guided by stable periodic orbits of the 5/2 MMR (see Fig. \ref{52_all}c) in the configuration ($\theta_1,\theta_2$)=($0,\pi$) on the planes \textbf{a} ($e_1,e_2$) and \textbf{b} ($\varpi_2,e_2$). Accordingly, in the panels {\bf c} and {\bf d}, the choice of the periodic orbit is made from the configuration ($\theta_1,\theta_2$)=($\pi,\pi$). Presented as in Fig. \ref{32_00pp0}}
\label{52_0p0}
\end{figure}

In Fig. \ref{52_0p0}, we present two DS-maps on the planes $(e_1,e_2)$, $(\varpi_2,e_2)$ (two columns) for two different configurations (two rows). In Figs. \ref{52_0p0}a,b, we are guided by a stable periodic orbit within the stable segment of the family evolving in the configuration ($0,\pi$), while in Figs. \ref{52_0p0}c,d, we select a stable periodic orbit from the configuration ($\pi,\pi$). In Figs. \ref{52_0p0}a,b, whenever an $R_L$ is observed, $(\theta_1,\theta_2)$ and $\Delta\varpi$ librate about $(0,\pi)$ and $\pi$, respectively. When a 3/1 secondary resonance is apparent, $\theta_1$ librates about 0. In Figs. \ref{52_0p0}c,d, when the $R_L$ is related to the periodic orbit, $(\theta_1,\theta_2)$ and $\Delta\varpi$ librate about $(\pi,\pi)$ and 0, respectively. When a 3/1 secondary resonance is apparent, $\theta_1$ librates about $\pi$. However, in Fig. \ref{52_0p0}d, when $\varpi_2$ is near $\pi/5$ (or $9\pi/5$ symmetrically), $\theta_1$ librates about 0, as observed in Fig. \ref{52_00p0}e. Finally, in Fig. \ref{52_0p0}d, we get an $R_L$ associated with a stable periodic orbit of the configuration ($0,\pi$), when $\varpi_2$ is near $3\pi/5$ (or $7\pi/5$ symmetrically) and $\pi$.

\clearpage
\subsection{3/1 MMR}

\begin{figure}[H]\centering
$\begin{array}{cp{-1.5cm}c}
\includegraphics[width=6.0cm]{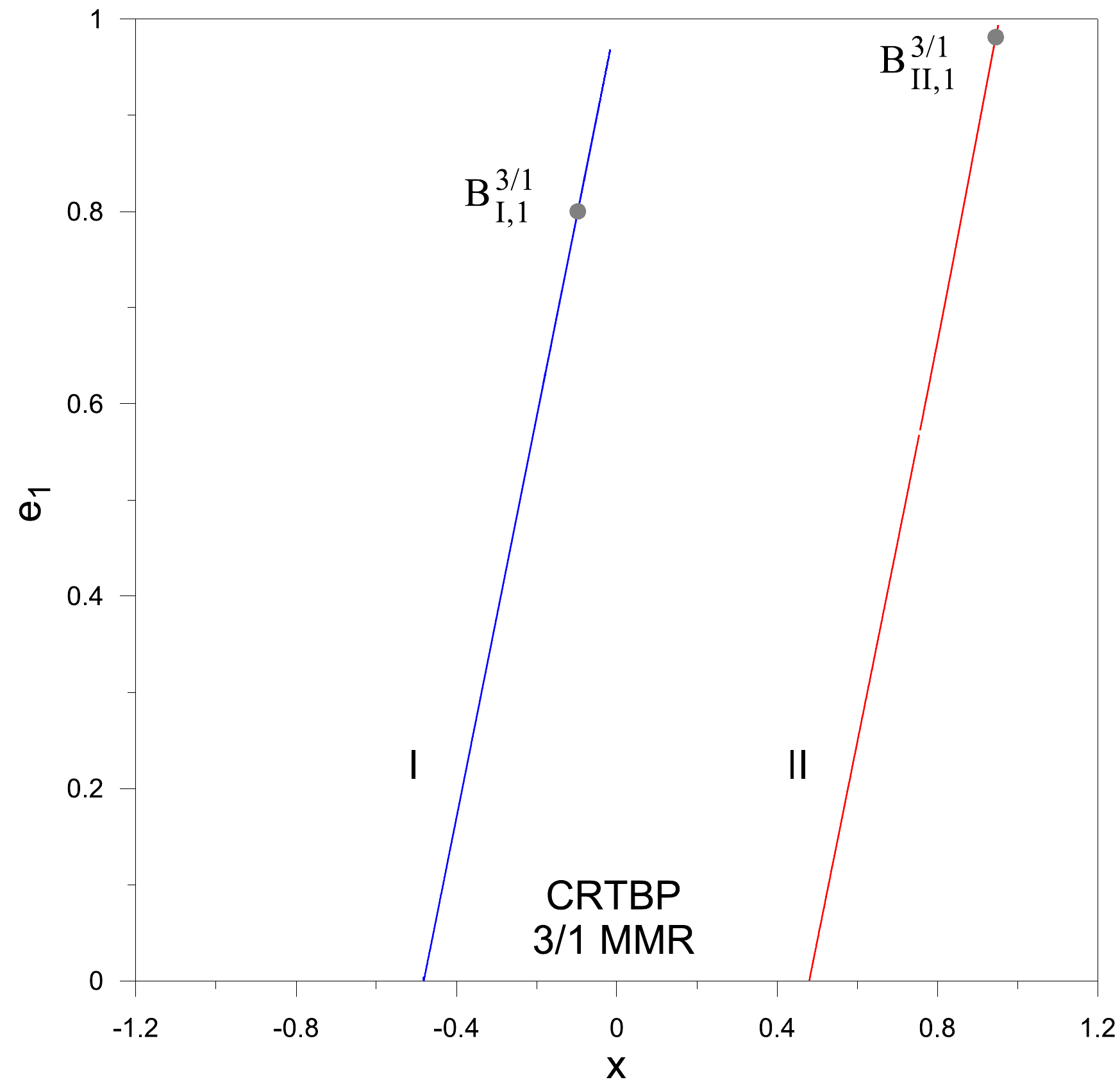}  & \qquad&
\includegraphics[width=5.8cm]{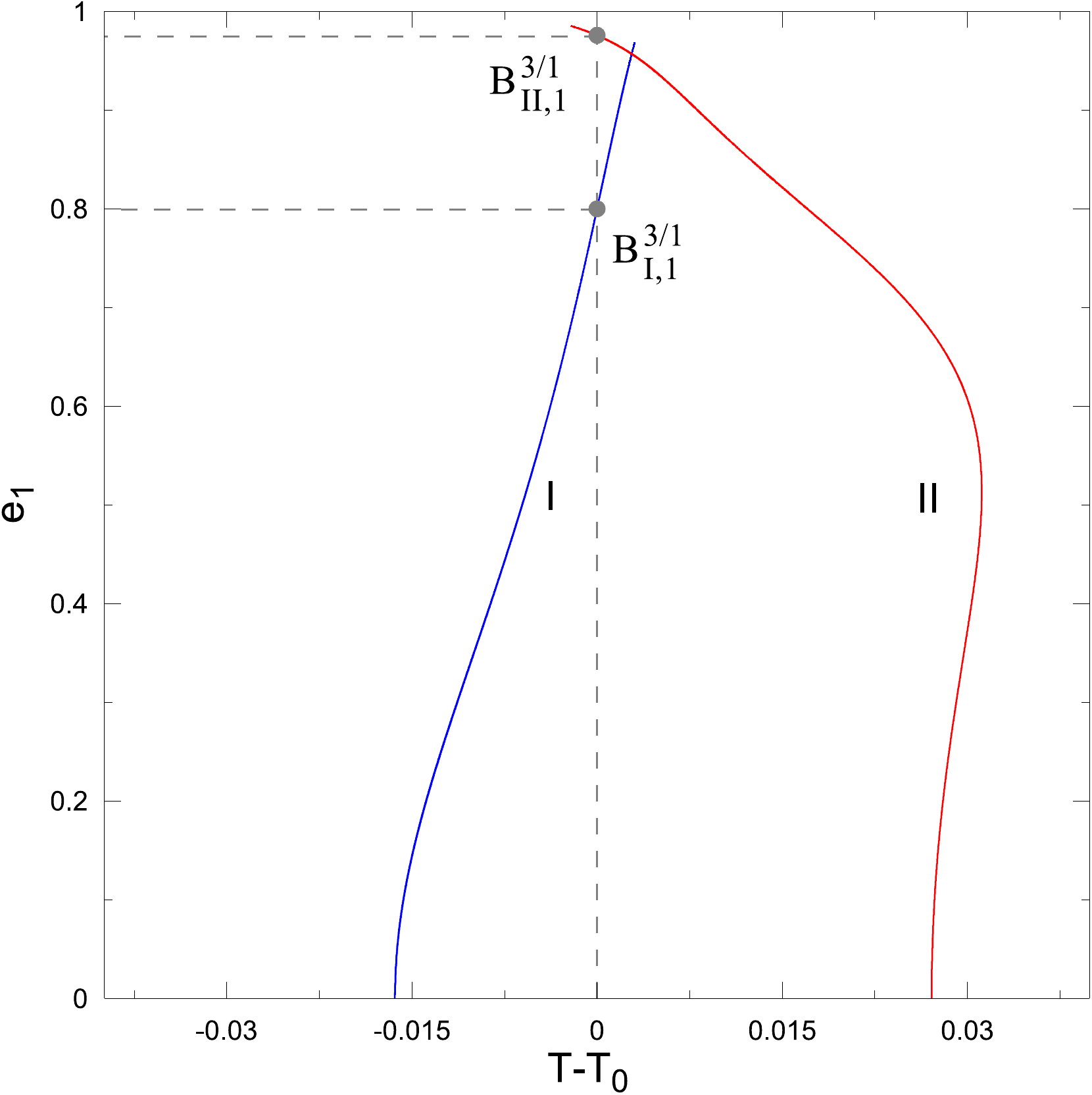} \\
\textnormal{(a)} & \qquad & \textnormal{(b)} 
\end{array} $
$\begin{array}{c}
\includegraphics[width=10cm]{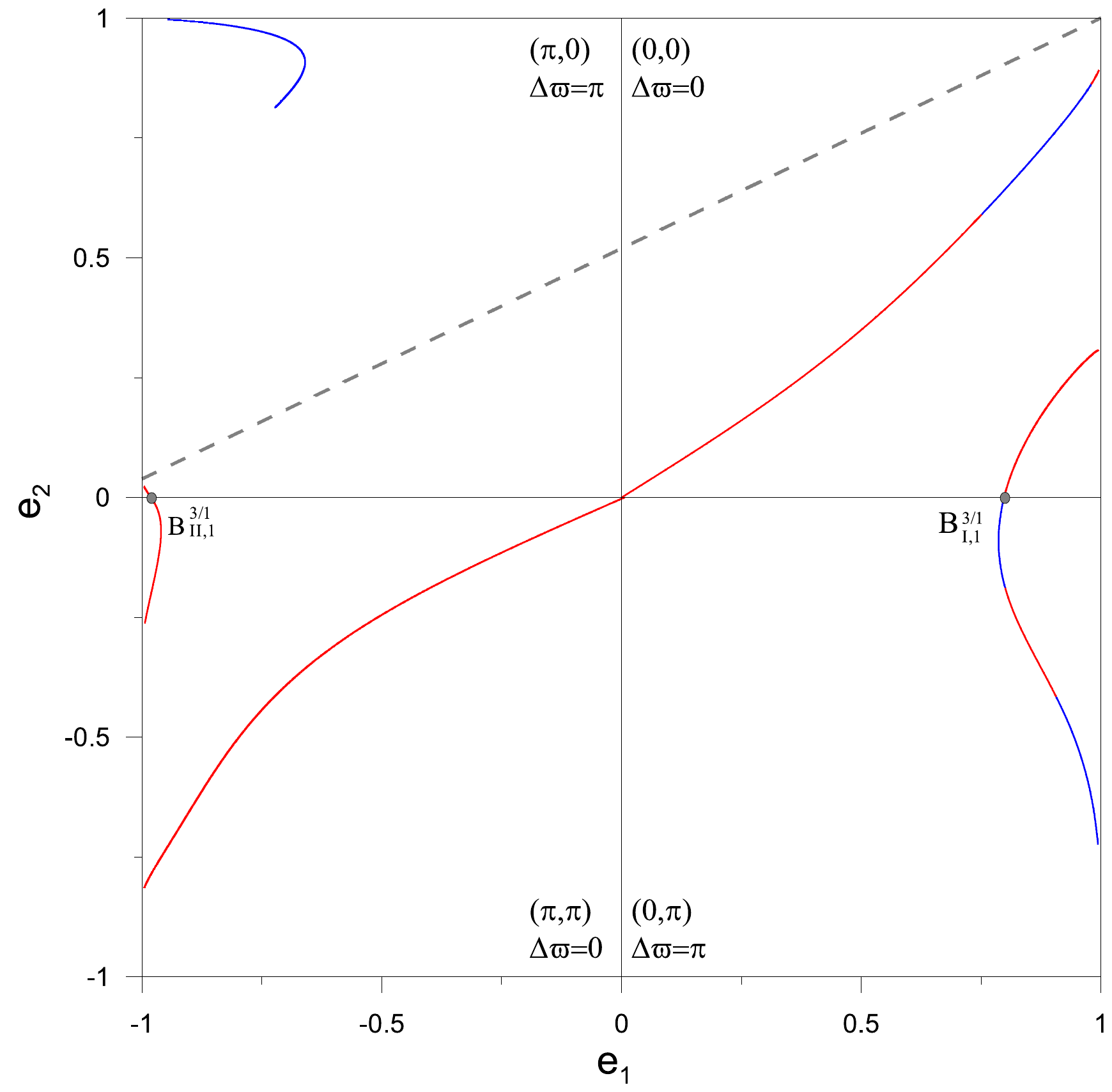} \\
\textnormal{(c)} 
\end{array} $
\caption{{\bf a} Families of periodic orbits in 3/1 MMR of the CRTBP presented as in Fig. \ref{32_all}a. {\bf b} Justification of existence of bifurcation points in the families of CRTBP in 3/1 MMR, where $T=2 T_0=2\pi$, that generate periodic orbits in the ERTBP. {\bf c} Families of periodic orbits in 3/1 MMR of the ERTBP, presented as in Fig. \ref{32_all}c. The angles in brackets represent the pair of resonant angles ($\theta_3,\theta_1$)}
\label{31_all}
\end{figure}

In Fig. \ref{31_all}a, we present the families of periodic orbits in 3/1 MMR in the CRTBP which are generated by the bifurcation points of the circular family. Along the circular family in the neighbourhood of this MMR, the periodic orbits are unstable and at the two endings of this unstable segment there exist critical orbits (the double eigenvalues are equal to -1). From those two orbits there bifurcate symmetric periodic orbits in the CRTBP with period $T_0=\pi$ which can be described twice, $T=2\;T_0$ (\textit{Scheme II}). Thus, two branches are formed: $I$, which consists of stable periodic orbits and $II$, where the orbits are unstable. These families were previously presented in \citet{hadj93,hadj93as}. We herein computed the unstable family for greater eccentricity values ($e_1$) and hence, we found another bifurcation point (see below).

Along each family, $I$ and $II$, there exists one bifurcation point when the period of the periodic orbits, $T$, equates to $2 T_0=2 \pi$. In Fig. \ref{31_all}b, we justify the existence of two such points, $B^{3/1}_{I,1}$ and $B^{3/1}_{II,1}$. 

In Fig. \ref{31_all}c, we provide the families of symmetric periodic orbits in the ERTBP. From the bifurcation point $B^{3/1}_{I,1}$, two families are generated: one being unstable (configuration $(\theta_3,\theta_1)=(0,0)$) and one starting with stable periodic orbits (configuration $(0,\pi)$). These families are respectively called $II_e$ and $I_e$ in \citet{hadj92,hadj93,hadj93as}. The family $I_e$ is continued herein for $e_2>0.3$ (in comparison with \citet{hadj93}) and hence, it is revealed that this family possesses an unstable segment before it becomes stable again. From the new bifurcation point $B^{3/1}_{II,1}$, two new families are generated, both of them are unstable and evolve to the configurations $(\pi,0)$ and $(\pi,\pi)$. All of the above-mentioned families were continued by following \textit{Scheme I}.
Along the circular family and within the unstable segment close to 3/1 MMR the period $T$ gets equal to $\pi$ (\textit{Scheme II}). This periodic orbit is continued to the ERTBP as doubly symmetric with period equal to $2\pi$ and the families that are being formed are unstable and evolve in the configurations $(0,0)$ and $(\pi,\pi)$ (see the families that are emanating from the point $(e_1,e_2)=(0,0)$). These families were mentioned in \citet{hadj92,hadj93,hadj93as} as $I_c$ and $II_c$, but were described as totally unstable since they were continued for $e_2<0.25$. We herein continued them for highly eccentric orbits for both bodies and it turned out that the family $I_c$ possesses stable periodic orbits when $0.75<e_1<0.98$.
Let us note that the families of periodic orbits studied by \citet{hadj92} had also been studied via the averaged Hamiltonian and its exact corotations by \citet{fetsukla92} as well. 
Finally, in the configuration $(\pi,0)$, we computed a new isolated family which possesses only stable periodic orbits where both the small body and the planet are highly eccentric.

\begin{figure}[H]
\begin{center}
$\begin{array}{cp{-2cm}c}
\includegraphics[width=6.0cm]{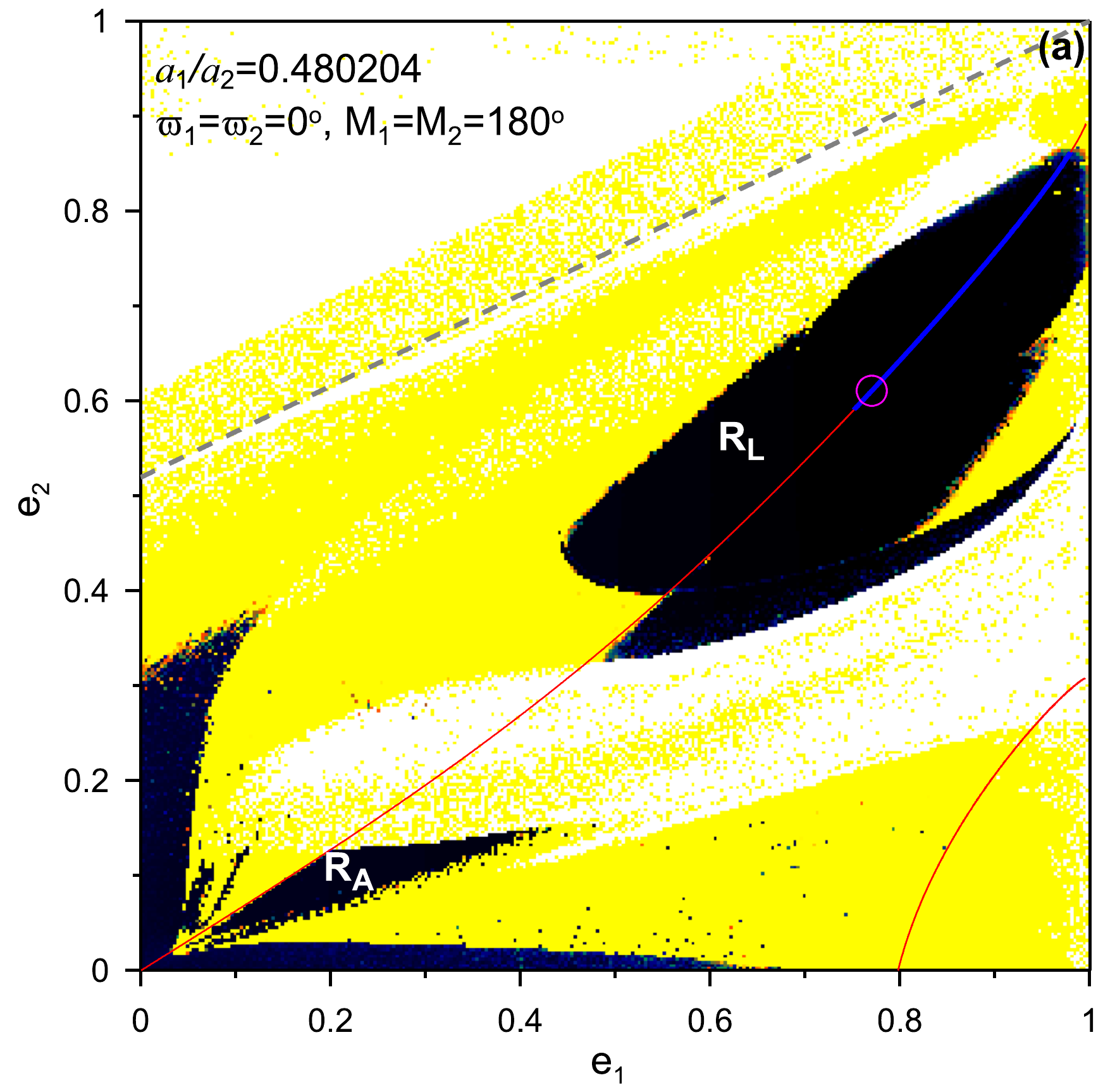} & \qquad&  \includegraphics[width=6.0cm]{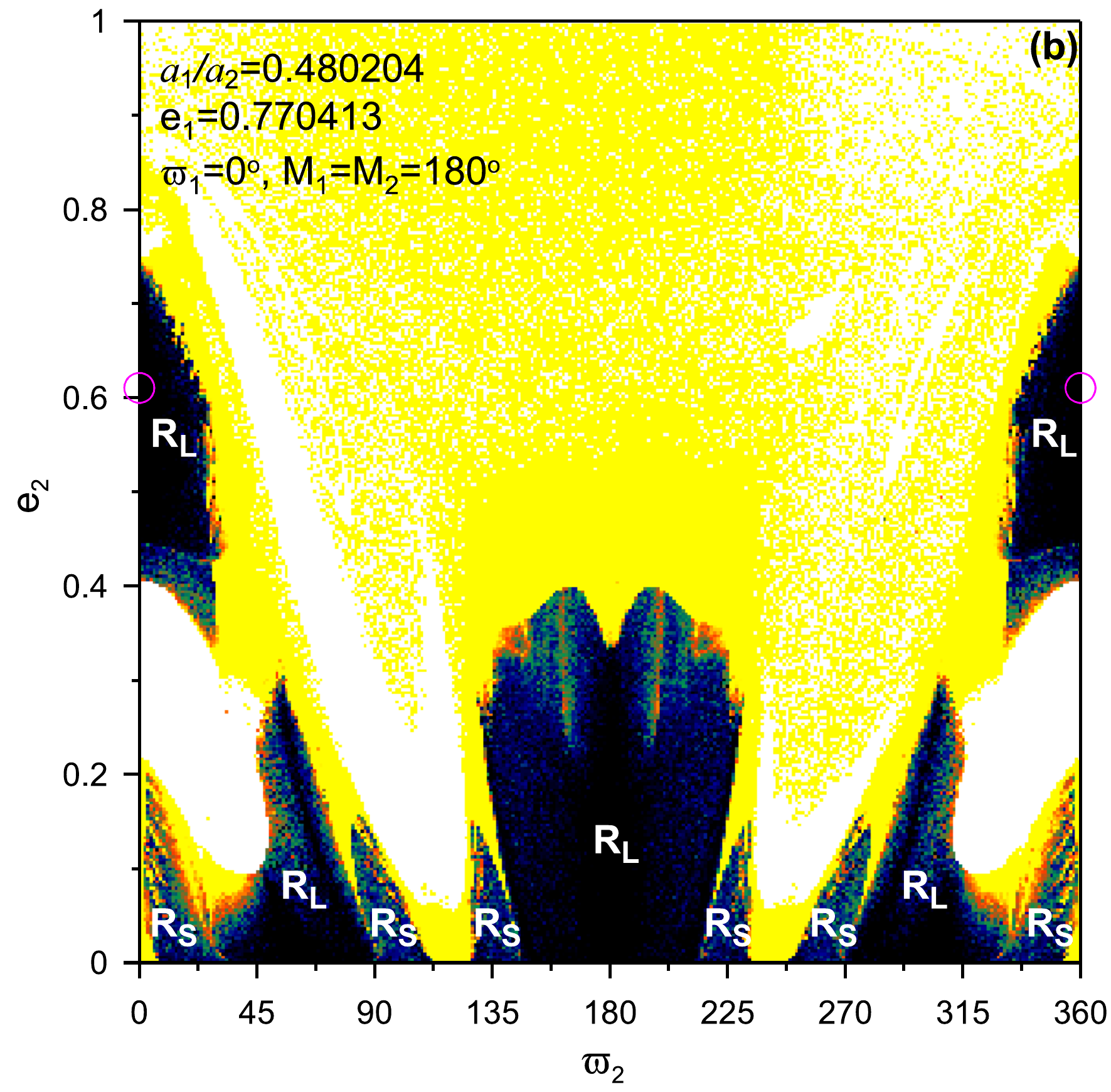}\\
\includegraphics[width=6cm]{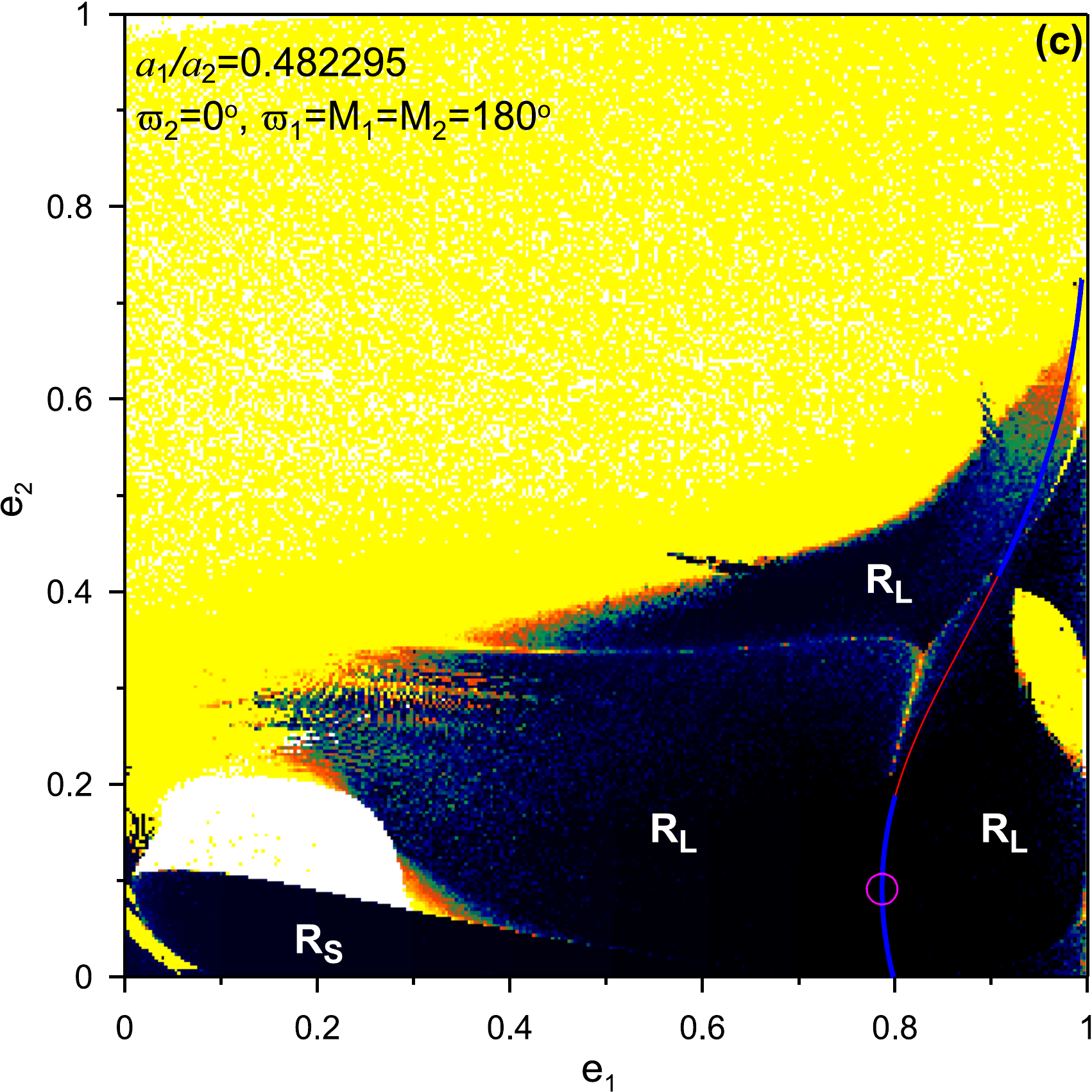} & \qquad& \includegraphics[width=6cm]{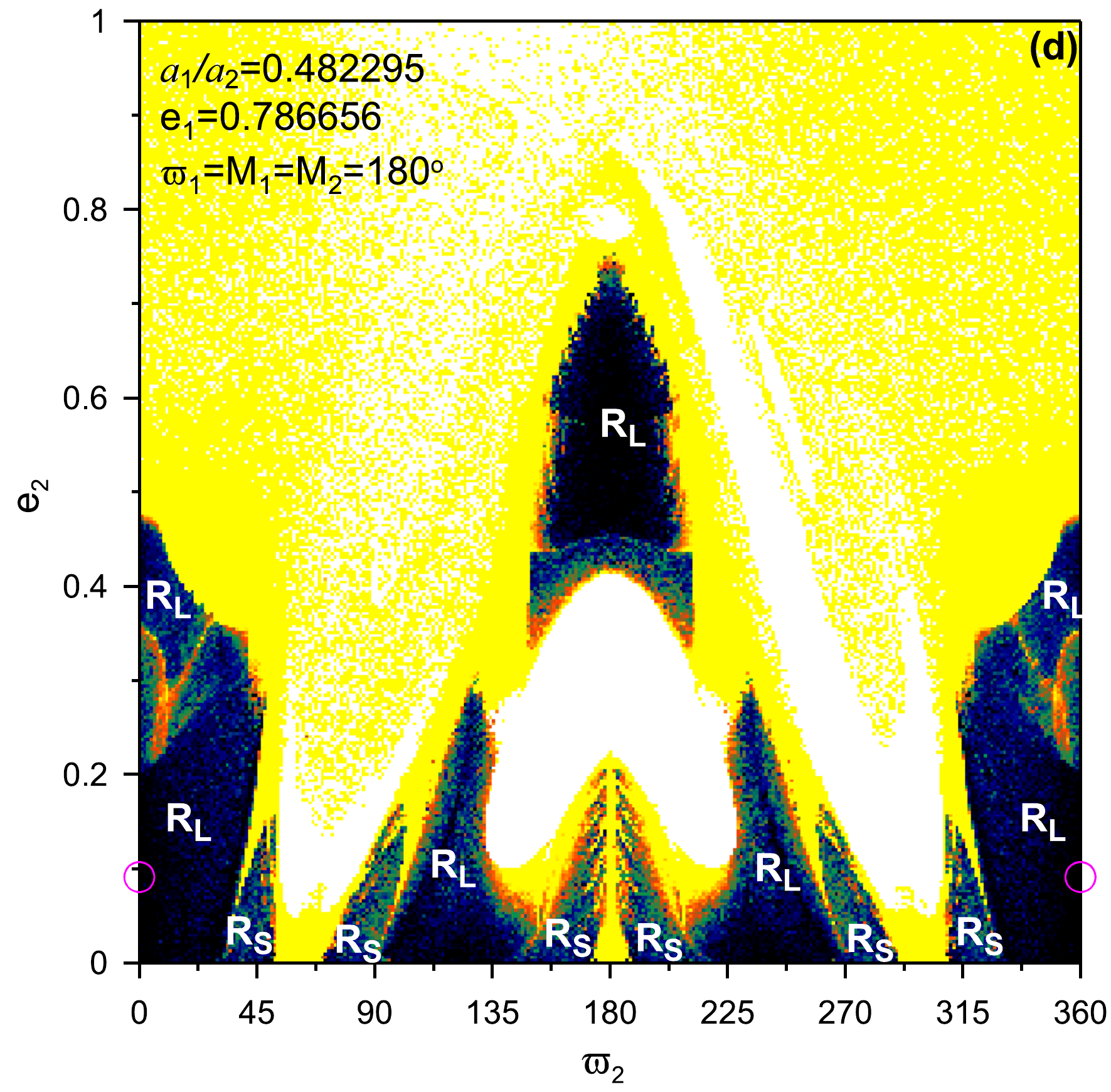} \\ 
\includegraphics[width=6cm]{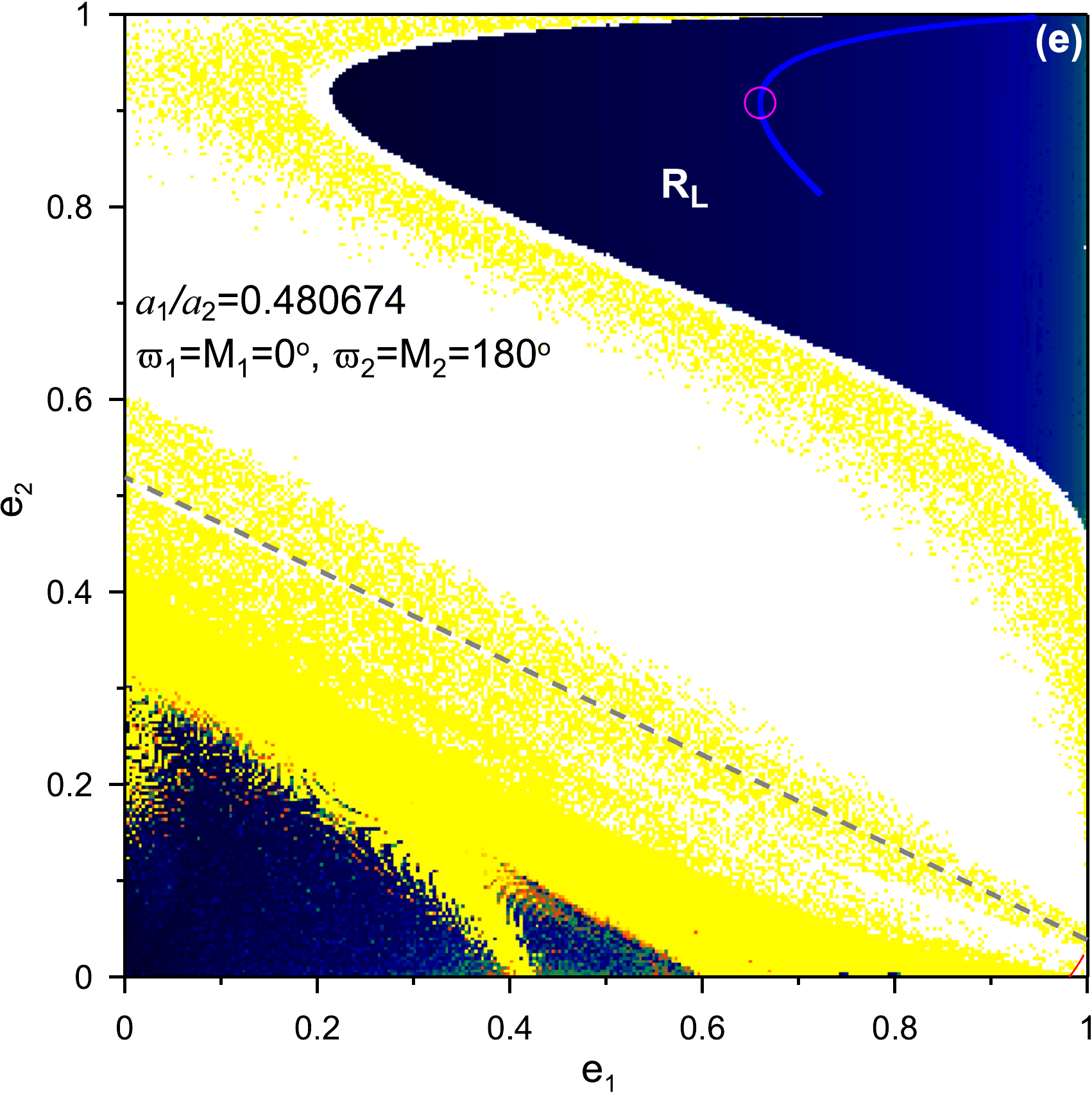} & \qquad& \includegraphics[width=6cm]{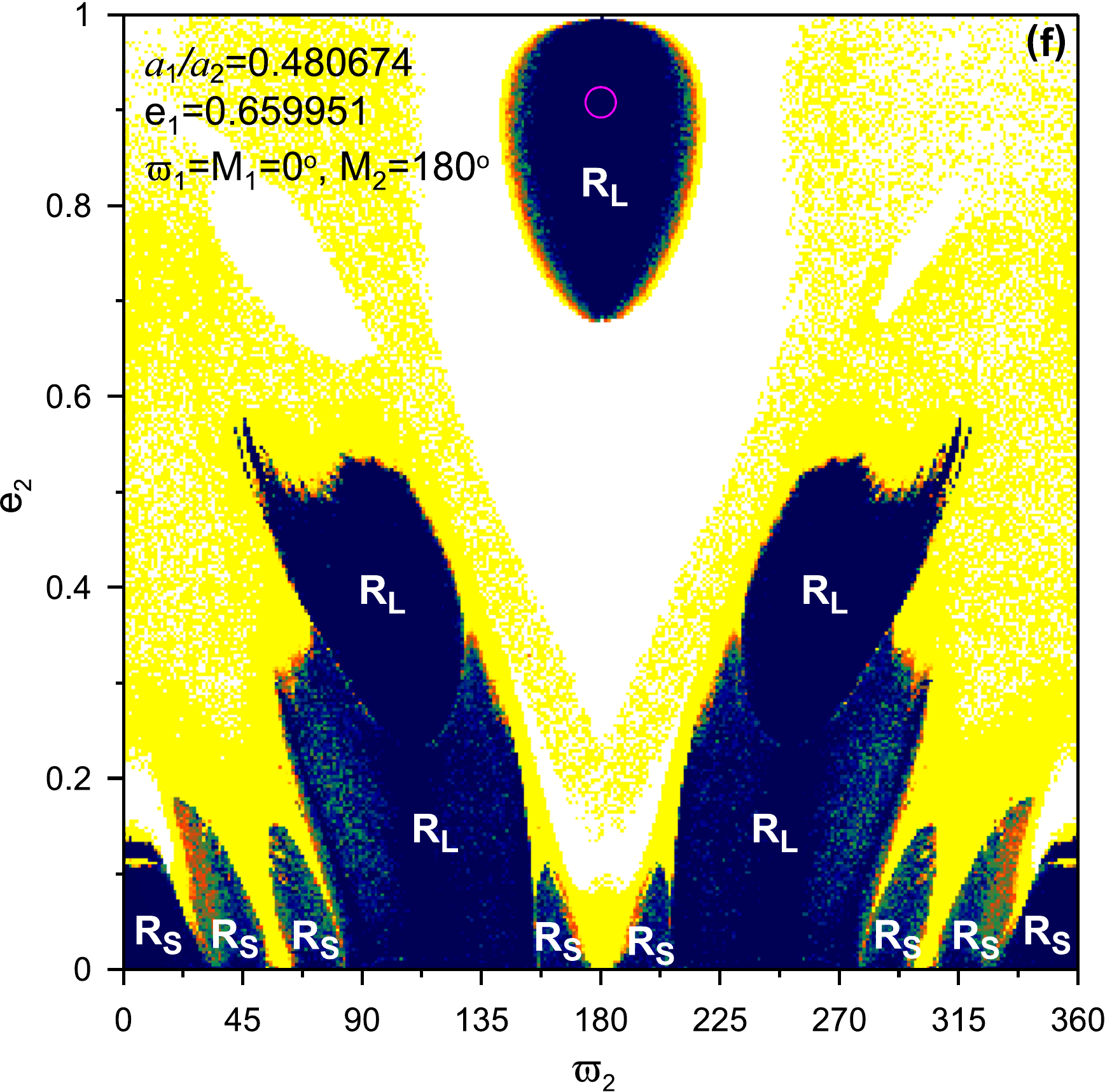}  
\end{array} $
\end{center}
\caption{DS-maps guided by stable periodic orbits of the 3/1 MMR (see Fig. \ref{31_all}c) in the configuration ($\theta_3,\theta_1$)=($0,0$) on the planes \textbf{a} ($e_1,e_2$) and \textbf{b} ($\varpi_2,e_2$) . Accordingly, in the panels {\bf c} and {\bf d}, the choice of the periodic orbit is made from the configuration ($\theta_3,\theta_1$)=($0,\pi$) and ($\theta_3,\theta_1$)=($\pi,0$) for the panels {\bf e} and {\bf f}. Presented as in Fig. \ref{32_00pp0}}
\label{31_00p0}
\end{figure}

In Fig. \ref{31_00p0}, we present two DS-maps on the planes $(e_1,e_2)$ and $(\varpi_2,e_2)$ (two columns) for three different configurations (three rows). In Figs. \ref{31_00p0}a,b, we are guided by a stable periodic orbit within the stable segment of the family evolving in the configuration ($\theta_3,\theta_1$)=($0,0$), while in Figs. \ref{31_00p0}c,d, we select a stable periodic orbit from the configuration ($0,\pi$) and in Figs. \ref{31_00p0}e,f, from the configuration ($\pi,0$). When an $R_A$ is observed, $\Delta\varpi$ oscillates about 0. The secondary resonance, $R_S$, that is observed, is 2/1, where $\theta_1$ librates about $\pi$.
In Figs. \ref{31_00p0}a,b, whenever an $R_L$ is observed in the regular domain built about the stable periodic orbit we selected (magenta circle), $(\theta_3,\theta_1)$ and $\Delta\varpi$ librate about $(0,0)$ and 0, respectively, whereas in Fig. \ref{31_00p0}b, when $\varpi_2$ is near $\pi/3$ (or $5\pi/3$ symmetrically), and $\pi$, the $R_L$ is due to a stable periodic orbit from the configuration $(\theta_3,\theta_1)$=$(\pi,0)$ and $\Delta\varpi=\pi$.
In Figs. \ref{31_00p0}c,d, whenever an $R_L$ is in the domain about the chosen stable periodic orbit, $(\theta_3,\theta_1)$ and $\Delta\varpi$ librate about $(0,\pi)$ and $\pi$, respectively. Same holds for Fig. \ref{31_00p0}d, when $\varpi_2$ is near 0 (magenta circle) and $2\pi/3$ (or $2\pi$ and $4\pi/3$ symmetrically). However, when $\varpi_2$ is near $\pi$, the $R_L$ is due to a stable periodic orbit from the configuration $(\theta_3,\theta_1)$=$(0,0)$ and $\Delta\varpi=0$.
In Figs. \ref{31_00p0}e,f, when the $R_L$ takes place in the region built about the showcased stable periodic orbit, $(\theta_3,\theta_1)$ and $\Delta\varpi$ librate about $(\pi,0)$ and $\pi$, respectively. However, in Fig. \ref{31_00p0}f, when $\varpi_2$ is near $2\pi/3$ (or $4\pi/3$ symmetrically), we have two different well-separated centres of libration; one corresponding to an asymmetric periodic orbit ($e_2>0.3$) and one to the symmetric periodic orbit of the configuration $(0,\pi)$. 


\clearpage
\subsection{4/1 MMR}

\begin{figure}[H]
\centering
$\begin{array}{cp{-1.5cm}c}
\includegraphics[width=6.0cm]{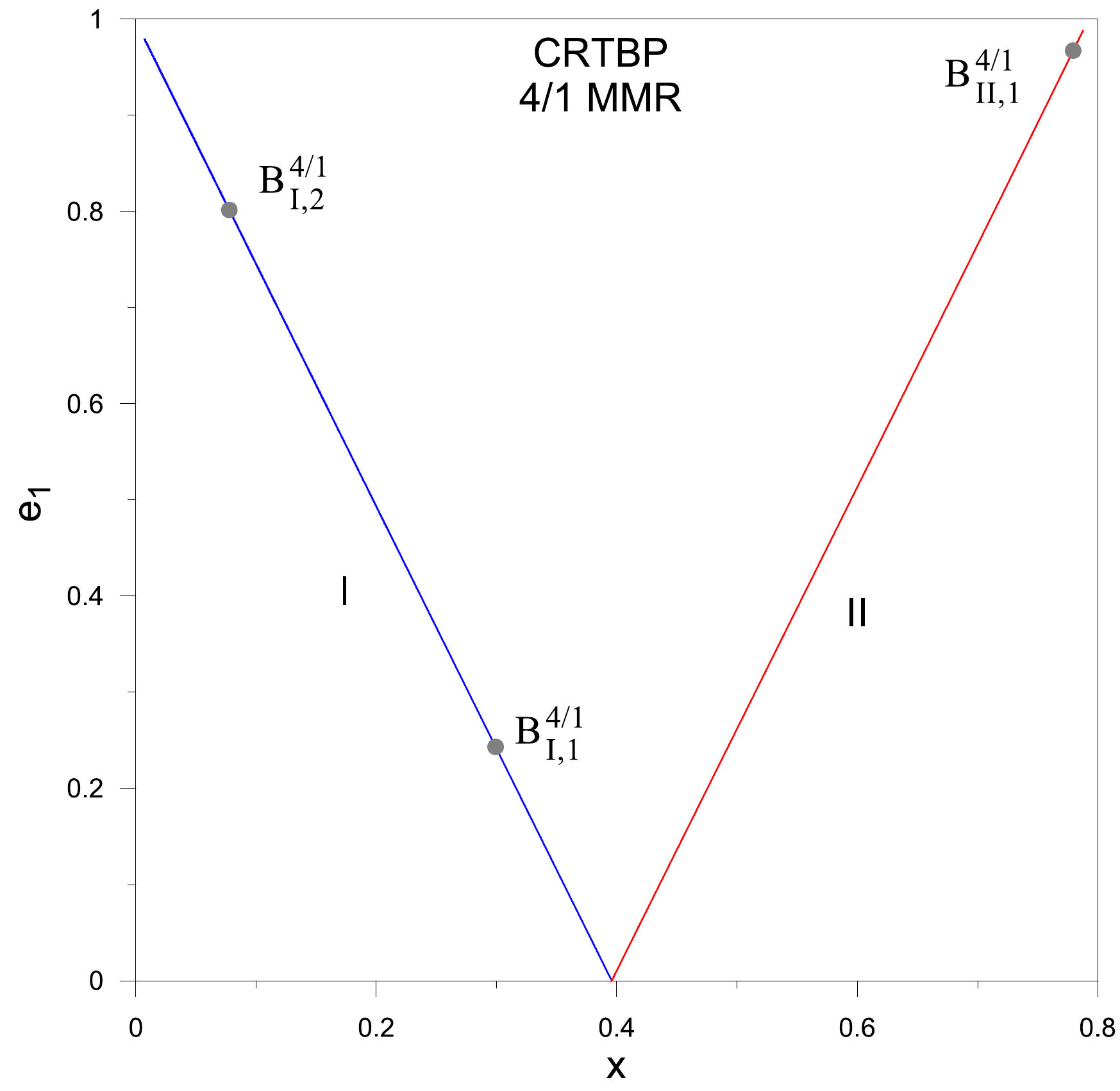}  & \qquad&
\includegraphics[width=5.8cm]{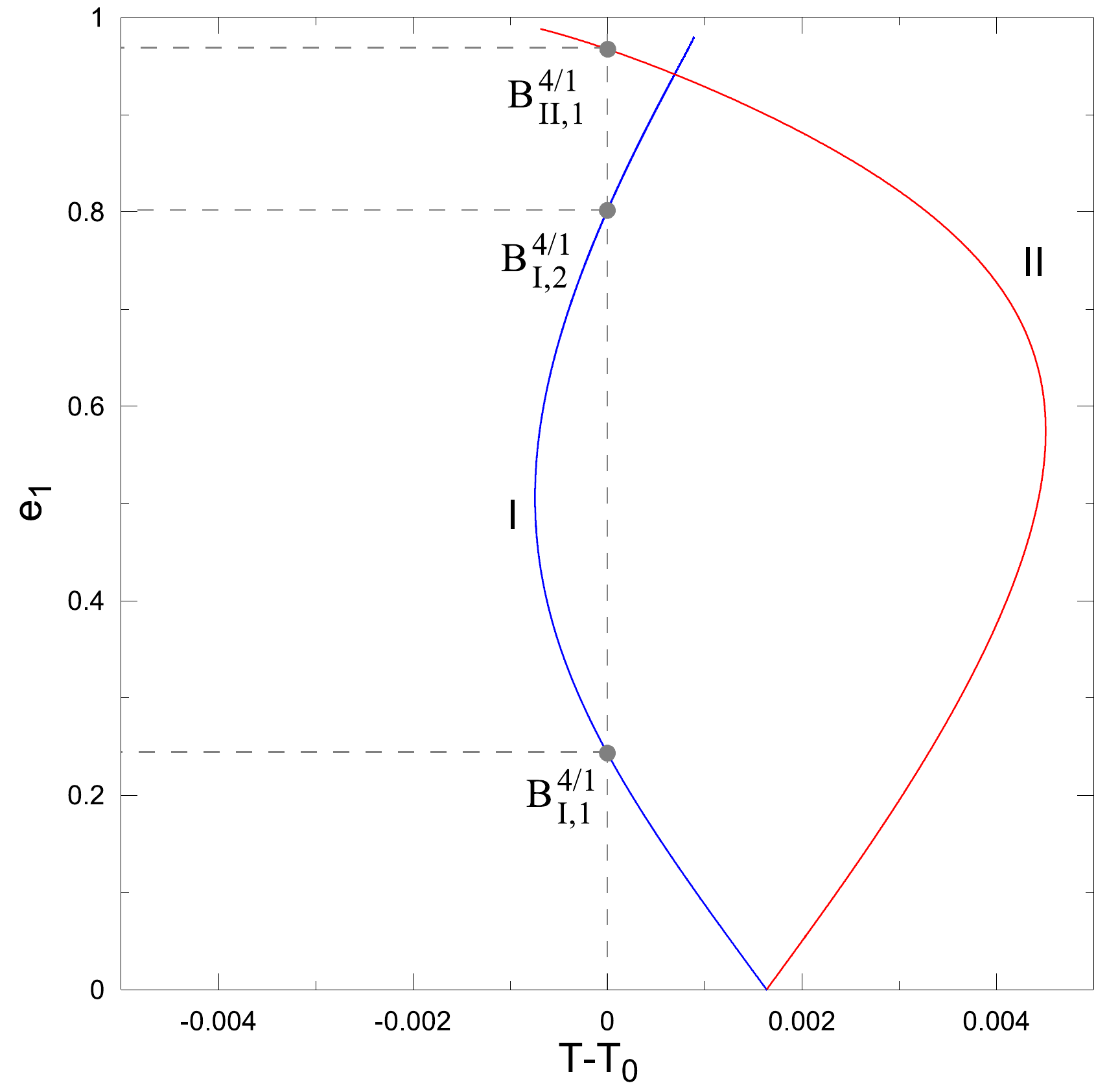} \\
\textnormal{(a)} & \qquad & \textnormal{(b)} 
\end{array} $
$\begin{array}{c}
\includegraphics[width=10cm]{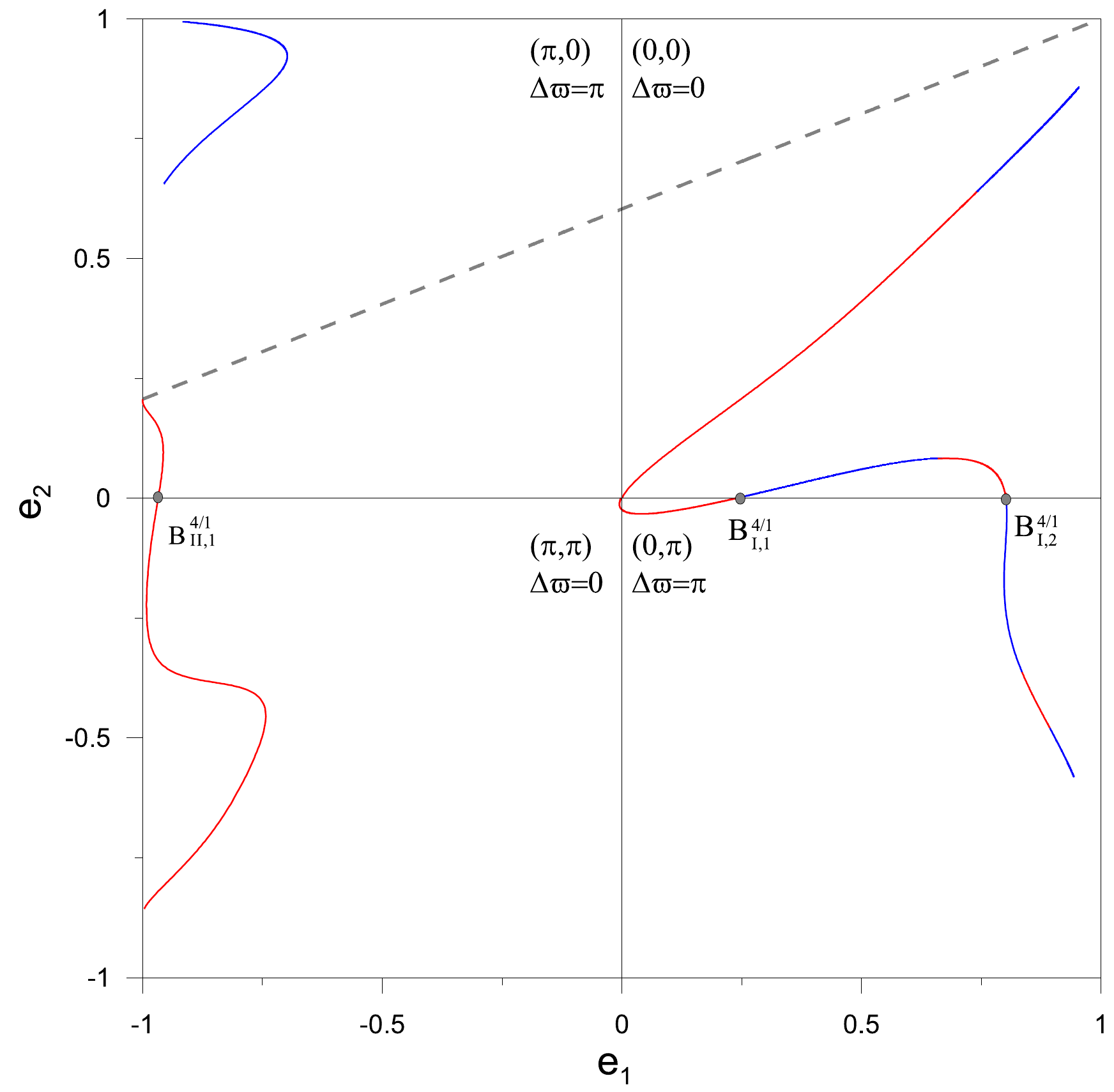} \\
\textnormal{(c)} 
\end{array} $
\caption{{\bf a} Families of periodic orbits in 4/1 MMR of the CRTBP presented as in Fig. \ref{32_all}a. {\bf b} Justification of existence of bifurcation points in the families of CRTBP in 4/1 MMR, where $T=T_0=2\pi/3$, that generate periodic orbits in the ERTBP. {\bf c} Families of periodic orbits in 4/1 MMR of the ERTBP, presented as in Fig. \ref{32_all}c}
\label{41_all}
\end{figure}

In Fig. \ref{41_all}a, we present the families of periodic orbits in 4/1 MMR in the CRTBP which are generated by the bifurcation points of the circular family with $a_1=\frac{4}{1}^{-2/3}$. At $x\approx 0.39685$  two branches are formed: $I$, which consists of stable periodic orbits and $II$, where the orbits are unstable. These families were previously presented also in \citet{hadj93}.

Along the families, $I$ and $II$, there exist bifurcation points when the period of the periodic orbits, $T$, equates to $T_0=2\pi/3$. In Fig. \ref{41_all}b, we justify the existence of three such points, $B^{4/1}_{I,1}$ and $B^{4/1}_{I,2}$ on the stable family and $B^{4/1}_{II,1}$ on the unstable family. 

In Fig. \ref{41_all}c, we provide the families of symmetric periodic orbits in the ERTBP. From the bifurcation points $B^{4/1}_{I,1}$ and $B^{4/1}_{I,2}$ two families are generated; one stable and one unstable. All of these families were continued by following \textit{Scheme I}. In \citet{hadj93} both of the families that bifurcate from $B^{4/1}_{I,1}$ ($I_e$ and $II_e$) were described as stable and no stability type was provided for the ones generated by the point $B^{4/1}_{I,2}$ ($III_e$ and $IV_e$). We herein extended them for greater eccentricity values. From each bifurcation point one family starts as stable and one as unstable. The stable family from $B^{4/1}_{I,1}$ (configuration $(\theta_1,\theta_2)=(0,0)$) joins smoothly with the unstable family from the $B^{4/1}_{I,2}$ (configuration $(0,0)$). The stable family from $B^{4/1}_{I,2}$ (configuration $(0,\pi)$) is continued herein for $e_2>0.3$ (in comparison with \citet{hadj93}) and hence, it is revealed that this family possesses an unstable segment before it becomes stable again. Along the unstable family from the $B^{4/1}_{I,1}$ there is a change of the configuration, when $P_1$ reaches $e_1=0$ and the family then evolves in the configuration $(\pi,\pi)$, and ends at the point $(e_1,e_2)=(0,0)$. From this point (circular family), there exists a bifurcation point to the ERTBP, since the period gets equal to $T=T_0$ and this orbit can generate periodic orbits described thrice (\textit{Scheme II}). This family (called $I_C$ by \citet{hadj93} and described as unstable therein) possesses stable periodic orbits for $e_1>0.74$. 
The families of periodic orbits studied by \citet{hadj93} had also been studied via the averaged Hamiltonian by \citet{fetsukla92} as well.
Moreover, we found a new bifurcation point that was not reported before. From $B^{4/1}_{II,1}$ two unstable families are generated; one evolves in the configuration $(\pi,0)$ and one in $(\pi,\pi)$.
Finally, in the configuration $(\pi,0)$, we computed a new isolated family which possesses only stable periodic orbits where both the small body and the planet are highly eccentric.

\begin{figure}[H]
\begin{center}
$\begin{array}{cp{-2cm}c}
\includegraphics[width=6.0cm]{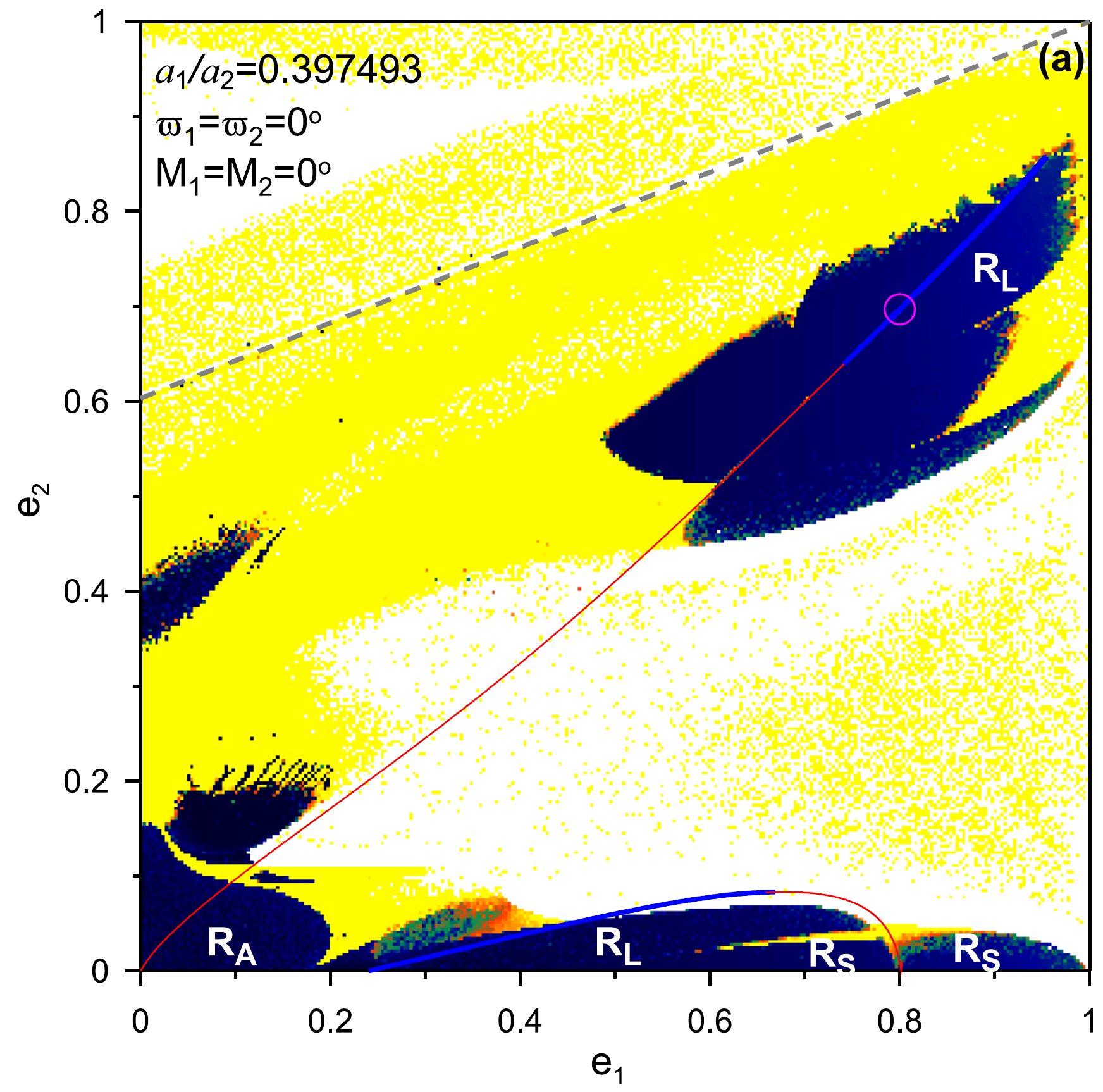} & \qquad&  \includegraphics[width=6.0cm]{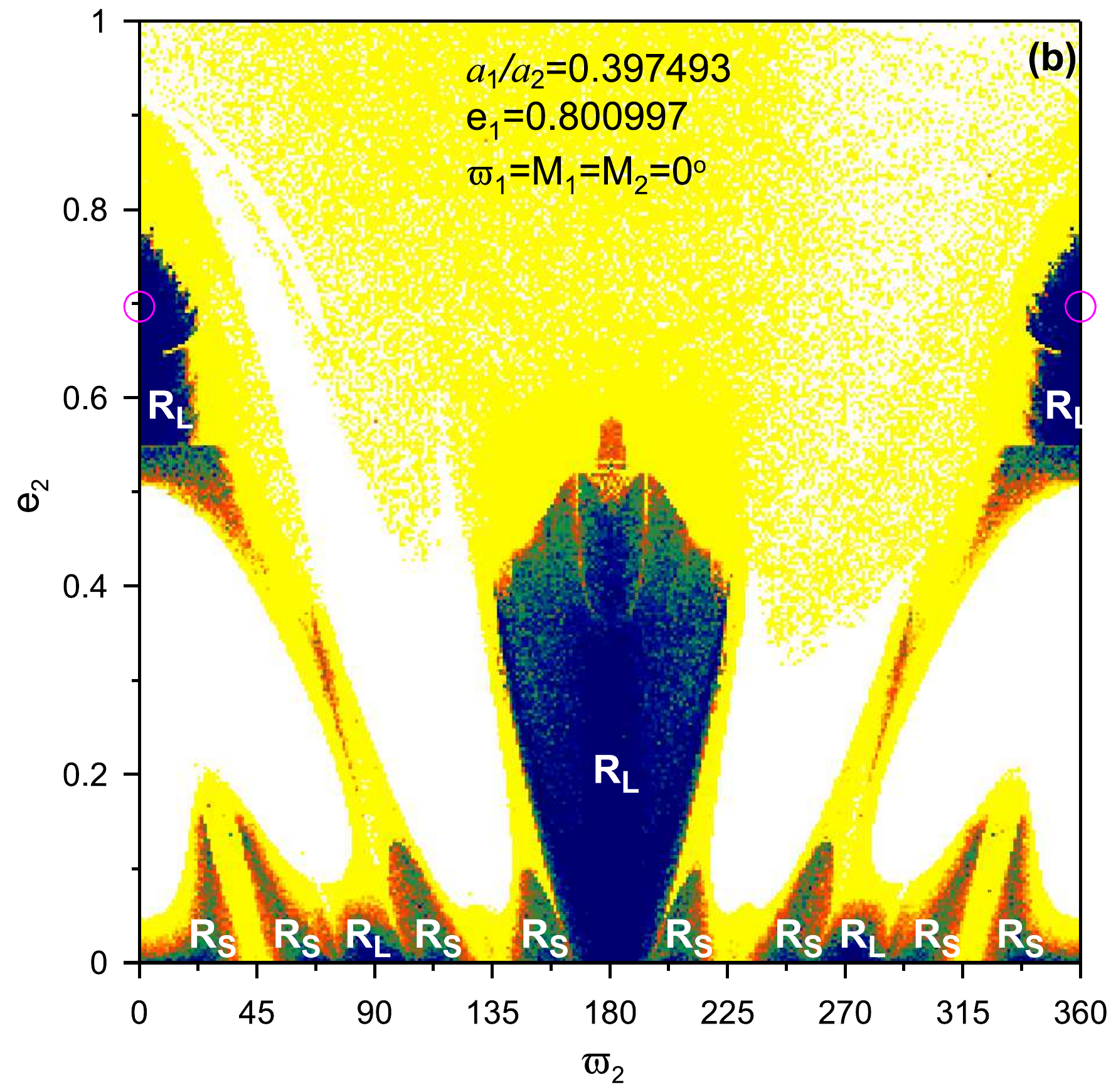}\\
\includegraphics[width=6.0cm]{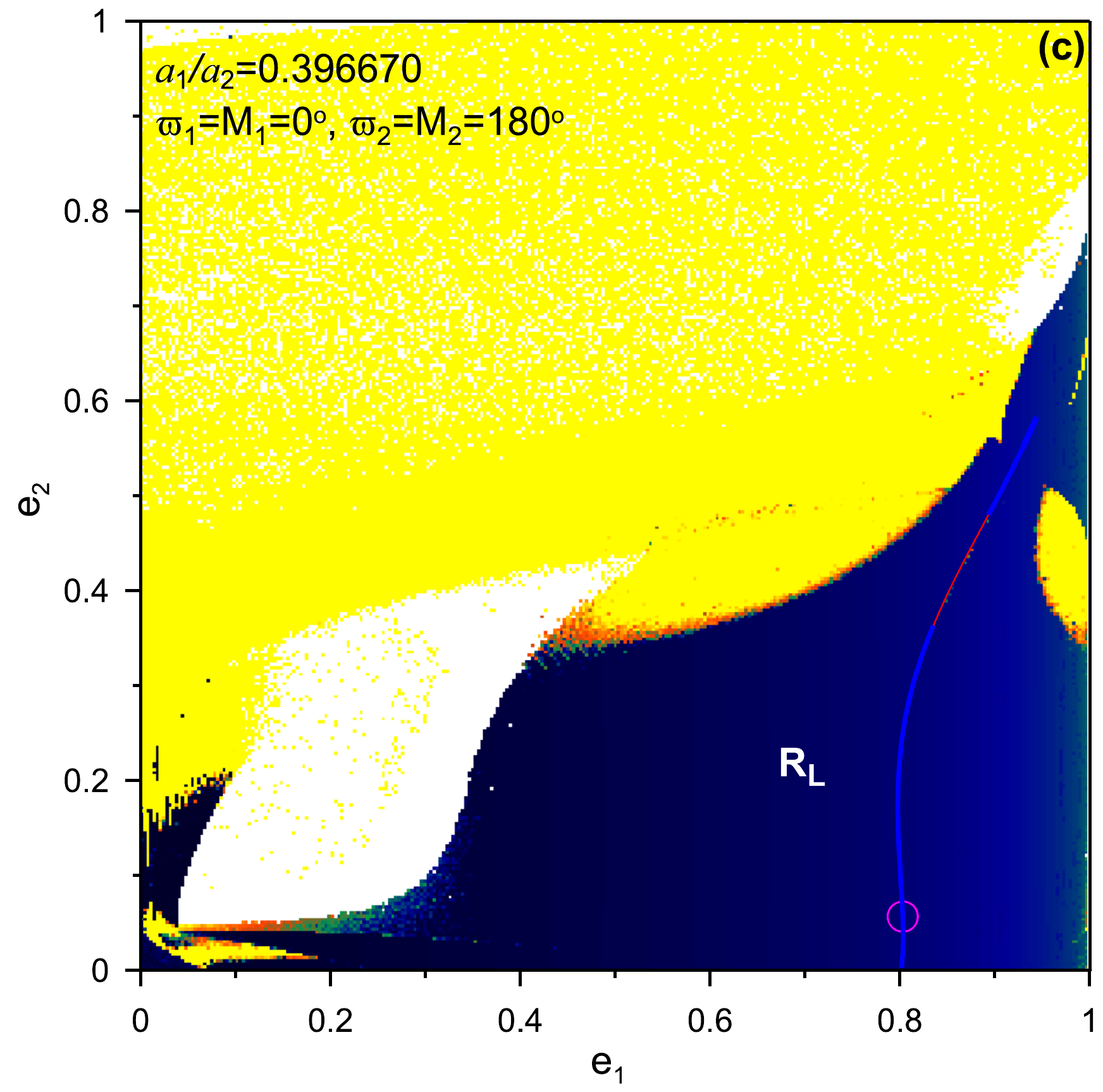} & \qquad&  \includegraphics[width=6.0cm]{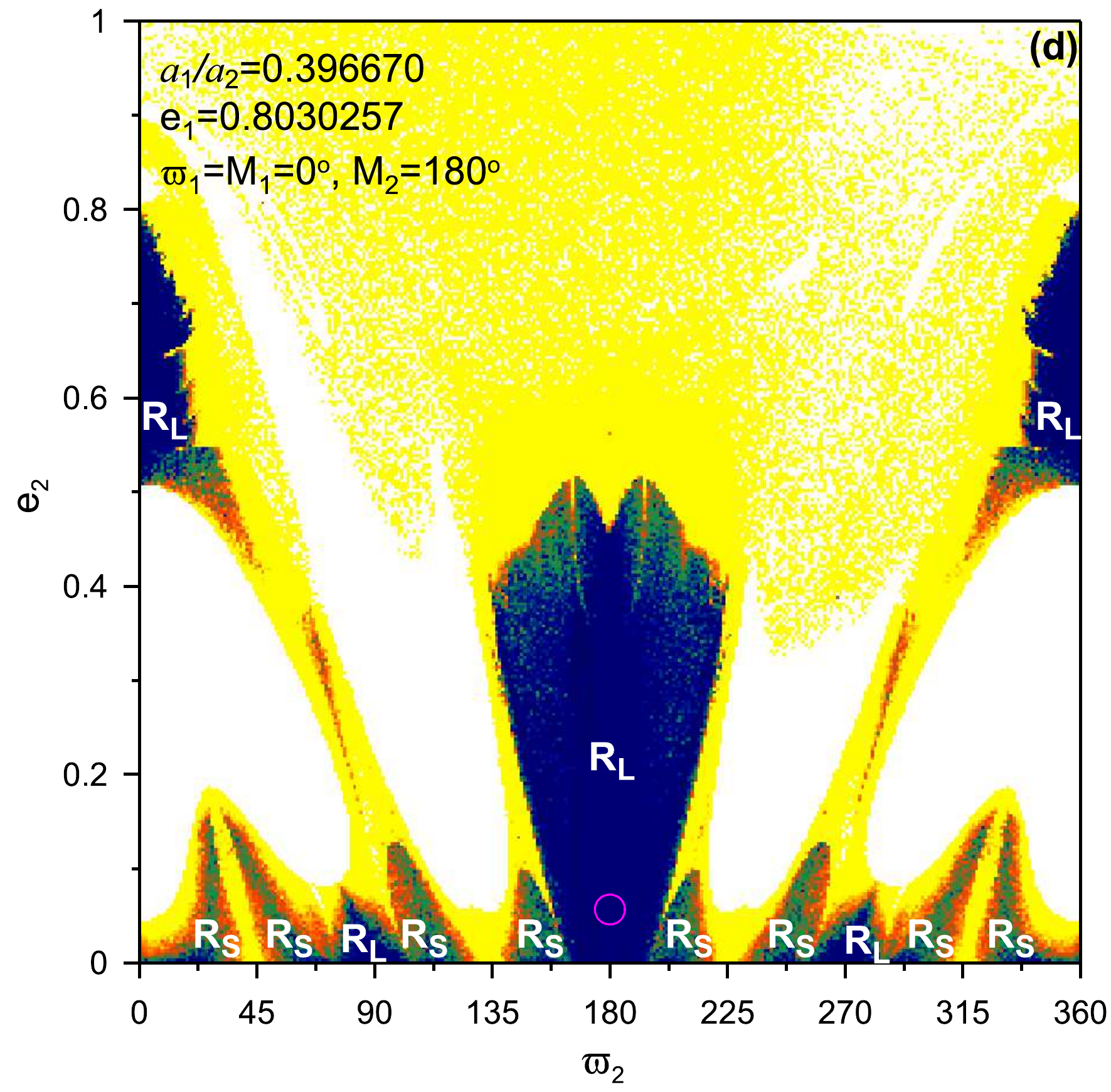}\\
\includegraphics[width=6cm]{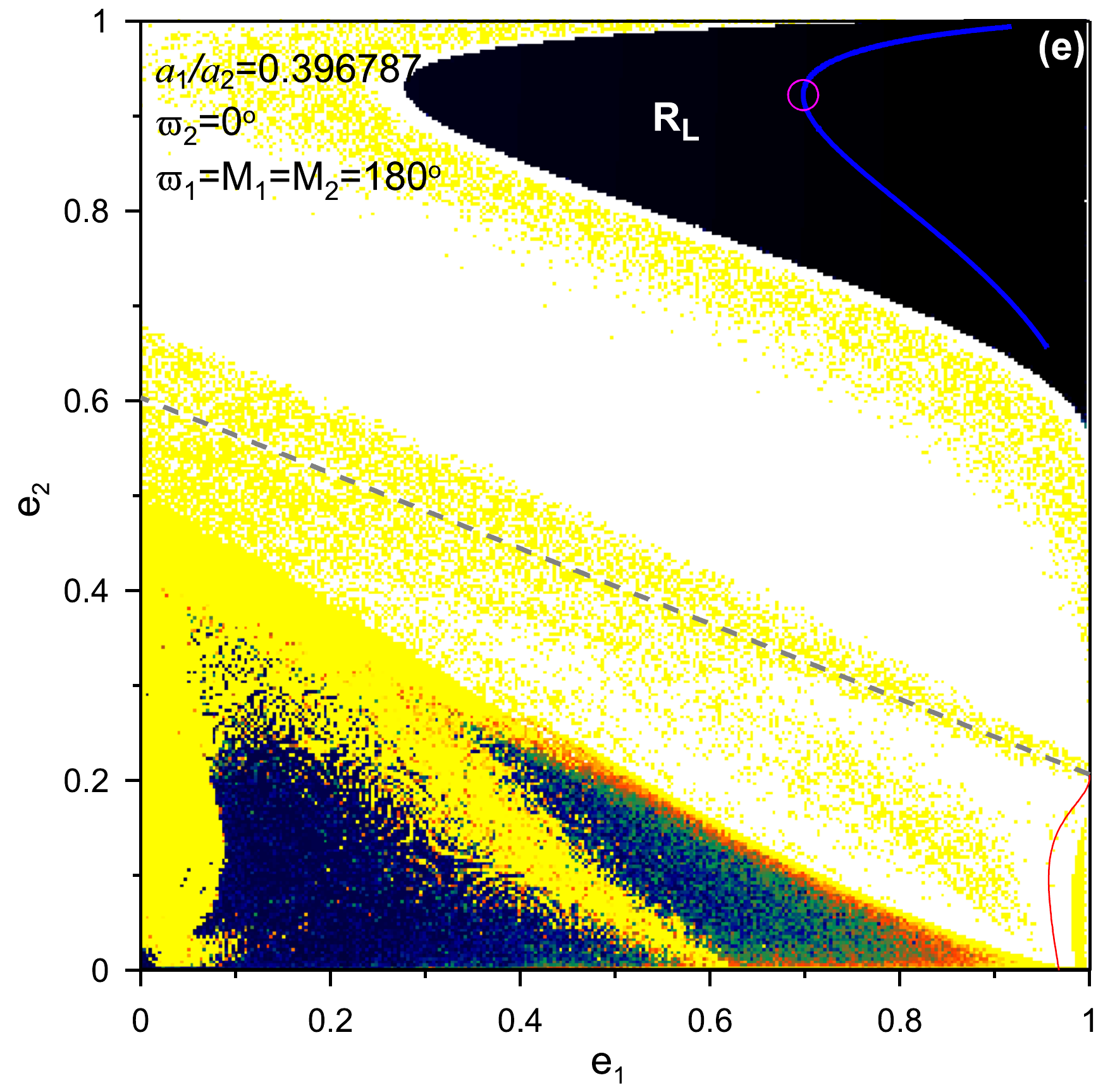} & \qquad& \includegraphics[width=6cm]{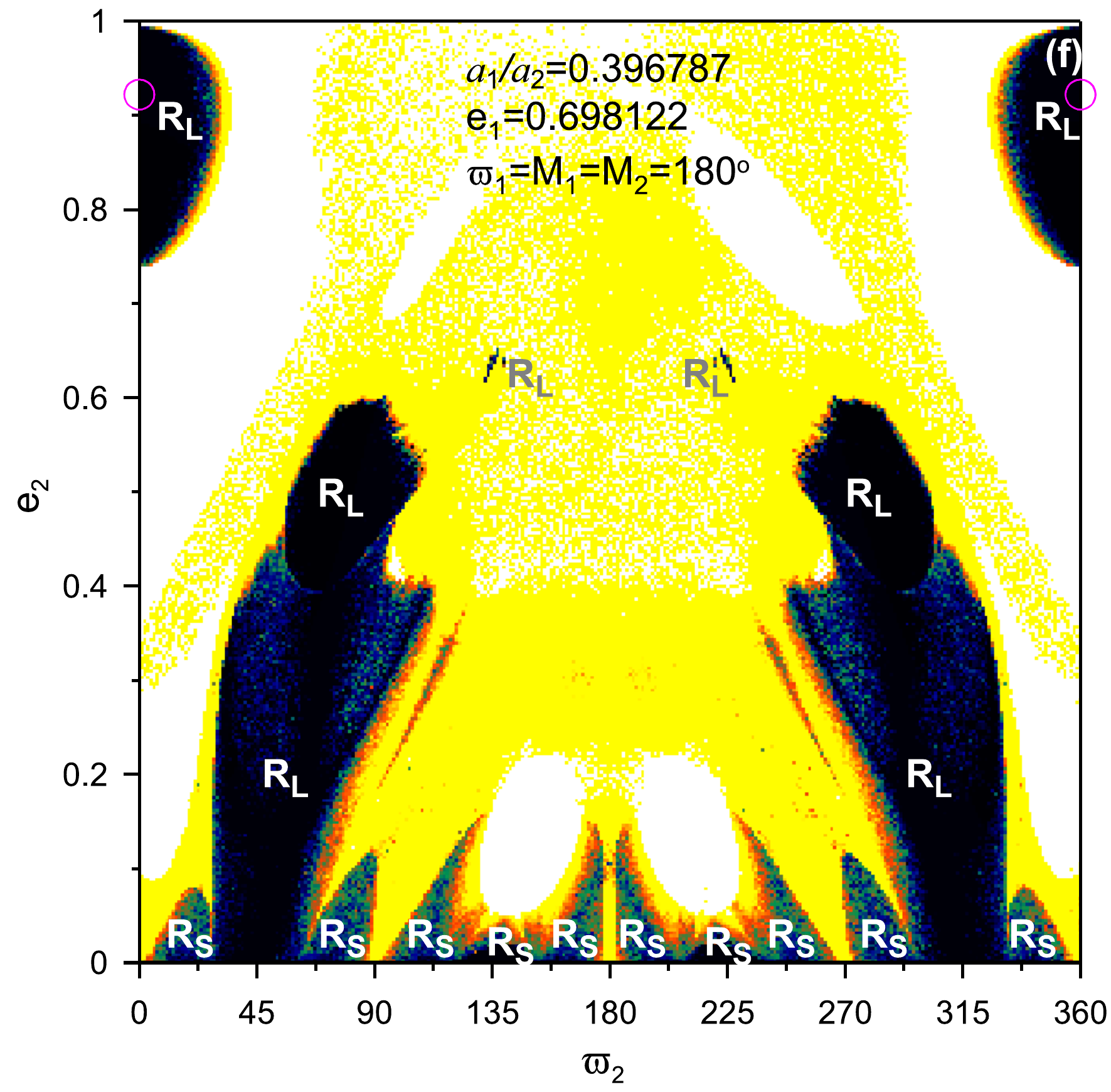}  
\end{array} $
\end{center}
\caption{DS-maps guided by stable periodic orbits of the 4/1 MMR (see Fig. \ref{41_all}c) in the configuration ($\theta_1,\theta_2$)=($0,0$) on the planes \textbf{a} ($e_1,e_2$) and \textbf{b} ($\varpi_2,e_2$). Accordingly, in the panels {\bf c} and {\bf d}, the choice of the periodic orbit is made from the configuration ($\theta_1,\theta_2$)=($0,\pi$) and ($\theta_1,\theta_2$)=($\pi,0$) for the panels {\bf e} and {\bf f}). Presented as in Fig. \ref{32_00pp0}}
\label{41_0pp0}
\end{figure}

In Fig. \ref{41_0pp0}, we present two DS-maps on the planes $(e_1,e_2)$ and $(\varpi_2,e_2)$ (two columns) for three different configurations (three rows). In Figs. \ref{41_0pp0}a,b, the designated stable periodic orbit belongs to the configuration ($\theta_1,\theta_2$)=($0,0$), in Figs. \ref{41_0pp0}c,d, to ($0,\pi$) and in Figs. \ref{41_0pp0}e,f, to ($\pi,0$). When an $R_A$ exists, $\Delta\varpi$ oscillates about 0. The secondary resonance, $R_S$, that is observed, is 3/1, where $\theta_1$ librates about 0.
In Figs. \ref{41_0pp0}a,b, when an $R_L$ is denoted in the regular domain built about the chosen stable periodic orbit, $(\theta_1,\theta_2)$ and $\Delta\varpi$ librate about $(0,0)$ and 0, respectively, whereas in Fig. \ref{41_0pp0}b, when $\varpi_2$ is near $\pi/2$ (or $3\pi/2$ symmetrically) and $\pi$, the $R_L$ is due to a stable periodic orbit from the configuration $(\theta_1,\theta_2)$=$(0,\pi)$. 
In Figs. \ref{41_0pp0}c,d, whenever an $R_L$ is in the domain about the chosen stable periodic orbit, $(\theta_1,\theta_2)$ and $\Delta\varpi$ librate about $(0,\pi)$ and $\pi$, respectively. Same holds for Fig. \ref{41_0pp0}d, when $\varpi_2$ is near $\pi$ (magenta circle) and $\pi/2$ (or $3\pi/2$ symmetrically), but when it is near 0 (or $2\pi$ symmetrically) the libration takes place about the angles $(0,0)$ and 0, accordingly.
In Figs. \ref{41_0pp0}e,f, when the $R_L$ is showcased, $(\theta_1,\theta_2)$ and $\Delta\varpi$ librate about $(\pi,0)$ and $\pi$, respectively. However, in Fig. \ref{41_0pp0}f, when $\varpi_2$ is near $\pi/4$ (or $7\pi/4$ symmetrically), we have two different well-separated centres of libration; one corresponding to an asymmetric periodic orbit ($e_2>0.4$) and one to the symmetric periodic orbit of the configuration $(0,\pi)$.

\clearpage
\subsection{5/1 MMR}

\begin{figure}
\centering
$\begin{array}{cp{-1.5cm}c}
\includegraphics[width=6.0cm]{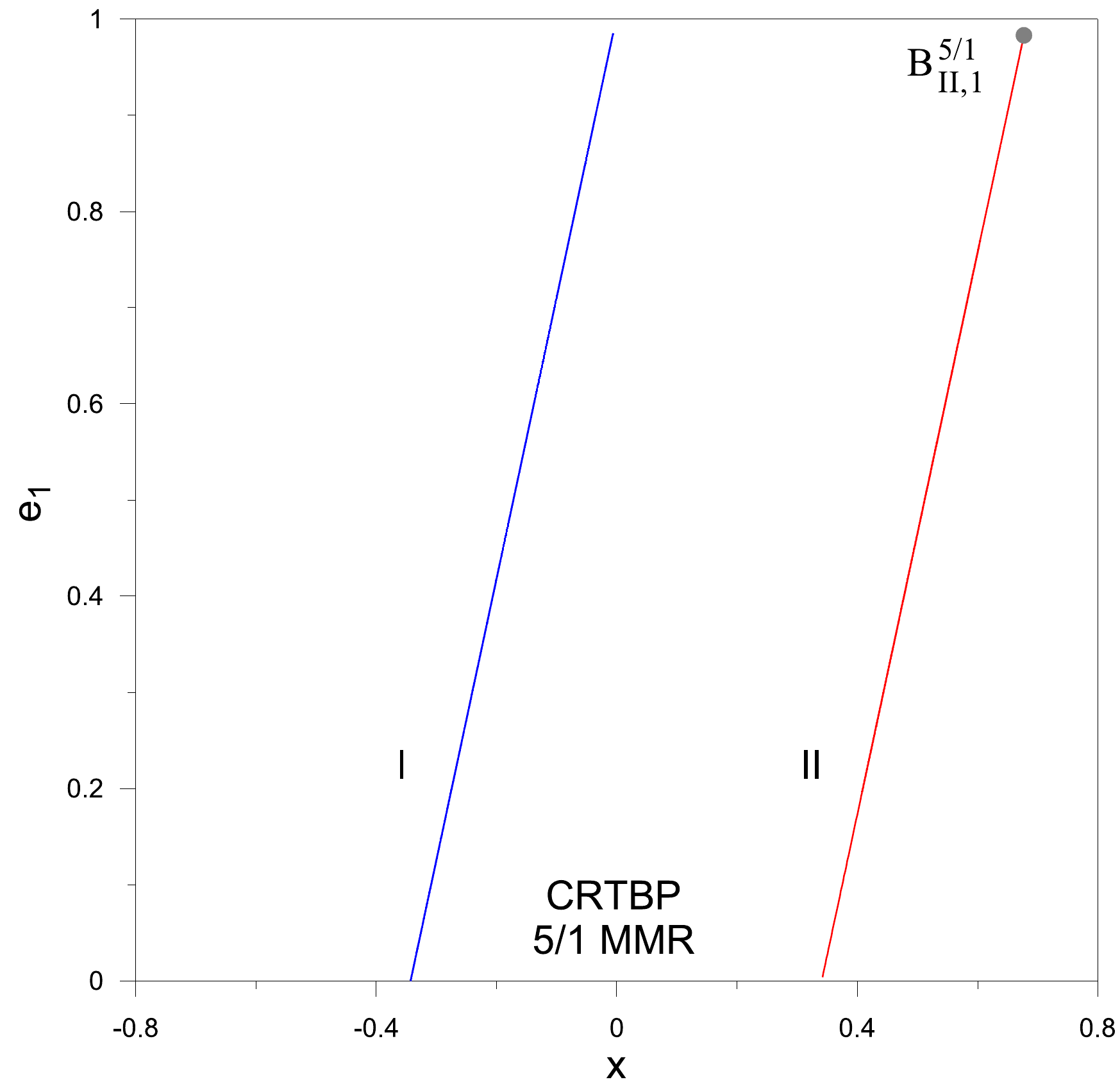}  & \qquad&
\includegraphics[width=5.8cm]{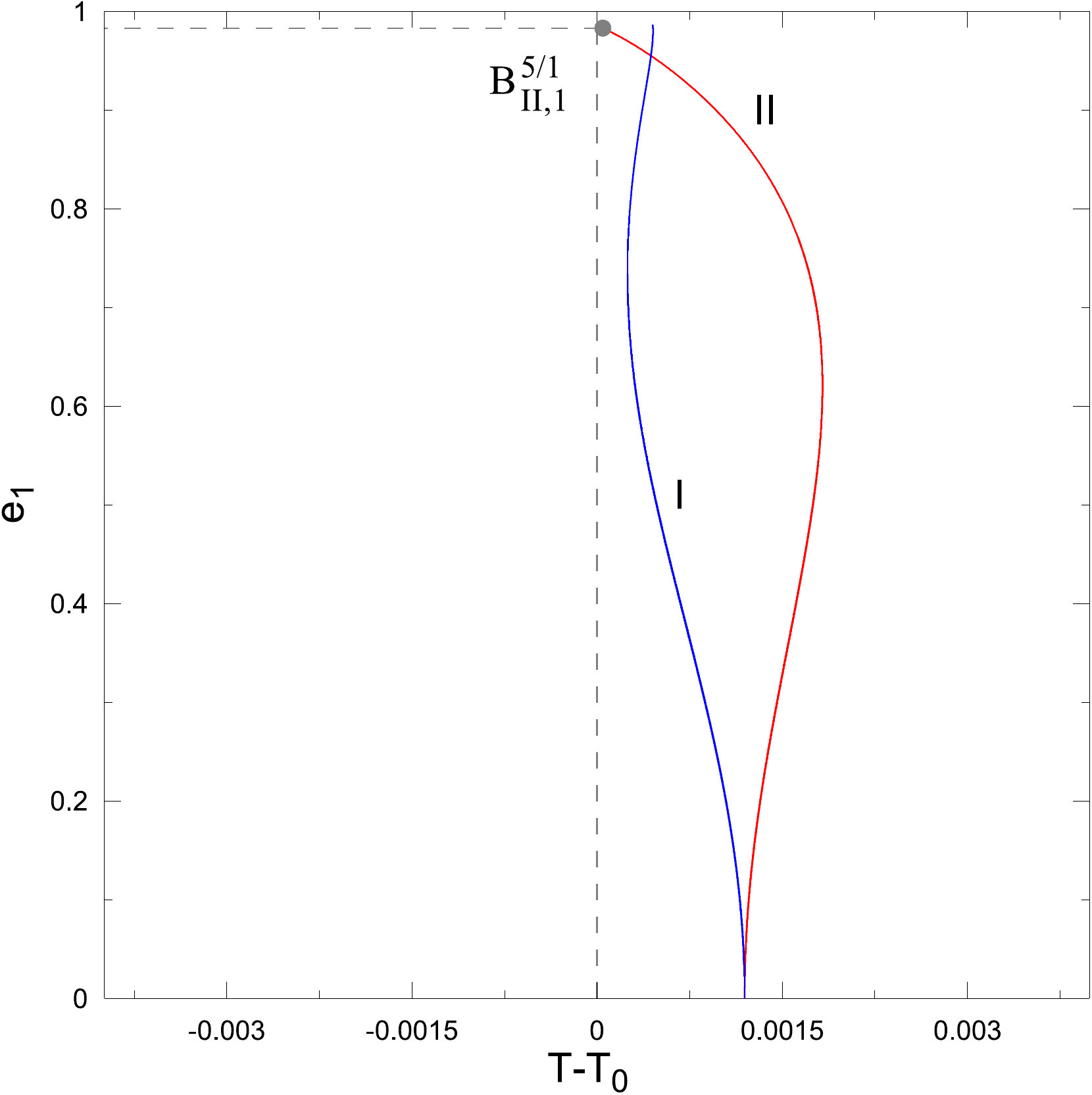} \\
\textnormal{(a)} & \qquad & \textnormal{(b)} 
\end{array} $
$\begin{array}{c}
\includegraphics[width=10cm]{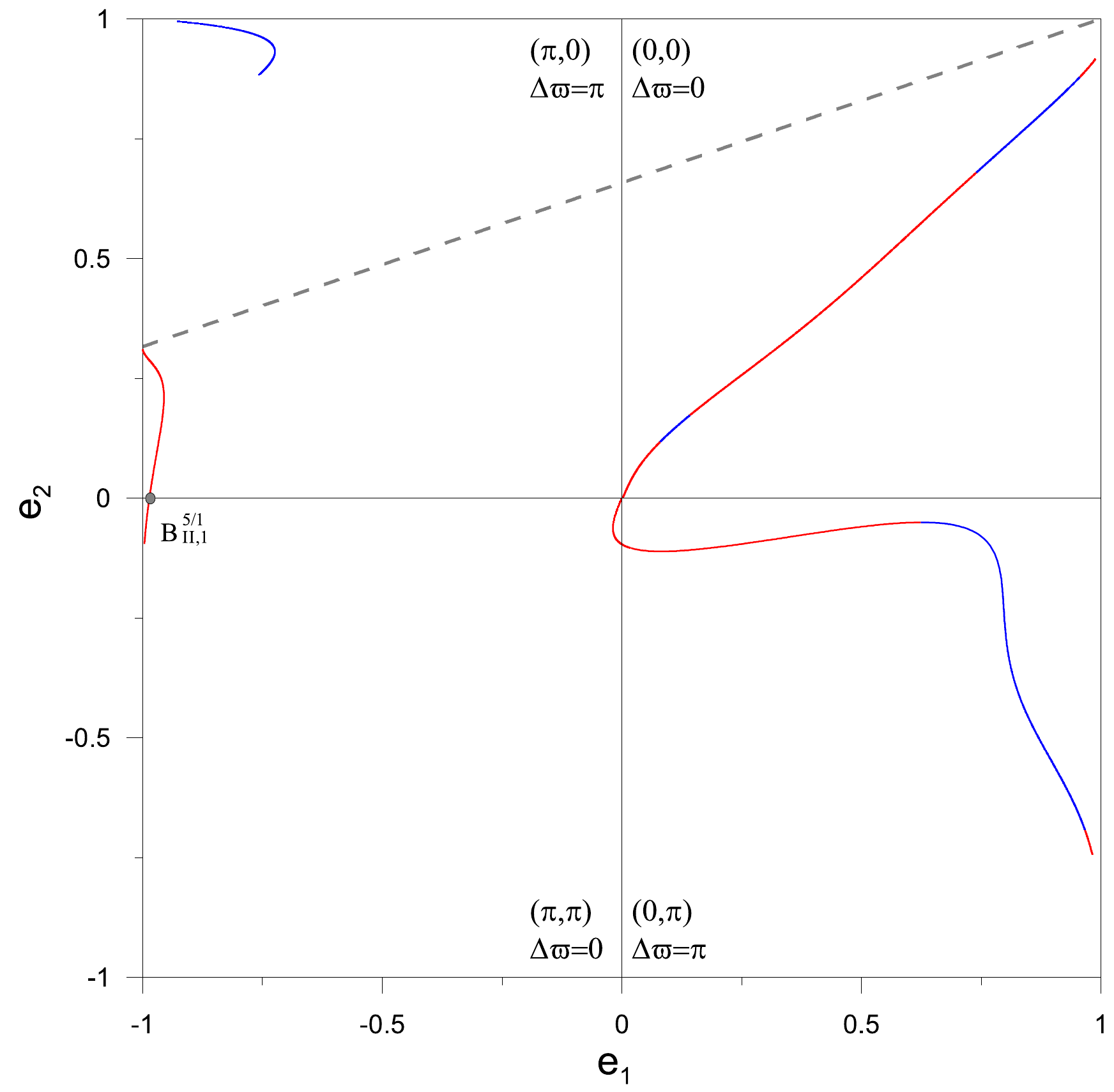} \\
\textnormal{(c)} 
\end{array} $
\caption{{\bf a} Families of periodic orbits in 5/1 MMR of the CRTBP presented as in Fig. \ref{32_all}a. {\bf b} Justification of existence of bifurcation points in the families of CRTBP in 5/1 MMR, where $T=4T_0=2\pi$, that generate periodic orbits in the ERTBP. {\bf c} Families of periodic orbits in 5/1 MMR of the ERTBP, presented as in Fig. \ref{32_all}c. The angles in brackets represent the pair of resonant angles ($\theta_4,\theta_1$)}
\label{51_all}
\end{figure}

In Fig. \ref{51_all}a, we present the families of periodic orbits in 5/1 MMR in the CRTBP which are generated by the bifurcation point of the circular family with period $T_0=\pi/2$ being described four times, $T=4\;T_0$. Thus, two branches are formed: $I$, which consists of stable periodic orbits and $II$, where the orbits are unstable.

Only along the unstable family, $II$, there exists one bifurcation point when the period of the periodic orbits, $T$ equates to $4 T_0=2\pi$. In Fig. \ref{51_all}b, we justify the existence of the bifurcation point, $B^{5/1}_{II,1}$. 

In Fig. \ref{51_all}c, we provide the families of symmetric periodic orbits in the ERTBP. From the bifurcation point $B^{5/1}_{II,1}$, two families are generated, both of them are unstable and one evolves at the configuration $(\theta_4,\theta_1)=(\pi,0)$ and the other at $(\pi,\pi)$. These families were continued by following \textit{Scheme I}.
Along the circular family the period $T$ gets equal to $\pi/2$ (\textit{Scheme II}). This periodic orbit is continued to the ERTBP with multiplicity four,  i.e. $T=4\;T_0$. Two families are generated and both of them start with unstable periodic orbits. One evolves in the configuration ($0,0$) and possesses two stable segments; one at low and one at high values of $e_{1,2}$. The other family evolves in the configuration $(\pi,\pi)$ and there is a change in the configuration as $e_1\rightarrow 0$ again. Then, the family evolves in the configuration $(0,\pi)$ and possesses mainly stable periodic orbits when the small body is highly eccentric. 
Finally, in the configuration $(\pi,0)$, we computed a new isolated family which possesses only stable periodic orbits where both the small body and the planet are highly eccentric.

\begin{figure}
\begin{center}
$\begin{array}{cp{-2cm}c}
\includegraphics[width=6.0cm]{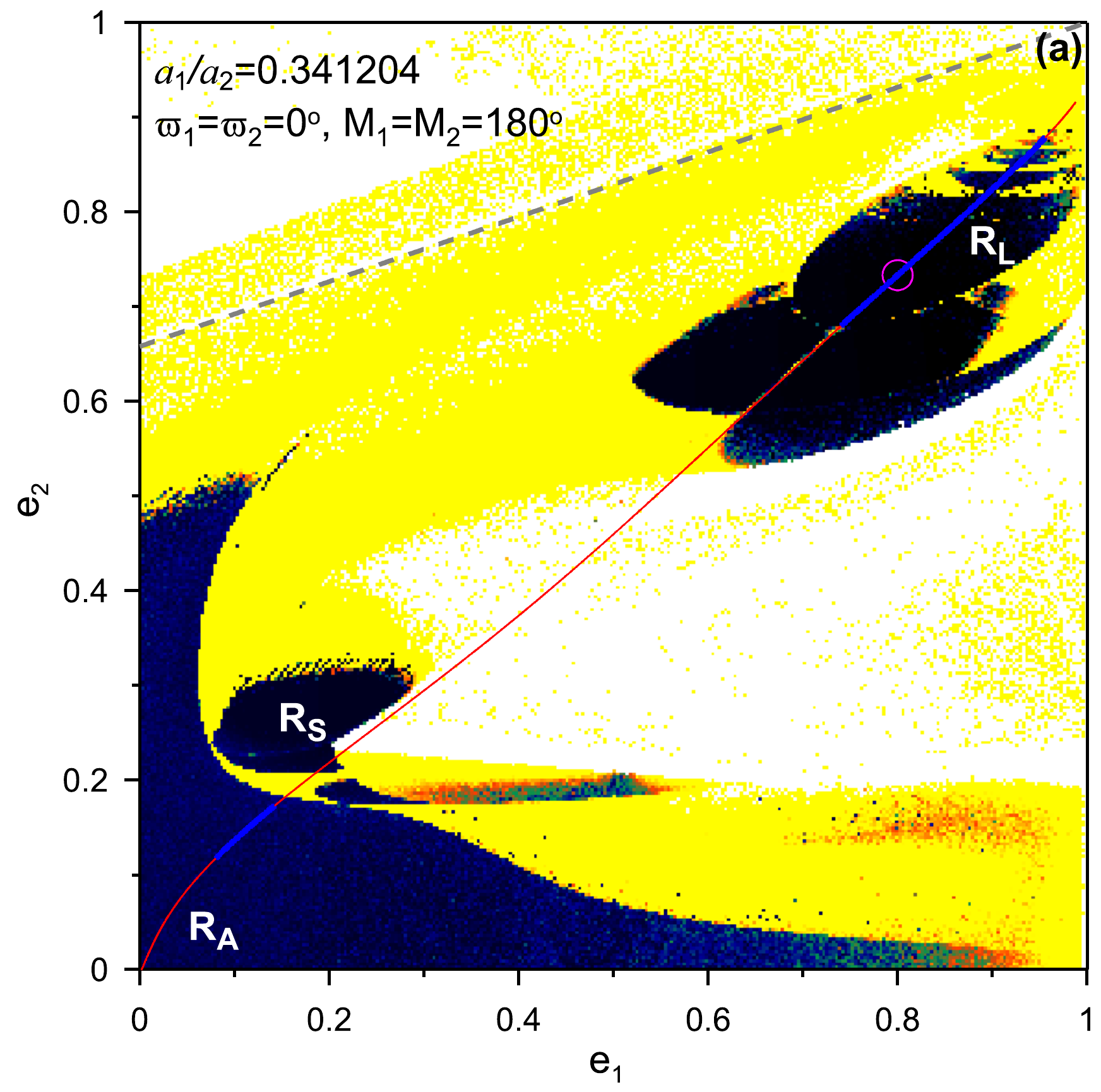} & \qquad&  \includegraphics[width=6.0cm]{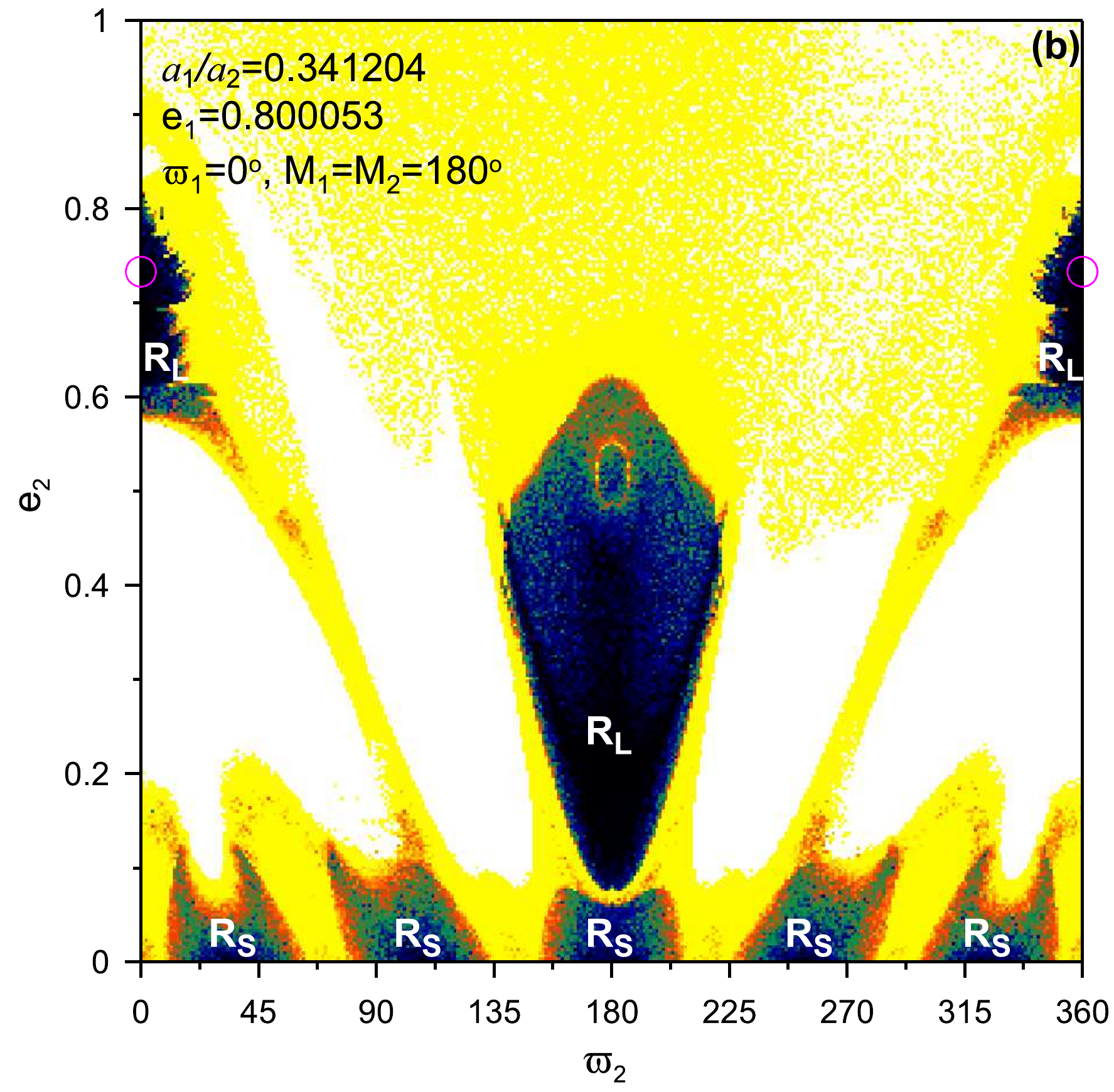}\\
\includegraphics[width=6cm]{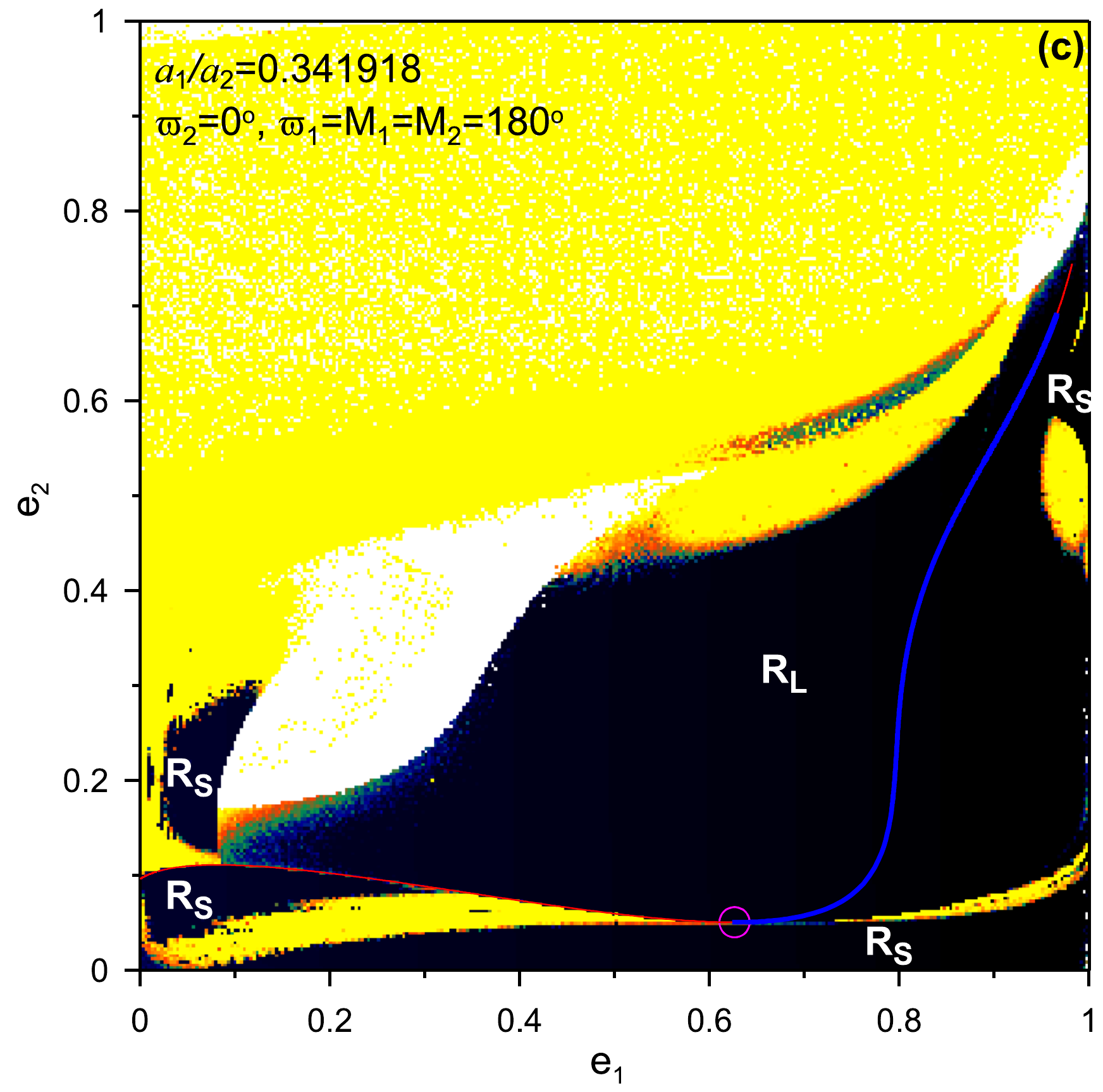} & \qquad& \includegraphics[width=6cm]{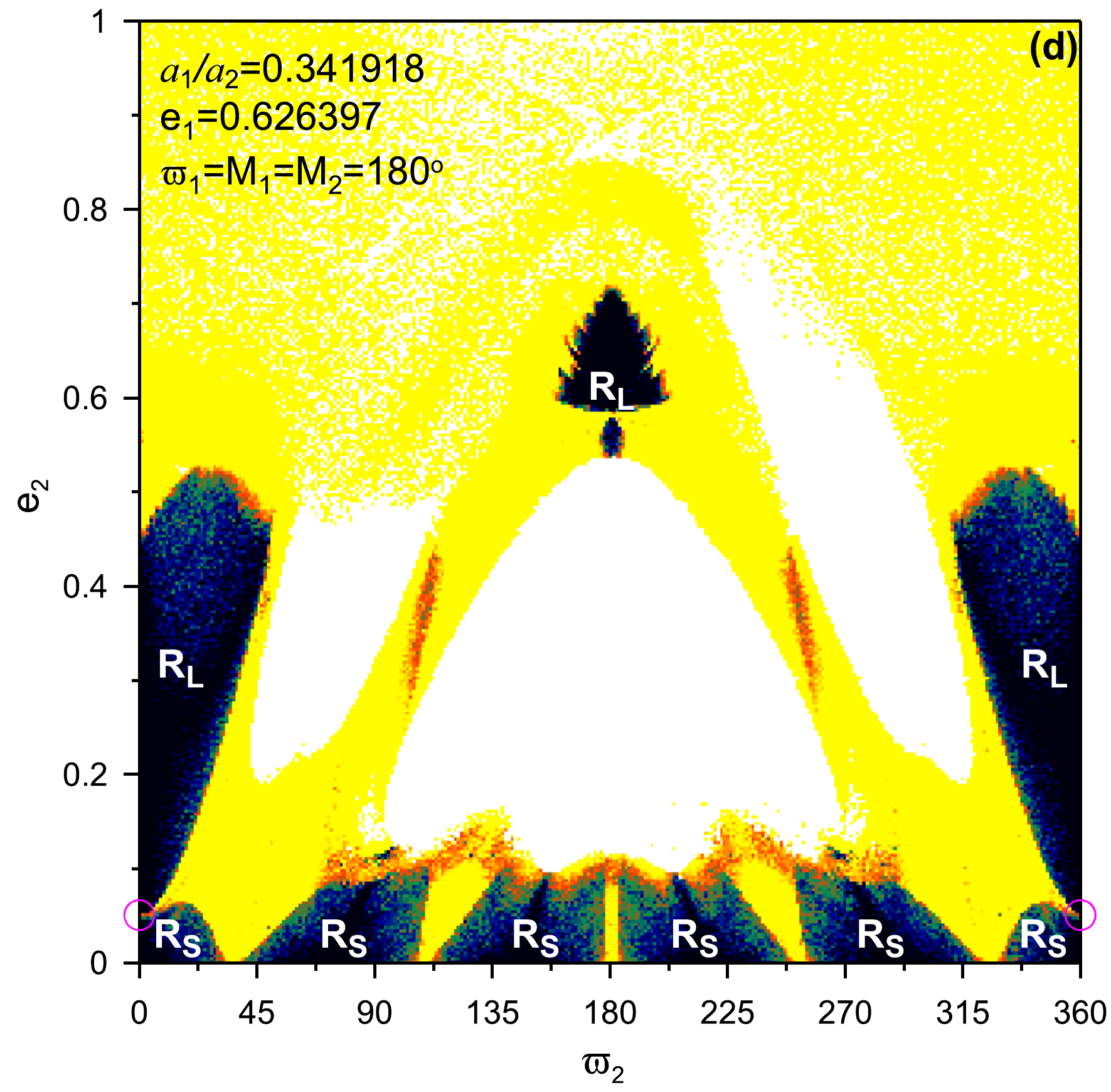} \\ 
\includegraphics[width=6cm]{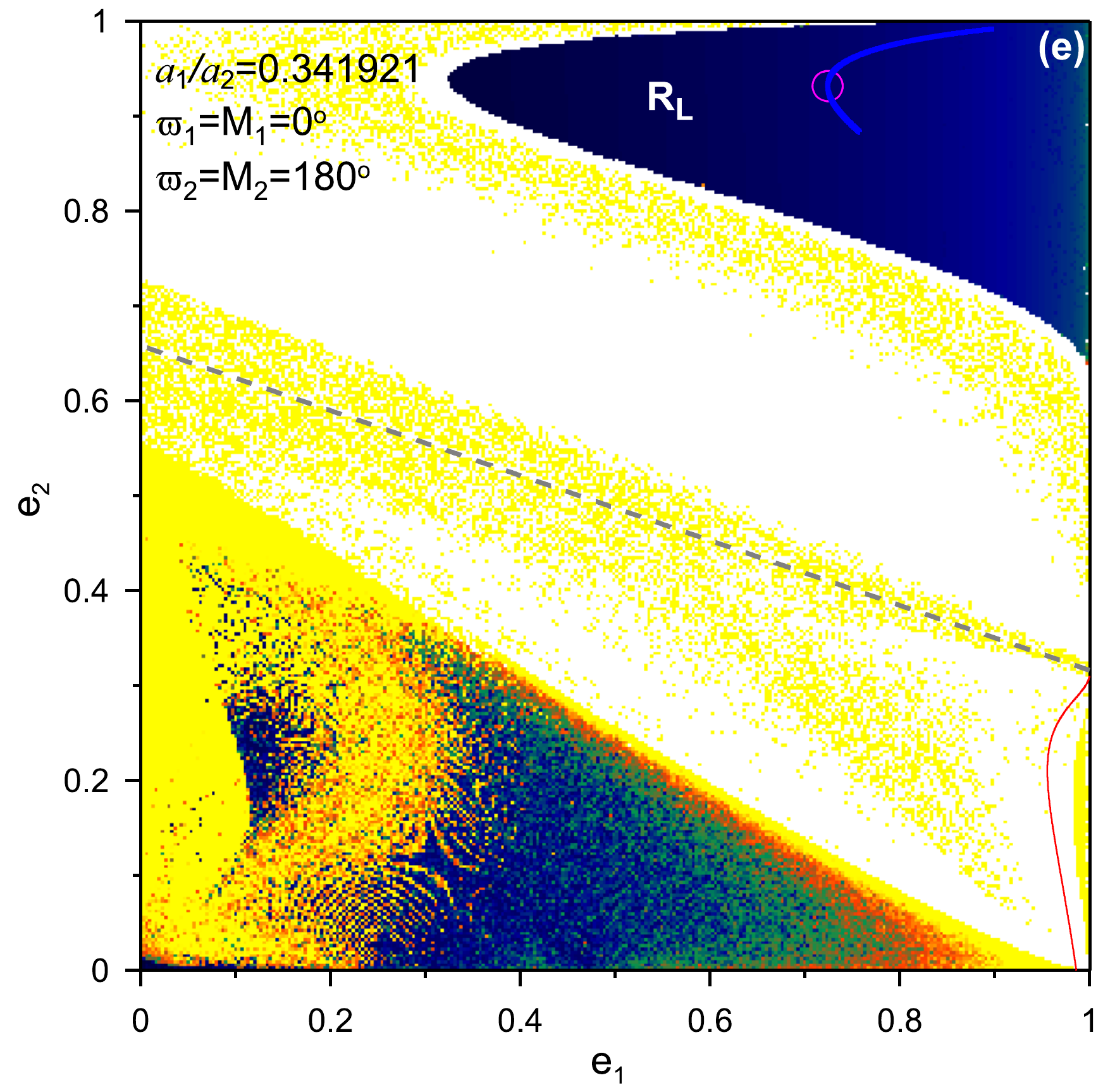} & \qquad& \includegraphics[width=6cm]{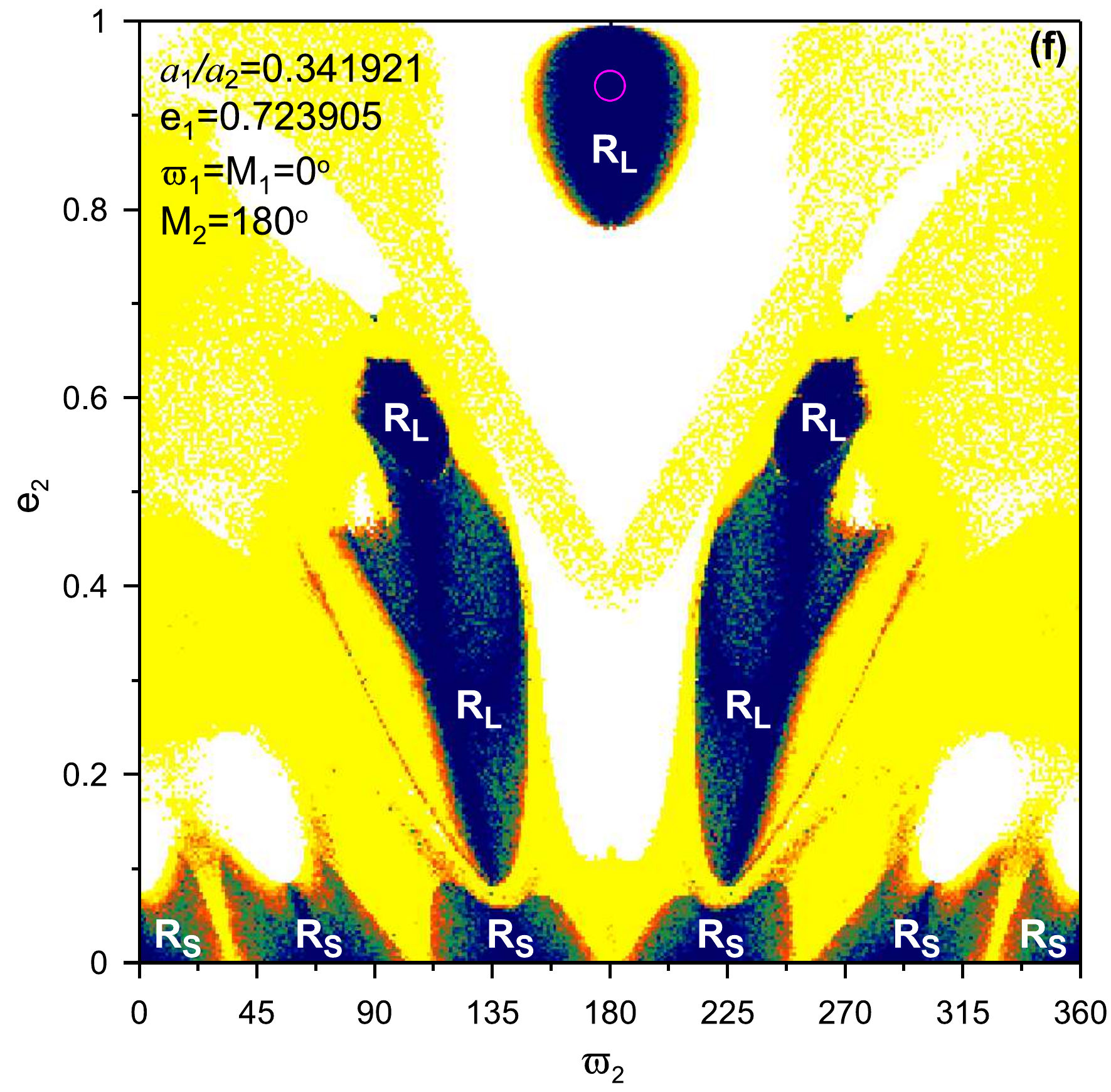}  
\end{array} $
\end{center}
\caption{DS-maps guided by stable periodic orbits of the 5/1 MMR (see Fig. \ref{41_all}c) in the configuration ($\theta_4,\theta_1$)=($0,0$) on the planes \textbf{a} ($e_1,e_2$) and \textbf{b} ($\varpi_2,e_2$). Accordingly, in the panels {\bf c} and {\bf d}, the choice of the periodic orbit is made from the configuration ($\theta_4,\theta_1$)=($0,\pi$) and ($\theta_4,\theta_1$)=($\pi,0$) for the panels {\bf e} and {\bf f}). Presented as in Fig. \ref{32_00pp0}}
\label{51_00p0}
\end{figure}

In Fig. \ref{51_00p0}, we present two DS-maps on the planes $(e_1,e_2)$ and $(\varpi_2,e_2)$ (two columns) for three different configurations (three rows). In Figs. \ref{51_00p0}a,b, we are guided by a stable periodic orbit within the stable segment of the family evolving in the configuration ($\theta_4,\theta_1$)=($0,0$), while in Figs. \ref{51_00p0}c,d, we select a stable periodic orbit from the configuration ($0,\pi$) and in Figs. \ref{51_00p0}e,f, from the configuration ($\pi,0$). When an $R_A$ is observed, $\Delta\varpi$ oscillates about 0. The secondary resonance, $R_S$, that is observed, is 4/1, where $\theta_1$ librates about $\pi$.
In Figs. \ref{51_00p0}a,b, whenever an $R_L$ is observed in the regular domain built about the stable periodic orbit we selected (magenta circle), $(\theta_4,\theta_1)$ and $\Delta\varpi$ librate about $(0,0)$ and 0, respectively, whereas in Fig. \ref{51_00p0}b, when $\varpi_2$ is near $\pi$, the $R_L$ is due to a stable periodic orbit from the configuration $(\theta_4,\theta_1)$=$(0,\pi)$.
In Figs. \ref{51_00p0}c,d, whenever an $R_L$ is in the domain about the stable periodic orbit we chose, $(\theta_4,\theta_1)$ and $\Delta\varpi$ librate about $(0,\pi)$ and $\pi$, respectively. Same holds for Fig. \ref{51_00p0}d, when $\varpi_2$ is near 0 (magenta circle) (or $2\pi$ symmetrically). However, when $\varpi_2$ is near $\pi$, $(\theta_4,\theta_1)$ and $\Delta\varpi$ librate about $(0,0)$ and $0$. 
In Figs. \ref{51_00p0}e,f, when the $R_L$ takes place in the region built about the showcased stable periodic orbit, $(\theta_4,\theta_1)$ and $\Delta\varpi$ librate about $(\pi,0)$ and $\pi$, respectively. However, in Fig. \ref{51_00p0}f, when $\varpi_2$ is near $4\pi/5$ (or $6\pi/5$ symmetrically), we have two different well-separated centres of libration; one corresponding to an asymmetric periodic orbit ($e_2>0.5$) and one to the symmetric of the configuration $(0,\pi)$.

\clearpage\relax
\section{Discussion and conclusions}\label{con}

In this work, we studied the long-term stability of a planetary system consisting of two primaries and a secondary modelled with the ERTBP for interior MMRs. We focused on 3/2, 5/2, 3/1, 4/1 and 5/1 MMRs and unravelled the regular domains in the phase space which encompass stable periodic orbits and are thus, built about them. Our results can be applied to any system, whose dynamics can be efficiently explored via the ERTBP, for instance star-terrestrial planet-giant planet, planet-spacecraft-satellite, binary stars-circumprimary (S-type) planet or star-asteroid-giant planet. 

Particularly, we started from the circular family and continued the orbits to the CRTBP. In this problem, we found new bifurcation points leading to the ERTBP and hence, new families emanating from them for every MMR studied. In this way, the families of each MMR that exist in the GTBP can be generated if we start from the periodic orbits of the ERTBP and through mono-parametric continuation increase the mass of the inner body \citep[see][]{mbf06,av13}. We also computed some isolated families, in particular at high eccentricities, for each MMR. As for the already known families in the ERTBP, considered as totally unstable, we continued them to higher eccentricity values and found stable segments in each of them.

The majority of the new families we found in each MMR has highly eccentric stable periodic orbits. However, the unstable ones could be useful in trajectory design for future missions by exploitation of their invariant manifolds. 

Additionally, by discovering those new families, we showcased that each MMR (not only first-order ones but also 5/2, 3/1, 4/1 and 5/1 MMRs) has broad regions of stability in the symmetric configurations $(0,0)$, $(0,\pi)$ and $(\pi,0)$, even when the semi-major axis ratio does not correspond to the exact MMR. In these regular domains, two types of resonance protect the phases and therefore, close encounters are avoided even for highly eccentric orbits: the MMR and the secondary resonance. The apsidal difference oscillation/circulation inside those MMRs can also account for the regular domains that appear, when no libration of the resonant angles is observed. When the non-resonant orbits are studied, e.g. on maps where the semi-major axis ratio varies, the apsidal resonance can protect the planets from close encounters. However, if the outer giant planet is highly eccentric, the inner body can only survive if it is locked in an MMR with it.

To highlight the extent of the stability regions and their agreement with the periodic orbits computed in this work, Fig. \ref{4m} (Appendix \ref{appendix}) sums up the 4 DS-maps merged together corresponding to the 4 symmetric configurations on the plane ($e_1,e_2$) in the ERTBP for different MMRs, when the semi-major axes are kept fixed corresponding to the exact 3/2, 2/1, 5/2, 3/1, 4/1 and 5/1 MMRs. We should note that along the families of periodic orbits in the ERTBP the semi-major axis ratio, $a_1/a_2$, varies slightly ($a_2=1$), but is close to the exact value. In Fig. \ref{4m}a, we overplot the families in 2/1 MMR studied in \citet{kiaasl}.

With regards to the search for possible terrestrial companions in single-planet giant systems observed to-date, based on those MMRs, observational astronomers can be driven to specific semi-major axis ratios, angles' values and eccentricities, so that a possible existence of a terrestrial planet could be revealed. Similarly to \citet{kiaasl}, Fig. \ref{02} can assist the prediction of terrestrial planets in stable dynamical neighbourhoods that could host them.

As for the potentially detected systems consisting of an inner terrestrial planet and an outer giant planet, observational data could likewise be  constrained or complemented. To this end, the DS-maps produced for each MMR aim to be utilised per configuration, i.e. once two celestial bodies, which can be simulated by the ERTBP for interior MMRs, are found evolving close to an MMR, then according to their alignment $(\Delta\varpi=0)$ or anti-alignment $(\Delta\varpi=\pi)$ the errors in the observational values of the eccentricities and the angles can be restricted within the dark-coloured regions. Therefore, the planetary system would be dynamically hosted in a neighbourhood enabling it to stably evolve for long-time spans. 

This work is a continuation of \citet{kiaasl}, where 93 single-planet giant systems discovered to-date were scrumptiously scrutinized. The computation of periodic orbits in the N-body problem for $N>3$ would certainly be imperative for the study of multiple exoplanetary systems being discovered, in order for a possible terrestrial planet to be well-constrained within a regular domain.

\vspace{1cm}
\noindent
{\bf Acknowledgements.} The work of KIA was supported by the Fonds de la Recherche Scientifique-FNRS under Grant No. T.0029.13 (“ExtraOrDynHa” research project) and the University of Namur. Computational resources have been provided by the Consortium des \'Equipements de Calcul Intensif (C\'ECI), funded by the Fonds de la Recherche Scientifique de Belgique (F.R.S.-FNRS) under Grant No. 2.5020.11.
 
\vspace{1cm}
\noindent
{\bf Conflict of Interest.} The authors declare that they have no conflict of interest. 
  
\bibliographystyle{plainnat}
\bibliography{nbib}


\begin{appendix}
\section{Maps of Dynamical stability for the exact MMRs}

\label{appendix}

\begin{figure}[H]
\begin{center}
$\begin{array}{cp{-2cm}c}
\includegraphics[width=6.0cm]{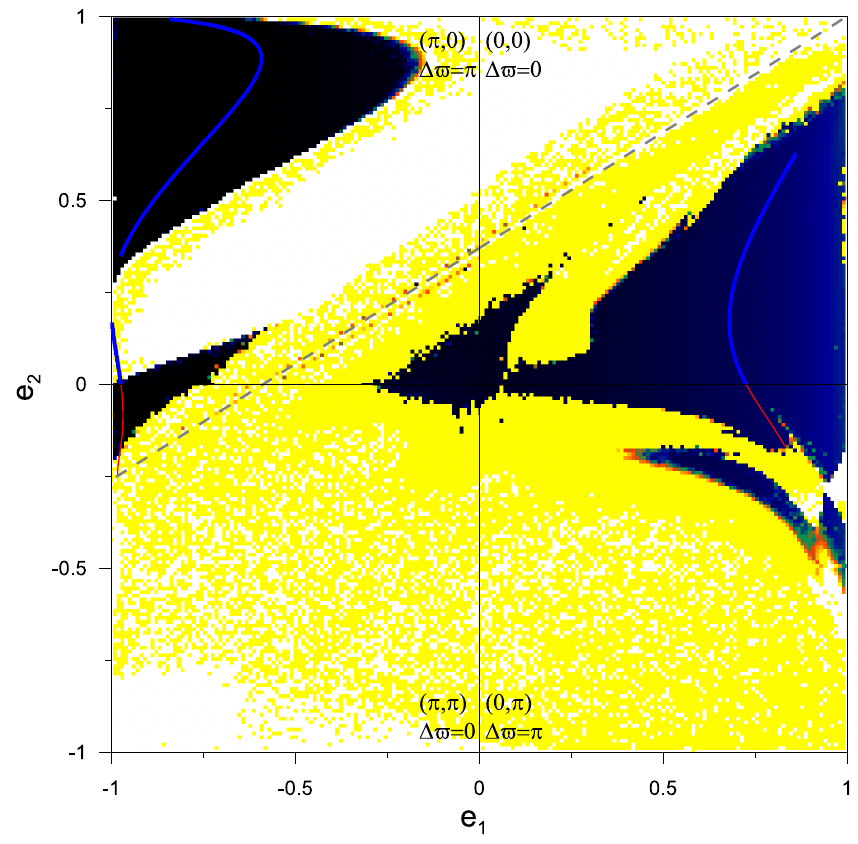} & \qquad&  \includegraphics[width=6.0cm]{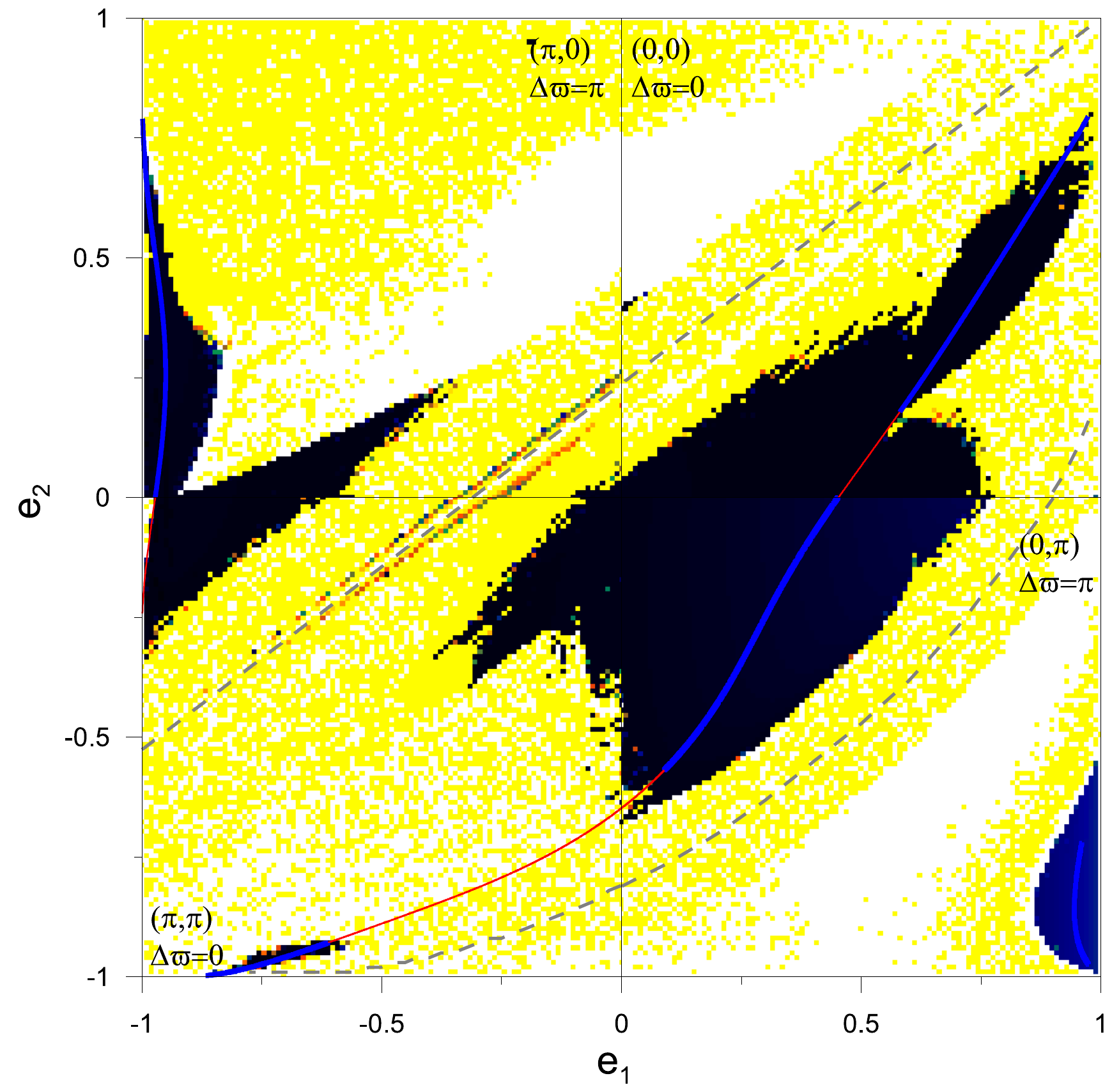}\\
\textnormal{(a)} & \qquad & \textnormal{(b)}\\ 
\includegraphics[width=6cm]{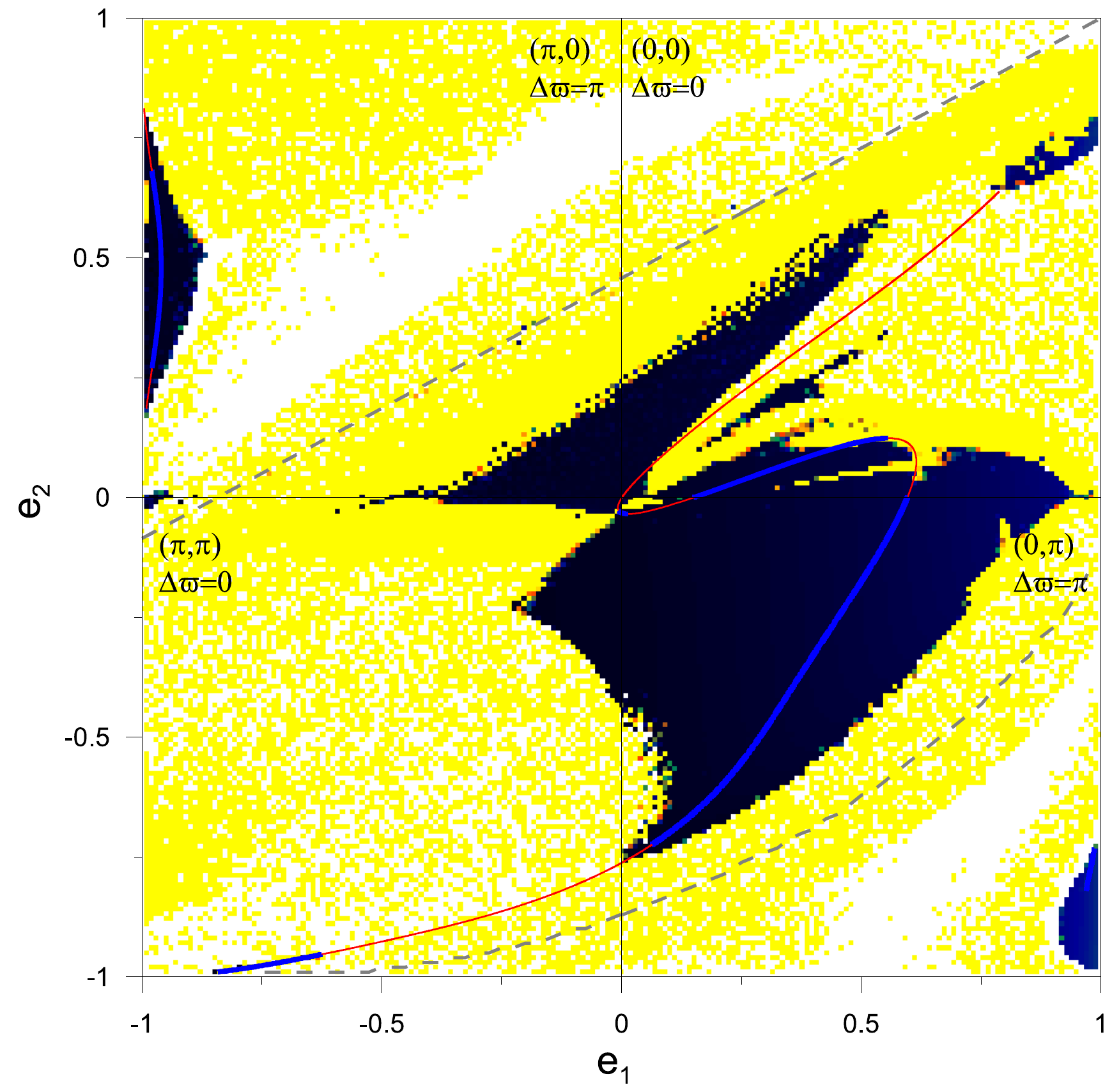} & \qquad& \includegraphics[width=6cm]{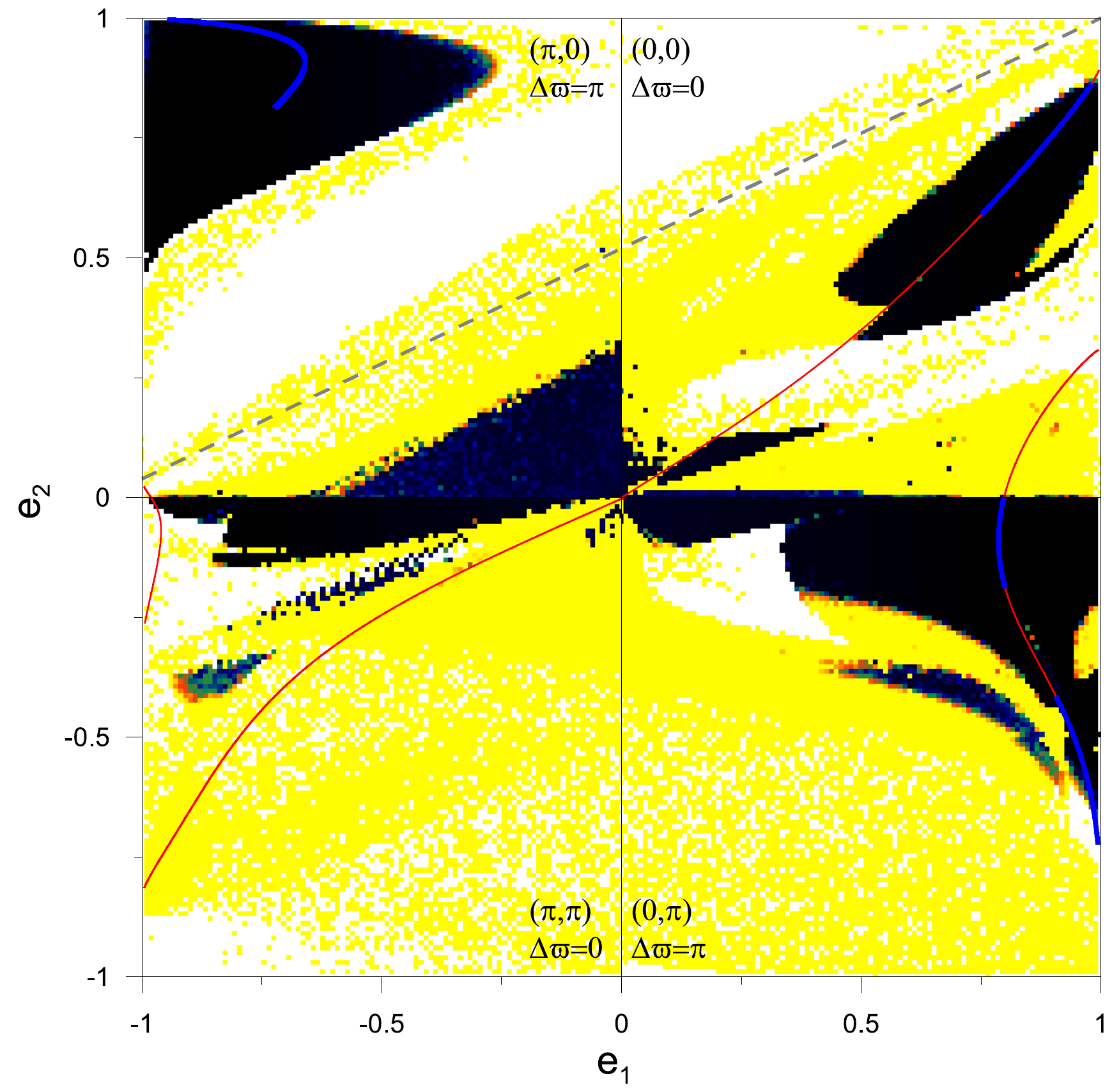} \\ 
\textnormal{(c)} & \qquad & \textnormal{(d)} \\
\includegraphics[width=6cm]{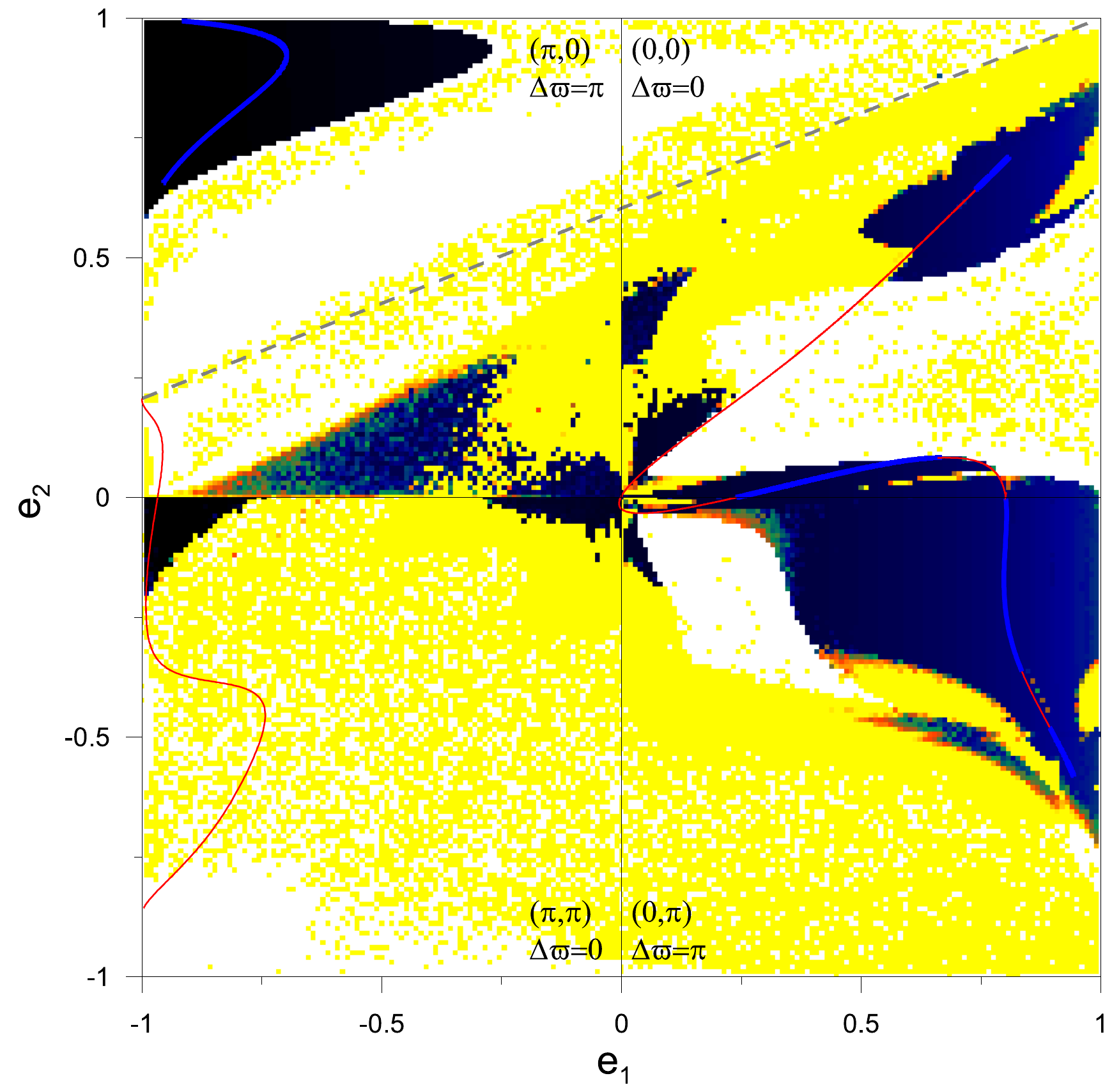} & \qquad& \includegraphics[width=6cm]{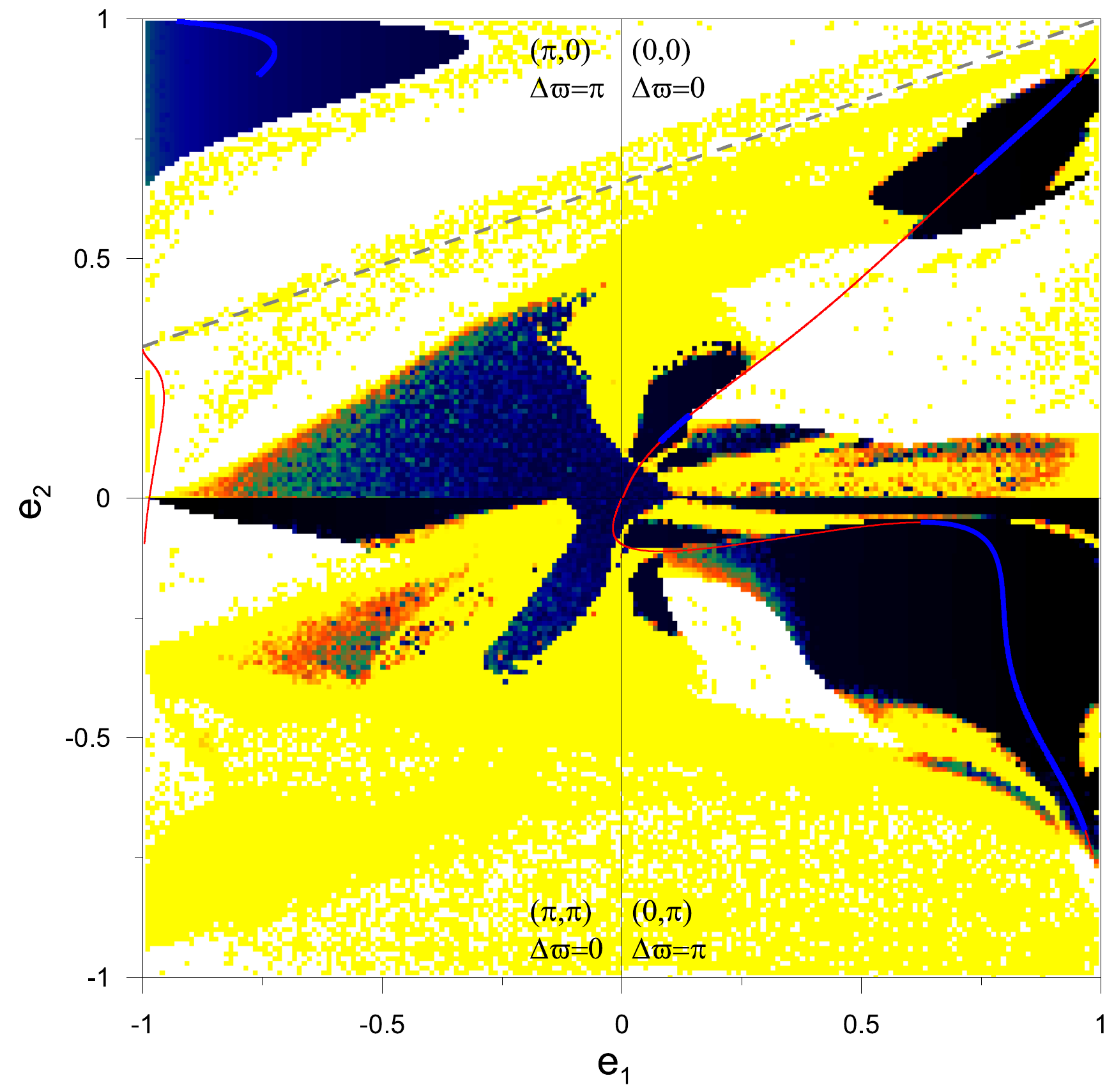}  \\
\textnormal{(e)} & \qquad & \textnormal{(f)} 
\end{array} $
\end{center}
\caption{DS-maps computed for the exact MMR  for the four symmetric configurations with \textbf{a} $a_1=2/1^{-2/3}$, \textbf{b} $a_1=3/2^{-2/3}$, \textbf{c} $a_1=5/2^{-2/3}$, \textbf{d} $a_1=3/1^{-2/3}$, \textbf{e} $a_1=4/1^{-2/3}$ and \textbf{f} $a_1=5/1^{-2/3}$. Overplotted are the families of periodic orbits in each MMR. The resolution of the initial conditions per configuration is $200\times200$}
\label{4m}
\end{figure} 
\end{appendix}

\end{document}